\documentclass[
reprint,
superscriptaddress,
 amsmath,amssymb,
 aps,
 prl,
floatfix,
]{revtex4-2}

\usepackage{graphicx}
\usepackage{pdfpages}
\usepackage{todonotes}
\usepackage{dcolumn}
\usepackage{bm}
\usepackage{hyperref}
\hypersetup{
    colorlinks=true,
    linkcolor=blue,
    filecolor=magenta,      
    urlcolor=blue,
    citecolor=blue,
}
\usepackage{siunitx}
\usepackage{xcolor}
\DeclareSIUnit\angstrom{\text{Å}}

\makeatletter
\AtBeginDocument{\let\LS@rot\@undefined}
\makeatother

\begin{document}

\preprint{Dispersing orbital excitations in 2D}

\title{
Collective nature of orbital excitations in layered cuprates\\
in the absence of apical oxygens 
}

\author{Leonardo Martinelli}
\altaffiliation[Present address:  ]{Physik-Institut, Universität Zürich, Winterthurerstrasse 190, CH-8057, Zürich, Switzerland\looseness=-1}
\email{leonardo.martinelli@polimi.it}
\affiliation{Dipartimento di Fisica, Politecnico di Milano, piazza Leonardo da Vinci 32, I-20133 Milano, Italy}
\author{Krzysztof Wohlfeld}%
\email{Krzysztof.Wohlfeld@fuw.edu.pl}
\affiliation{Institute of Theoretical Physics, Faculty of Physics, University of Warsaw, Pasteura 5, PL-02093 Warsaw, Poland}
\author{Jonathan Pelliciari}
\affiliation{National Synchrotron Light Source II, Brookhaven National Laboratory, Upton, New York 11973, USA}
\affiliation{Department of Physics, Massachusetts Institute of Technology, Cambridge, MA, USA}
\author{Riccardo Arpaia}
\affiliation{Quantum Device Physics Laboratory, Department of Microtechnology and Nanoscience, Chalmers University of Technology, SE-41296 Göteborg, Sweden}
\author{Nicholas B. Brookes}%
\affiliation{ESRF—The European Synchrotron, 71 Avenue des Martyrs, CS 40220, F-38043 Grenoble, France\looseness=-1}
\author{Daniele Di Castro}
\affiliation{CNR-SPIN and Dipartimento di Ingegneria Civile e Ingegneria Informatica, Università di Roma Tor Vergata, Via del Politecnico 1, I-00133 Roma, Italy}
\author{Mirian G. Fernandez}
\affiliation{Diamond Light Source, Harwell Campus, Didcot OX11 0DE, United Kingdom\looseness=-1}
\author{Mingu Kang}
\affiliation{Department of Physics, Massachusetts Institute of Technology, Cambridge, MA, USA}
\author{Yoshiharu Krockenberger}
\affiliation{NTT Basic Research Laboratories, NTT Corporation, Atsugi, Kanagawa, 243-0198, Japan}
\author{Kurt Kummer}%
\affiliation{ESRF—The European Synchrotron, 71 Avenue des Martyrs, CS 40220, F-38043 Grenoble, France\looseness=-1}
\author{Daniel E. McNally}
\affiliation{Photon Science Division, Paul Scherrer Institut, 5232 Villigen PSI, Switzerland}
\author{Eugenio Paris}
\affiliation{Photon Science Division, Paul Scherrer Institut, 5232 Villigen PSI, Switzerland}
\author{Thorsten Schmitt} 
\affiliation{Photon Science Division, Paul Scherrer Institut, 5232 Villigen PSI, Switzerland}
\author{Hideki Yamamoto}
\affiliation{NTT Basic Research Laboratories, NTT Corporation, Atsugi, Kanagawa, 243-0198, Japan}
\author{Andrew Walters}
\affiliation{Diamond Light Source, Harwell Campus, Didcot OX11 0DE, United Kingdom\looseness=-1}
\author{Ke-Jin Zhou}
\affiliation{Diamond Light Source, Harwell Campus, Didcot OX11 0DE, United Kingdom\looseness=-1}
\author{Lucio Braicovich}
\affiliation{Dipartimento di Fisica, Politecnico di Milano, piazza Leonardo da Vinci 32, I-20133 Milano, Italy}
\affiliation{ESRF—The European Synchrotron, 71 Avenue des Martyrs, CS 40220, F-38043 Grenoble, France\looseness=-1}
\author{Riccardo Comin}
\affiliation{Department of Physics, Massachusetts Institute of Technology, Cambridge, MA, USA}
\author{Marco Moretti Sala}
\affiliation{Dipartimento di Fisica, Politecnico di Milano, piazza Leonardo da Vinci 32, I-20133 Milano, Italy}
\author{Thomas P. Devereaux}%
\affiliation{Stanford Institute for Materials and Energy Sciences, SLAC, Menlo Park, California 94025, USA}
\affiliation{Department of Materials Science and Engineering, Stanford University, Stanford, California 94305, USA}
\affiliation{Geballe Laboratory for Advanced Materials, Stanford University, Stanford, California 94305, USA}
\author{Maria Daghofer}%
\affiliation{Institute for Functional Matter and Quantum Technologies, University of Stuttgart, Pfaffenwaldring 57, D-70550 Stuttgart, Germany}
\affiliation{Center for Integrated Quantum Science and Technology, University of Stuttgart, Pfaffenwaldring 57, D-70550 Stuttgart, Germany}
\author{Giacomo Ghiringhelli}%
\email{giacomo.ghiringhelli@polimi.it}
\affiliation{Dipartimento di Fisica, Politecnico di Milano, piazza Leonardo da Vinci 32, I-20133 Milano, Italy}
\affiliation{CNR-SPIN, Dipartimento di Fisica, Politecnico di Milano, I-20133 Milano, Italy}

\date{\today}

\begin{abstract}
We have investigated the $3d$ orbital excitations in CaCuO$_2$ (CCO), Nd$_2$CuO$_4$ (NCO), and La$_2$CuO$_4$ (LCO) using high-resolution resonant inelastic x-ray scattering. In LCO they behave as well-localized excitations, similarly to several other cuprates. On the contrary, in CCO and NCO the 
$d_{xy}$ orbital clearly disperse, pointing to a collective character of this excitation (orbiton) in compounds without apical oxygen. 
We ascribe the origin of the dispersion as stemming from
a substantial next-nearest-neighbor (NNN) orbital superexchange.
Such an exchange leads to the liberation of the orbiton from its coupling to magnons, which is associated with the orbiton hopping between {\it nearest neighbor} copper sites.
Finally, we show that the exceptionally large NNN orbital superexchange can be traced back to the absence of apical oxygens suppressing the charge transfer energy.
\end{abstract}

\maketitle


\emph{Introduction} -- Strongly correlated electron materials display simultaneously the presence of a strong, localizing repulsion between the $d$-electrons of the metal cations and large hopping integrals mediated by the ligand anions, which instead tend to delocalize the carriers~
\cite{Khomskii2014}.
These opposite tendencies lead to the appearance of a plethora of electronic orders and broken symmetries, as well as the
the emergence of collective excitations with various quantum numbers and complex origin~\cite{Auerbach1994, Venema2016, Powell2020}. 

Understanding the possible onset of such collective excitations is particularly challenging in the case of 
the $d$-orbital excitations. On one hand, the large hopping elements give rise to dispersive bands enumerated by the $d$-orbital quantum numbers. On the other hand, 
the strong Coulomb repulsion suppresses charge mobility and favors local, atomic-like, orbital excitations. In fact, 
it has been widely believed that the orbital ($dd$) excitations 
in the Mott insulating two-dimensional (2D) cuprates are purely local and well-described using the single-ion picture \cite{sala2011energy}.
Here we show that, similarly to electronic charge \cite{Ishii2014, Lee2014} and spin \cite{Coldea2001, LeTacon2011, Dean2013}, also the cuprate $d$-orbital degree of freedom {\it can} display a collective nature -- for
we observe the long-sought collective orbital excitations (orbitons) \cite{kugel1982jahn, tokura2000orbital} in CaCuO$_2$ (CCO) and Nd$_2$CuO$_4$ (NCO).
\begin{figure*}[thbp]
    \centering
    \includegraphics[width = \textwidth]{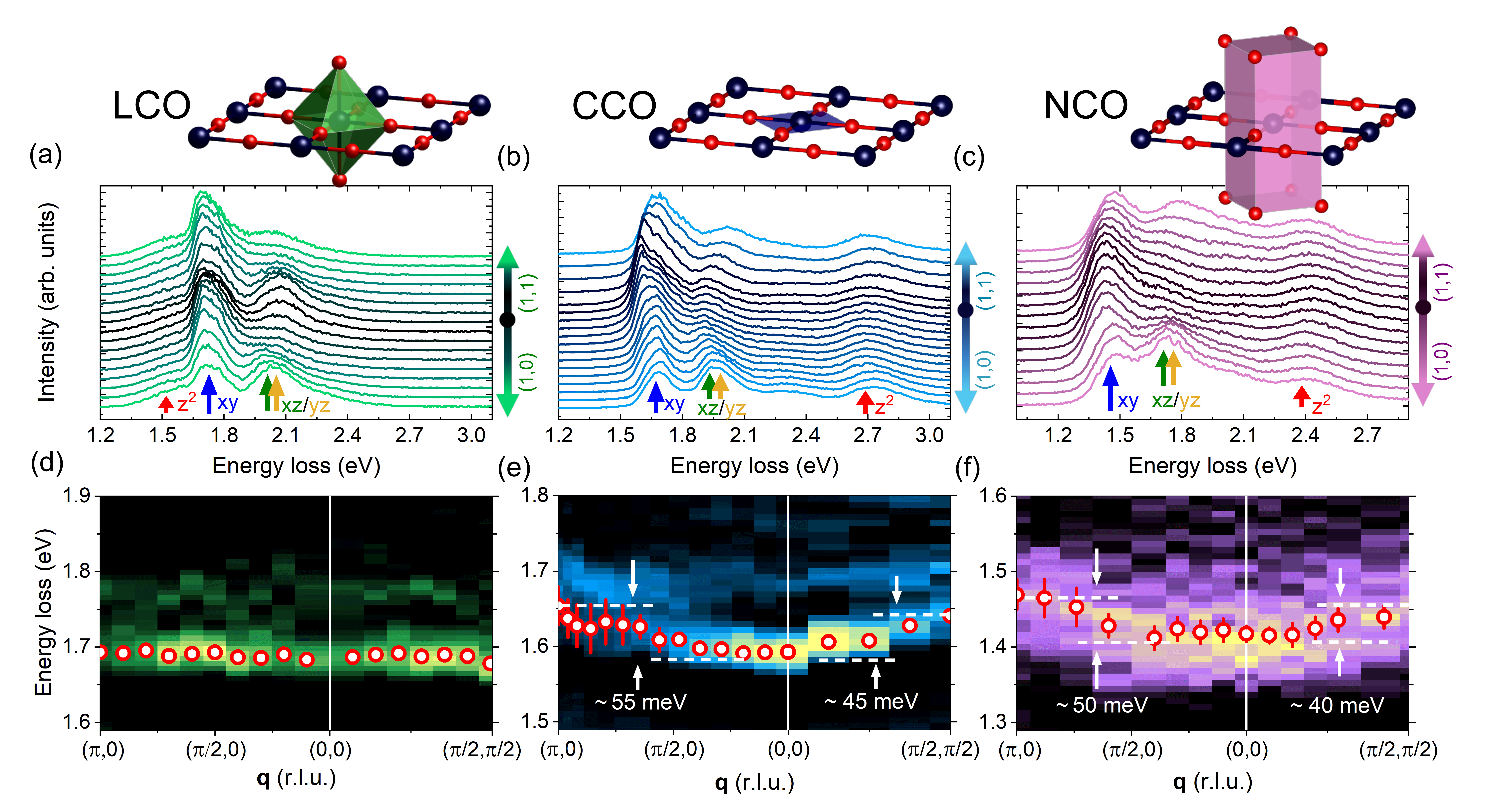}
    \caption{Overview of the Cu L$_3$ RIXS data, measured with $\sigma$ incident polarization, for LCO, CCO and NCO. Panels (a)-(c): below the respective scheme of the Cu coordination, we show the stack of RIXS spectra as a function of momentum $\mathbf{q}$ along the $(1,0)$ and $(1,1)$ directions. The labels indicate the symmetry of the peaks as determined from comparison with single-ion calculations \cite{sala2011energy, SupplementaryMat}. 
    (d)-(f): map of the second derivative of the scans of panels (a)-(c), 
    zoomed on the energy range of the $xy$ excitation. Red dots highlight the position of local maxima, which correspond to a peak in the original RIXS scans. Note that the energy scale is the same in the three panels although centered at different absolute energies.}
    \label{fig:Fig1}
\end{figure*}

Evidence of collective orbital excitations (orbitons) has been found in one-dimensional (1D) cuprates, where the reduced dimensionality leads to the fractionalization of the electron, i.e., the effective separation of its charge, spin, and orbital components that can propagate independently from each other. 
Experiments on chain and ladder cuprates \cite{schlappa2012spin, bisogni2015orbital, fumagalli2020mobile}) as well as calculations \cite{wohlfeld2011intrinsic, Wohlfeld2013, heverhagen2018spinon} have shown that the exciton formed by the promotion of an electron to a higher-energy $d$ orbital can propagate coherently through the 
antiferromagnetic (AFM) 
background, 
in a very similar fashion to a single hole in the AFM state~\cite{Kim1996, Kim2006}. 
The resulting excitation leads to a lens-like dispersion map, with a dominant signal at its lower edge displaying a maximum in energy at the $\Gamma$ point and a minimum at the magnetic zone boundary \cite{schlappa2012spin, fumagalli2020mobile}. 

In two-dimensional (2D) and three-dimensional (3D) compounds,
the fractionalization of electrons is still under debate.
In AFM systems, the motion of the orbiton generates a `trail' of magnetic excitations that  prevents the independent propagation of spinons and orbitons~\cite{Wohlfeld2012dispersion, kim2014excitonic}.
In Kugel-Khomskii systems with ferromagnetic (FM) order, and therefore no coupling between orbitons and magnons,
a theoretical prediction that {\it disregards} coupling to the lattice suggests the possibility of itinerant orbitons \cite{vandenbrink1999orbital}. However, the coupling to the lattice, which is stronger in 2D and 3D, further hinders the orbiton motion~\cite{Nasu2013vibronic, tanaka2004search, li2021unraveling}. 
In fact, the extensive  experimental search for orbitons in the FM manganites and titanates
\cite{tanaka2004search, saitoh2001observation, gruninger2002experimental, ulrich2009momentum} has so far been inconclusive.
In layered cuprates, the parent compounds of high-T$_\text{c}$ superconductors with  2D AFM order, orbital excitations have always been found to be localized in nature and interpreted as atomic transitions in the presence of a ligand field~\cite{sala2011energy, fumagalli2019polarization}. More in general, apart from the `special' 1D case mentioned above, dispersing orbitons 
in $3d$ transition-metal oxides have remained elusive 
\footnote{The observed orbiton dispersion in the {\it doped} cuprates~\cite{Ellis2015} has a distinct nature than the one reported here and probably can be attributed to the large itinerancy of the doped systems.},
resulting in a striking asymmetry between magnetic and orbital excitations. \footnote{Dispersing spin-orbitons have been found in $5d$ iridates \cite{kim2012magnetic, kim2014excitonic}, where however the orbital quantum number is no more a good one because of the spin-orbit coupling of iridium.}
In this letter, we report Cu $L_3$-edge resonant inelastic x-ray scattering (RIXS) measurements of the orbital excitations in three cuprate families: the single-layer La$_2$CuO$_4$ (LCO) and Nd$_2$CuO$_4$ (NCO), and the infinite-layer CaCuO$_2$ (CCO). 
While the orbital excitations in La$_2$CuO$_4$ show no sign of dispersion, in agreement with previous data \cite{sala2011energy}, in CCO and NCO we observe a clear collective nature of the $dd$ excitations with a dispersion larger than $50$\,meV along the $(1, 0)$ and the $(1, 1)$ crystallographic directions. 
%

%
\begin{figure}[thbp]
    \centering
    \includegraphics[width=0.45\textwidth]{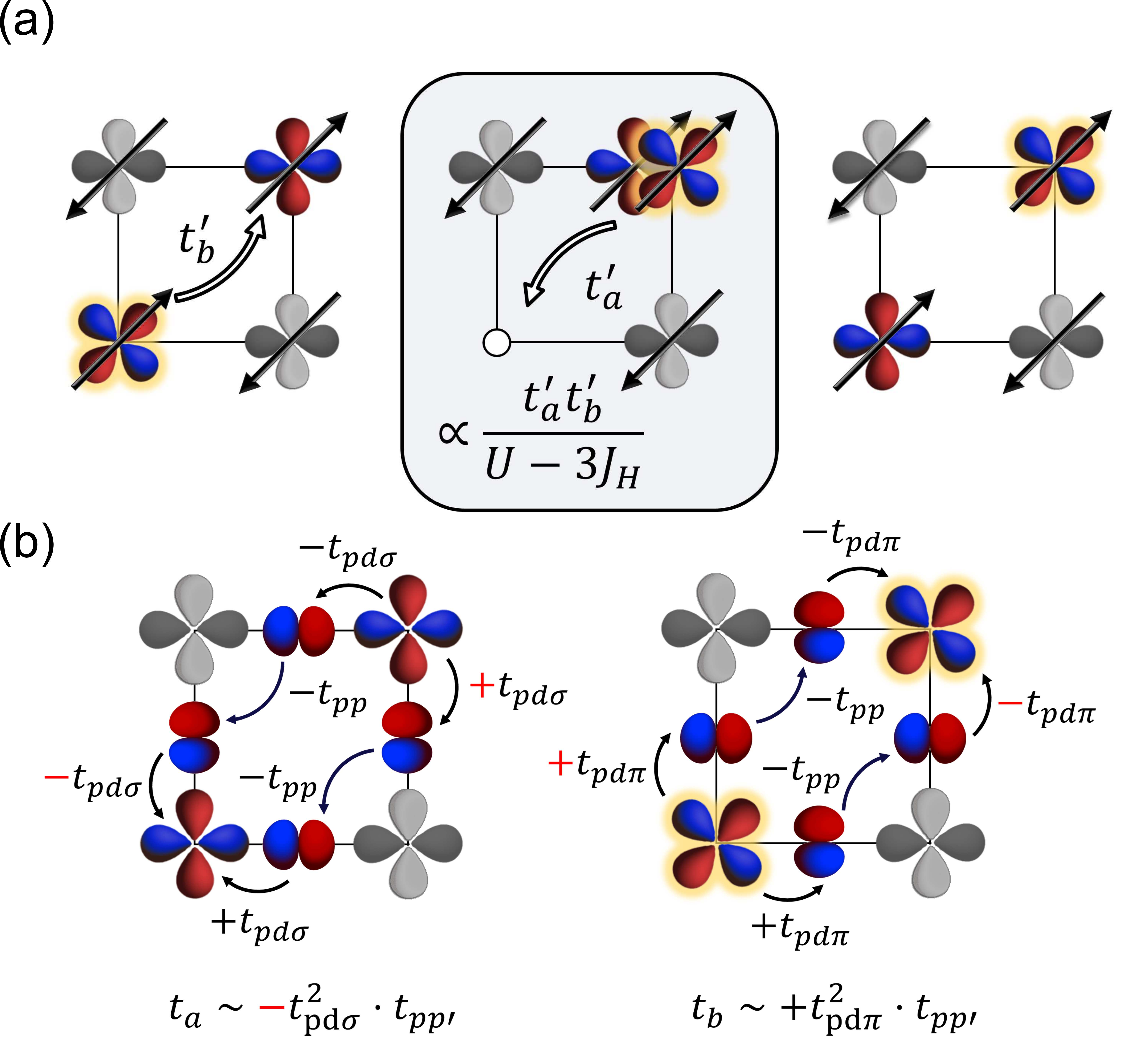}
    \caption{(a): Schematic representation of the next-nearest neighbour (NNN) $xy$ orbiton propagation in the hole language
    in 2D cuprates.
    The sketch on the left corresponds to the final state of the RIXS process. 
    The total amplitude of the orbiton superexchange process is divided by 
the energy $U-3J_H$ of the spin-triplet intermediate state.
    (b): Cartoon of the effective hopping between NNN $x^2-y^2$ orbitals ($t_a$, left panel) and $xy$ orbitals ($t_b$, right panel). The `Slater-Koster' hopping terms~\cite{Slater1954} between copper and oxygen orbitals ($t_{pd\sigma}$, $t_{pd\pi}$ and $t_{pp}$) are defined in the usual way, cf.~\cite{Jefferson1992, Wohlfeld2013, SupplementaryMat}. In both cases, the two different paths across the plaquette constructively interfere, but the number of negative signs (highlighted in red) is different in the two cases, so that $t_a t_b < 0$. }
    \label{fig:KKmodel}
\end{figure}

{\it Experimental results} -- The experiment has been performed on high-quality films of La$_2$CuO$_4$, CaCuO$_2$, and Nd$_2$CuO$_4$. Details of the preparation are described in \cite{bozovic2002epitaxial, Di_Castro_2014superconductivity, dicastro2015high, krockenberger2013emerging,krockenberger2012unconventional, SupplementaryMat}. Spectra of LCO and CCO were measured at the beamline ID32 \cite{brookes2018beamline} of the European Synchrotron ESRF, France, and those of NCO at the beamline I21 of the Diamond Light Source, UK 
\cite{zhou2022I21}. The scattering angle was kept fixed at \ang{150}, and in-plane momentum scans were performed by changing the incident angle $\theta$. The energy was fixed at the copper L$_3$ resonance ($\sim931$ eV). We employed incident linear-vertical ($\sigma$) and linear-horizontal $(\pi$) polarizations. 
The total experimental line width was $\sim\!40$ meV for CCO and LCO, $\sim\!50$ meV for NCO. 
The temperature was kept fixed at 20 K.
The three compounds share the same in-plane CuO$_2$ square lattice structure but differ in the out-of-plane Cu-O coordination, as shown in the insets of Fig.\,\ref{fig:Fig1}. In LCO the Cu$^{2+}$ ion is directly bound to two out-of-plane (apical) oxygen atoms to form elongated octahedra.
The infinite-layer CCO has no out-of-plane oxygen at all. And in the T' structure of NCO the CuO$_2$ planes are separated by Nd-O layers with oxygen not in apical position 
but above and below the in-plane oxygens. 
Panels (a)-(c) of Fig.\,\ref{fig:Fig1} display the momentum dependence of the orbital excitations in LCO (left), CCO (centre), and NCO (right) along the high-symmetry directions $(1,0)$ and $(1,1)$, acquired with $\sigma$ incident polarization. 
The orbital spectra are composed by three main features, which correspond to the transitions between the ($x^2\!-\!y^2$) ground state and the other $3d$ orbitals ($xy, xz/yz, z^2$) split by the tetragonal crystal field. The excitation energy is similar among the 3 cases for the ($xy$) case,  and differs more for the other orbitals, more influenced by the out-of-plane coordination \cite{sala2011energy}.
As the first ligand-field excited state is at very high energy ($\sim\!1.5$\,eV) all three compounds are characterized by a rigid ferro-orbital order with ($x^2\!-\!y^2$) symmetry.
In turn, the virtual hopping of the single hole of the Cu atom determines strong AFM interactions between neighbouring sites (in agreement with the Goodenough-Kanamori rules), and indeed all samples display dispersing magnetic excitations at energies $<0.4$ eV (not shown in Fig. \ref{fig:Fig1}). 
Their intensity evolution with momentum is mostly due to the RIXS matrix elements, which depend on photon polarization and scattering angles \cite{sala2011energy}. 

A closer inspection reveals that, while the energy of the peaks in LCO is independent of $\mathbf{q}$, the energy of the $xy$ and $xz/yx$ excitations shows an appreciable dispersion for CCO and NCO. For a quantitative estimation we extracted the second derivative of the RIXS spectra [Fig. \ref{fig:Fig1}(d-f)]. For LCO the $xy$ excitation shows no dispersion within error-bars, while for CCO it exhibits a dispersion of $55\!\pm\!15$ meV and $50\!\pm\!5$ meV along the (1,0) and (1,1) directions, respectively. 
The energy is maximum at $(\pi,0)$ and $(\frac{\pi}{2},\frac{\pi}{2})$ and minimum at the $\Gamma$ point (0,0). A similar, though smaller, dispersion is observed for NCO ($50\pm20$ meV and $40\pm10$ meV, respectively). This behavior is at odds to what has been observed \cite{schlappa2012spin, fumagalli2020mobile} and calculated \cite{wohlfeld2011intrinsic} in 1D compounds, where the $dd$ excitations' lowest energy edge disperses with a $\pi$-periodicity and reaches its maximum energy at $\Gamma$. For the three samples, a satellite of the main dispersing $xy$ peak, about 50-70 meV higher in energy, shows very limited or zero dispersion and an asymmetric lineshape. 
The ($xz/yz$) orbital also shows a marked dispersion especially along the $(1,1)$ direction, with again a minimum at the $\Gamma$ point (see panels (b) and (c)).
Interestingly, for both CCO and NCO an additional broad feature is evident at $\sim 300-400$\,meV above the ($xz/yz$) peak. The momentum dependence of their integrated spectral weight is similar to that of the main ($xz/yx$) peak, and both agree with the single-ion calculations for that final state (see Supplementary Material \cite{SupplementaryMat}). These observations indicate that the ($xz/yz$) excitations are very broad in energy and cannot be identified with just a single, though dispersing, peak. The ($z^2$) excitation is quite broad too 
and it is more difficult to identify an associated dispersion.
In previous investigations of CCO by RIXS \cite{sala2011energy, peng2017influence} the dispersion of orbital excitations was overlooked either because of insufficient energy resolution (240 meV in Ref. \onlinecite{sala2011energy}) or because the focus of the articles was on on spin excitations in the mid-IR energy range, not the $dd$ excitation in the eV range  \cite{peng2017influence, martinelli2022fractional}. Here \cite{SupplementaryMat} we have exploited a better combination of scattering geometries (grazing-incidence and reflection) and incident polarizations ($\sigma$ and $\pi$) to assess the dispersion of $dd$ excitations, and excluded that the observed effect is the result of multiple peaks at adjacent energies following different intensity dependencies on momentum.

{\it Orbital superexchange model --}
The dependence on the wave-vector of the $dd$ excitation energy is so far unreported in layered cuprates, and in general in two-dimensional $3d$ transition metal compounds. Phenomenologically, the different behaviour of LCO and CCO/NCO clearly correlates to the presence or absence of apical oxygens. On the theoretical level, our observations can be linked to the reduced value of $\Delta$, the charge-transfer energy between Cu ($3d$, $4s$) and in-plane O ($2p$) states, in the absence of apical ligands \cite{ohta1991charge, Weber2012,pavarini2001band}. A smaller $\Delta$, in turn, increases the nearest-neighbour and longer-range Cu-Cu hopping amplitudes $t$, $t'$ \cite{peng2017influence, pavarini2001band, Weber2012}.
In fact, as discussed below, a similar situation concerns also the orbiton hopping elements -- which turn out to be significantly increased in CCO / NCO  w.r.t. LCO.
A simple 2-site model can reproduce the measured dispersion, but fails to give a proper physical description as it intrinsically neglects the presence of the underlying 2D AFM lattice (see \cite{SupplementaryMat}).

Our starting point is, therefore,
the Kugel-Khomskii--type spin-orbital model~\footnote{For simplicity, we focus below on the $xy$ excitation leaving an even more complex~\cite{SupplementaryMat} $xz/yz$ case for further studies.}, where orbital excitations can move from site to site thanks to a perturbative three-step superexchange (SE) processes \cite{Khomskii2014}. 
In 2D and 3D AFM systems, the nearest-neighbor (NN) SE is impeded by the strong interaction between orbitons and magnons \cite{Wohlfeld2012dispersion}, what is known as magnetic string effect (see \cite{SupplementaryMat} for an intuitive explanation). The orbiton dispersion through this mechanism is 
effectively
forbidden.  
Therefore, 
and further 
stimulated by the experimental evidence, 
we propose an orbital SE process between next-nearest-neighbor (NNN) sites, see Fig.~\ref{fig:KKmodel}(a).
%
We focus our analysis on the $xy$ excitation. The $xy$ hole can move to a NNN site through a hopping integral $t'_b$; the NNN site becomes thus occupied by two holes with different symmetry ($x^2\!-\!y^2$ and $xy$) and parallel spin, which costs energy $U-3J_H$ \cite{Wohlfeld2013}; finally the  $x^2\!-\!y^2$ hole can move by hopping $t'_a$ to the original site, resulting in the SE parameter $J^{\rm orb}_{\rm NNN}\propto t'_a t'_b / (U-3J_H)$.
Altogether, such an orbital SE model leads to an orbiton dispersion relation unaffected by magnons: 
\begin{align}\label{eq:orb}
\varepsilon_{\bf k} =  2 J^{\rm orb}_{\rm NNN} \cos k_x \cos k_y.
\end{align}
%

%

Interestingly, 
the sign of the 
$J^{\rm orb}_{\rm NNN}$
orbital exchange is always {\it negative} in the 2D cuprates, though its modulus depends on the crystalline structure (see below).
Indeed, the sign is determined by the relative phase factors of the considered oxygen and copper orbitals, as schematized in Fig.~\ref{fig:KKmodel}\textcolor{blue}{(b)}. The Cu-Cu hopping integrals $t'_a$ and $t'_b$ have opposite sign, due to the distinct signs before the $t_{pd \sigma}$ and $t_{pd \pi}$ hoppings for the respective bonding copper and oxygen orbitals. 
Ultimately, the sign depends on the fact that the $xy$ orbital has its lobes, and therefore phases, rotated by \ang{45} with respect to the ($x^2\!-\!y^2$) orbital.
It is important to note that the opposite sign of $t'_a$ and $t'_b$ is obtained in the cuprate charge-transfer (Emery) model, whereas in a canonical multi-orbital Hubbard model this sign is assumed to be the same~\cite{Khomskii2014}.

\begin{figure}[tbp]
    \centering
    \includegraphics[width =0.95\columnwidth]{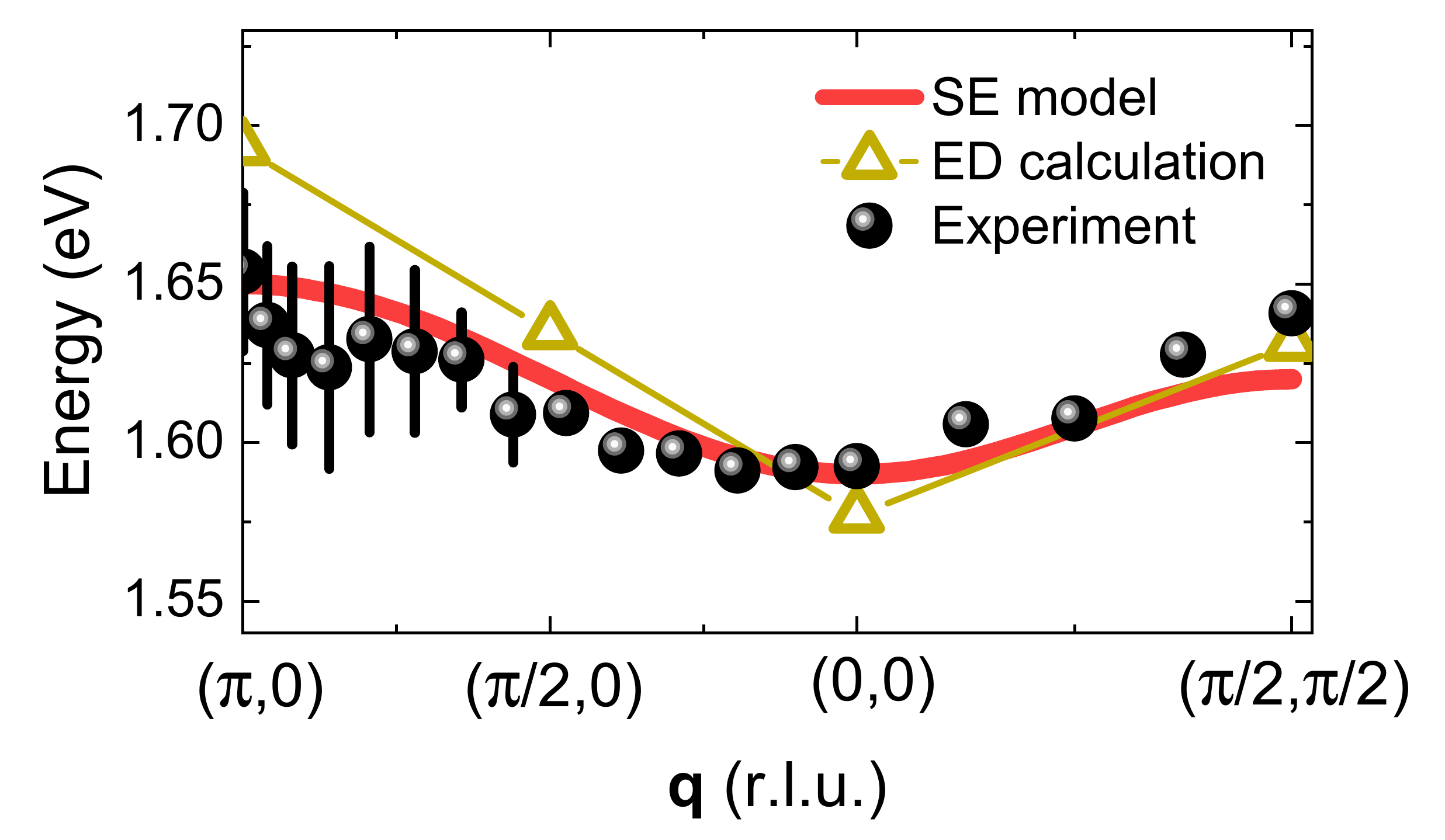}
    \caption{Dispersion relation $\varepsilon_k$, for the ($xy$) orbiton in CCO, obtained from the RIXS experiment (points), the SE model (red line)
    Eq.~(\ref{eq:orb}),
    and from the numerical exact diagonalization (ED) of the multi-orbital Hubbard model on a $4 \times 4$ cluster (yellow points and line). See text for further details.}
    \label{fig:theory}
\end{figure}
%

{\it Validity of the model -- }
As shown in Fig.~\ref{fig:theory}(a), the calculated dispersion of the ($xy$) orbiton using Eq.~\eqref{eq:orb} agrees very well with the experimental RIXS data of CCO. 
Note that for CCO, the relatively large oxygen-oxygen hopping $t_{pp} \approx 0.7 $ eV~\cite{barivsic2022high, pavarini2001band} and small charge transfer energy $\Delta \approx 1.8 $ eV~\cite{perucchi2018electrodynamic, Weber2012, kan2004preparation, yoshida1992two} lead to the estimation $J^{\rm orb}_{\rm NNN} \approx -15 $ meV by state-of-the-art cell perturbation theory~\cite{Jefferson1992} applied to the charge transfer model
(see SM \cite{SupplementaryMat} for details, which includes Refs.~\cite{Wohlfeld2013,Botana2020, Weber2012, neudert2000four, barivsic2022high, pavarini2001band, kim2003unusual}).
Moreover, our model can account for the difference between LCO and the other two compounds. In fact, the presence of apical oxygens in LCO raises the charge-transfer energy to $\Delta\approx 2.6$ eV \cite{Weber2012, watanabe2021unified}, decreases the covalency and leads to $|J^{\rm orb}_\text{NNN}| < 6$ meV, i.e. 
at least two-and-a-half times
smaller than in  CCO \cite{SupplementaryMat}. 
The presence of apicals should also decrease the oxygen-oxygen $t_{pp}$~\cite{SupplementaryMat}, further reducing $|J^{\rm orb}_\text{NNN}|$.
This explains why the orbiton dispersion in LCO falls below the current experimental sensitivity.
%

To verify the assumption that orbiton and magnons effectively decouple [which has lead to Eq.~(\ref{eq:orb})], we performed exact diagonalization (ED) of a realistic multiorbital Hubbard model \cite{SupplementaryMat}. Such calculations inherently account for all possible SE processes, including magnon-orbiton coupling. The latter could lead to spectral weight transfer from the orbiton quasiparticle to the continuum and to a renormalization of the orbiton dispersion ~\cite{kim2012magnetic, kim2014excitonic, Wohlfeld2012dispersion, wohlfeld2011intrinsic}. Both effects, as shown in the Supplementary Material \cite{SupplementaryMat}, turn out to be very small.
The dispersion calculated with the ED calculations follows semi-quantitatively the one predicted with the SE model, and is reported with yellow triangles in Fig.~\ref{fig:theory}.
Moreover, the surprisingly small incoherent part of the ED spectrum is barely visible in the asymmetric shape of the line spectra, as reported in the Supplementary Material \cite{SupplementaryMat}.
This good agreement is due to the different kinetic energies of magnon and orbiton (the former being faster than the latter) and to the dominant NNN 
orbital exchange.

As explained above, the experimental spectra also reveal the presence of a non-dispersing peak $\sim$\SI{70}{\milli\electronvolt} above the dispersing $xy$ excitation. The ED calculation shows indeed the presence of a continuum of magnetic origin caused by the NN SE interaction, but underestimate its intensity as described in the Supplementary Material \cite{SupplementaryMat}.
Another probable explanation is in terms of a ladder spectrum consisting of orbitons coupled to phonon satellites, 
which are always present in the $dd$ excitation spectra \cite{marciniakVibrationalCoherentControl2021}.
The energy separation between dispersing and non-dispersing $xy$ peak indeed agrees with the energy of bond-stretching oxygen modes \cite{andersen1996out}, which posses the strongest electron-phonon coupling among all branches and indeed dominate the low-energy RIXS response (see Supplementary Material \cite{SupplementaryMat}).
The dispersionless nature of the satellite peak can be rationalised by noting that, not only coupling the mobile orbitons to a massive local excitation would strongly reduce its dispersion \cite{Bieniasz2022,Bulaevskii1968, Martinez1991, Bohrdt2020}, but even in case of coupling to dispersive (e.g.~acoustic) phonons the observed dispersion is strongly renormalized and below our senstitivity ---as realistic calculations for coupling of cuprate electrons to phonons reveal~\cite{Devereaux_Shvaika2016, Lee2013}. Moreover, the unavoidable presence of multiple phonon harmonics (ladder spectrum) explains the asymmetric lineshape of the non-dispersing peak.

{\it Conclusions --} The observation of a sizable dispersion of orbital excitations in cuprates without apical oxygens
demonstrates that $dd$ excitations can have a collective nature beyond the 1D case.
The large orbiton dispersion is a consequence of a significant NNN orbital SE. Unlike the NN one, the NNN exchange takes place on the same AFM sublattice and allows for an almost free orbiton hopping, i.e.~without coupling between orbitons and magnons.
This mechanism is different from the 1D case, where the orbiton hops between NN copper sites and decouples from the magnetic excitations
solely due to the fractionalization of electrons in the 1D interacting systems~\cite{wohlfeld2011intrinsic, schlappa2012spin}. Such fractionalization in the orbital sector is instead absent in our 2D case.
The narrower orbiton bandwidth in 2D with respect to the 1D cases might be due to a smaller superexchange interaction~\cite{Walters2009, schlappa2012spin, martinelli2022fractional} following a larger  charge-transfer energy~\cite{Maiti1998}, but further studies are required for an exhaustive understanding of the phenomenon.
%
%

The exceptionally large NNN orbital superexchange can be traced back to the strong copper-oxygen covalency in the copper oxides without apical oxygens~\cite{Weber2012}. We note that longer-range hopping is at the origin of the peculiar properties of spin excitations of CCO, with spinon-like behavior emerging close to $(\pi,0)$, 
in analogy to 1D spin systems \cite{martinelli2022fractional}.
It turns out that a canonical Hubbard-like description cannot yield the observed sign of the orbital superexchange.  One has to go `back' to the charge transfer (Emery) model to get this sign correct, highlighting the fundamental role played by oxygen bands in the physics of 2D cuprates.
Lastly,
the 
observed
mobility of orbitons might explain the measured coupling between $dd$ excitations and doped holes \cite{fumagalli2019polarization, barantani2022resonant}. 
Recently, it has been demonstrated that such coupling can lead to an attractive interaction between holes, positively contributing to superconducting pairing \cite{barantani2023can}.
It would be therefore of interest to explore the effect of doping on the dispersion of orbital excitations.


{\it Acknowledgements --}
We would like to thank Claude Monney and Takami Tohyama for insightful discussions.
L.M., M. M. S. and G. G. acknowledge support by the project PRIN2017 “Quantum-2D” ID 2017Z8TS5B of the Ministry for University and Research (MIUR) of Italy. 
R.A. acknowledges support  by the Swedish Research Council (VR) under the Project 2020-04945. K.W. acknowledges support by Narodowe Centrum Nauki (NCN, National Science Center) under Project No. 2016/22/E/ST3/00560. J.~P. acknowledges financial support by the Swiss National Science Foundation Early Postdoc Mobility Fellowship Project No. P2FRP2\_171824 and P400P2\_180744. J.~P. was partially supported by the Laboratory Directed Research and Development project of Brookhaven National Laboratory No. 21-037. Work at MIT was supported by the Air Force Office of Scientific Research Young Investigator Program under grant FA9550-19-1-0063, and by the MIT-Italy Progetto Roberto Rocca. 
M.D., E.P. and T.S. acknowledge support by the Swiss National Science Foundation through SNSF research grant no. 160765. We acknowledge Diamond Light Source for providing the beam time at the I21-RIXS beamline under proposal SP20690.
%





%


\bibliography{orbitonsCCO}

\providecommand{\noopsort}[1]{}\providecommand{\singleletter}[1]{#1}%
\begin{thebibliography}{70}%
\makeatletter
\providecommand \@ifxundefined [1]{%
 \@ifx{#1\undefined}
}%
\providecommand \@ifnum [1]{%
 \ifnum #1\expandafter \@firstoftwo
 \else \expandafter \@secondoftwo
 \fi
}%
\providecommand \@ifx [1]{%
 \ifx #1\expandafter \@firstoftwo
 \else \expandafter \@secondoftwo
 \fi
}%
\providecommand \natexlab [1]{#1}%
\providecommand \enquote  [1]{``#1''}%
\providecommand \bibnamefont  [1]{#1}%
\providecommand \bibfnamefont [1]{#1}%
\providecommand \citenamefont [1]{#1}%
\providecommand \href@noop [0]{\@secondoftwo}%
\providecommand \href [0]{\begingroup \@sanitize@url \@href}%
\providecommand \@href[1]{\@@startlink{#1}\@@href}%
\providecommand \@@href[1]{\endgroup#1\@@endlink}%
\providecommand \@sanitize@url [0]{\catcode `\\12\catcode `\$12\catcode
  `\&12\catcode `\#12\catcode `\^12\catcode `\_12\catcode `\%12\relax}%
\providecommand \@@startlink[1]{}%
\providecommand \@@endlink[0]{}%
\providecommand \url  [0]{\begingroup\@sanitize@url \@url }%
\providecommand \@url [1]{\endgroup\@href {#1}{\urlprefix }}%
\providecommand \urlprefix  [0]{URL }%
\providecommand \Eprint [0]{\href }%
\providecommand \doibase [0]{https://doi.org/}%
\providecommand \selectlanguage [0]{\@gobble}%
\providecommand \bibinfo  [0]{\@secondoftwo}%
\providecommand \bibfield  [0]{\@secondoftwo}%
\providecommand \translation [1]{[#1]}%
\providecommand \BibitemOpen [0]{}%
\providecommand \bibitemStop [0]{}%
\providecommand \bibitemNoStop [0]{.\EOS\space}%
\providecommand \EOS [0]{\spacefactor3000\relax}%
\providecommand \BibitemShut  [1]{\csname bibitem#1\endcsname}%
\let\auto@bib@innerbib\@empty
\bibitem [{\citenamefont {Khomskii}(2014)}]{Khomskii2014}%
  \BibitemOpen
  \bibfield  {author} {\bibinfo {author} {\bibfnamefont {D.~I.}\ \bibnamefont
  {Khomskii}},\ }\href {https://doi.org/10.1017/CBO9781139096782} {\emph
  {\bibinfo {title} {Transition Metal Compounds}}}\ (\bibinfo  {publisher}
  {Cambridge University Press},\ \bibinfo {year} {2014})\BibitemShut {NoStop}%
\bibitem [{\citenamefont {Auerbach}(1994)}]{Auerbach1994}%
  \BibitemOpen
  \bibfield  {author} {\bibinfo {author} {\bibfnamefont {A.}~\bibnamefont
  {Auerbach}},\ }\href@noop {} {\emph {\bibinfo {title} {Interacting Electrons
  and Quantum Magnetism}}}\ (\bibinfo  {publisher} {Springer-Verlag},\ \bibinfo
  {year} {1994})\BibitemShut {NoStop}%
\bibitem [{\citenamefont {Venema}\ \emph {et~al.}(2016)\citenamefont {Venema},
  \citenamefont {Verberck}, \citenamefont {Georgescu}, \citenamefont {Prando},
  \citenamefont {Couderc}, \citenamefont {Milana}, \citenamefont {Maragkou},
  \citenamefont {Persechini}, \citenamefont {Pacchioni},\ and\ \citenamefont
  {Fleet}}]{Venema2016}%
  \BibitemOpen
  \bibfield  {author} {\bibinfo {author} {\bibfnamefont {L.}~\bibnamefont
  {Venema}}, \bibinfo {author} {\bibfnamefont {B.}~\bibnamefont {Verberck}},
  \bibinfo {author} {\bibfnamefont {I.}~\bibnamefont {Georgescu}}, \bibinfo
  {author} {\bibfnamefont {G.}~\bibnamefont {Prando}}, \bibinfo {author}
  {\bibfnamefont {E.}~\bibnamefont {Couderc}}, \bibinfo {author} {\bibfnamefont
  {S.}~\bibnamefont {Milana}}, \bibinfo {author} {\bibfnamefont
  {M.}~\bibnamefont {Maragkou}}, \bibinfo {author} {\bibfnamefont
  {L.}~\bibnamefont {Persechini}}, \bibinfo {author} {\bibfnamefont
  {G.}~\bibnamefont {Pacchioni}},\ and\ \bibinfo {author} {\bibfnamefont
  {L.}~\bibnamefont {Fleet}},\ }\bibfield  {title} {\bibinfo {title} {The
  quasiparticle zoo},\ }\href {https://doi.org/10.1038/nphys3977} {\bibfield
  {journal} {\bibinfo  {journal} {Nature Physics}\ }\textbf {\bibinfo {volume}
  {12}},\ \bibinfo {pages} {1085} (\bibinfo {year} {2016})}\BibitemShut
  {NoStop}%
\bibitem [{\citenamefont {Powell}(2020)}]{Powell2020}%
  \BibitemOpen
  \bibfield  {author} {\bibinfo {author} {\bibfnamefont {B.~J.}\ \bibnamefont
  {Powell}},\ }\bibfield  {title} {\bibinfo {title} {Emergent particles and
  gauge fields in quantum matter},\ }\href
  {https://doi.org/10.1080/00107514.2020.1832350} {\bibfield  {journal}
  {\bibinfo  {journal} {Contemporary Physics}\ }\textbf {\bibinfo {volume}
  {61}},\ \bibinfo {pages} {96} (\bibinfo {year} {2020})},\ \Eprint
  {https://arxiv.org/abs/https://doi.org/10.1080/00107514.2020.1832350}
  {https://doi.org/10.1080/00107514.2020.1832350} \BibitemShut {NoStop}%
\bibitem [{\citenamefont {Sala}\ \emph {et~al.}(2011)\citenamefont {Sala},
  \citenamefont {Bisogni}, \citenamefont {Aruta}, \citenamefont {Balestrino},
  \citenamefont {Berger}, \citenamefont {Brookes}, \citenamefont {De~Luca},
  \citenamefont {Di~Castro}, \citenamefont {Grioni}, \citenamefont {Guarise}
  \emph {et~al.}}]{sala2011energy}%
  \BibitemOpen
  \bibfield  {author} {\bibinfo {author} {\bibfnamefont {M.~M.}\ \bibnamefont
  {Sala}}, \bibinfo {author} {\bibfnamefont {V.}~\bibnamefont {Bisogni}},
  \bibinfo {author} {\bibfnamefont {C.}~\bibnamefont {Aruta}}, \bibinfo
  {author} {\bibfnamefont {G.}~\bibnamefont {Balestrino}}, \bibinfo {author}
  {\bibfnamefont {H.}~\bibnamefont {Berger}}, \bibinfo {author} {\bibfnamefont
  {N.~B.}\ \bibnamefont {Brookes}}, \bibinfo {author} {\bibfnamefont
  {G.}~\bibnamefont {De~Luca}}, \bibinfo {author} {\bibfnamefont
  {D.}~\bibnamefont {Di~Castro}}, \bibinfo {author} {\bibfnamefont
  {M.}~\bibnamefont {Grioni}}, \bibinfo {author} {\bibfnamefont
  {M.}~\bibnamefont {Guarise}}, \emph {et~al.},\ }\bibfield  {title} {\bibinfo
  {title} {{Energy and symmetry of dd excitations in undoped layered cuprates
  measured by Cu L3 resonant inelastic x-ray scattering}},\ }\href
  {https://iopscience.iop.org/article/10.1088/1367-2630/13/4/043026/meta}
  {\bibfield  {journal} {\bibinfo  {journal} {New Journal of Physics}\ }\textbf
  {\bibinfo {volume} {13}},\ \bibinfo {pages} {043026} (\bibinfo {year}
  {2011})}\BibitemShut {NoStop}%
\bibitem [{\citenamefont {Ishii}\ \emph {et~al.}(2014)\citenamefont {Ishii},
  \citenamefont {Fujita}, \citenamefont {Sasaki}, \citenamefont {Minola},
  \citenamefont {Dellea}, \citenamefont {Mazzoli}, \citenamefont {Kummer},
  \citenamefont {Ghiringhelli}, \citenamefont {Braicovich}, \citenamefont
  {Tohyama}, \citenamefont {Tsutsumi}, \citenamefont {Sato}, \citenamefont
  {Kajimoto}, \citenamefont {Ikeuchi}, \citenamefont {Yamada}, \citenamefont
  {Yoshida}, \citenamefont {Kurooka},\ and\ \citenamefont
  {Mizuki}}]{Ishii2014}%
  \BibitemOpen
  \bibfield  {author} {\bibinfo {author} {\bibfnamefont {K.}~\bibnamefont
  {Ishii}}, \bibinfo {author} {\bibfnamefont {M.}~\bibnamefont {Fujita}},
  \bibinfo {author} {\bibfnamefont {T.}~\bibnamefont {Sasaki}}, \bibinfo
  {author} {\bibfnamefont {M.}~\bibnamefont {Minola}}, \bibinfo {author}
  {\bibfnamefont {G.}~\bibnamefont {Dellea}}, \bibinfo {author} {\bibfnamefont
  {C.}~\bibnamefont {Mazzoli}}, \bibinfo {author} {\bibfnamefont
  {K.}~\bibnamefont {Kummer}}, \bibinfo {author} {\bibfnamefont
  {G.}~\bibnamefont {Ghiringhelli}}, \bibinfo {author} {\bibfnamefont
  {L.}~\bibnamefont {Braicovich}}, \bibinfo {author} {\bibfnamefont
  {T.}~\bibnamefont {Tohyama}}, \bibinfo {author} {\bibfnamefont
  {K.}~\bibnamefont {Tsutsumi}}, \bibinfo {author} {\bibfnamefont
  {K.}~\bibnamefont {Sato}}, \bibinfo {author} {\bibfnamefont {R.}~\bibnamefont
  {Kajimoto}}, \bibinfo {author} {\bibfnamefont {K.}~\bibnamefont {Ikeuchi}},
  \bibinfo {author} {\bibfnamefont {K.}~\bibnamefont {Yamada}}, \bibinfo
  {author} {\bibfnamefont {M.}~\bibnamefont {Yoshida}}, \bibinfo {author}
  {\bibfnamefont {M.}~\bibnamefont {Kurooka}},\ and\ \bibinfo {author}
  {\bibfnamefont {J.}~\bibnamefont {Mizuki}},\ }\bibfield  {title} {\bibinfo
  {title} {High-energy spin and charge excitations in electron-doped copper
  oxide superconductors},\ }\href {https://doi.org/10.1038/ncomms4714}
  {\bibfield  {journal} {\bibinfo  {journal} {Nature Communications}\ }\textbf
  {\bibinfo {volume} {5}},\ \bibinfo {pages} {3714} (\bibinfo {year}
  {2014})}\BibitemShut {NoStop}%
\bibitem [{\citenamefont {Lee}\ \emph {et~al.}(2014)\citenamefont {Lee},
  \citenamefont {Lee}, \citenamefont {Nowadnick}, \citenamefont {Gerber},
  \citenamefont {Tabis}, \citenamefont {Huang}, \citenamefont {Strocov},
  \citenamefont {Motoyama}, \citenamefont {Yu}, \citenamefont {Moritz},
  \citenamefont {Huang}, \citenamefont {Wang}, \citenamefont {Huang},
  \citenamefont {Wu}, \citenamefont {Chen}, \citenamefont {Huang},
  \citenamefont {Greven}, \citenamefont {Schmitt}, \citenamefont {Shen},\ and\
  \citenamefont {Devereaux}}]{Lee2014}%
  \BibitemOpen
  \bibfield  {author} {\bibinfo {author} {\bibfnamefont {W.~S.}\ \bibnamefont
  {Lee}}, \bibinfo {author} {\bibfnamefont {J.~J.}\ \bibnamefont {Lee}},
  \bibinfo {author} {\bibfnamefont {E.~A.}\ \bibnamefont {Nowadnick}}, \bibinfo
  {author} {\bibfnamefont {S.}~\bibnamefont {Gerber}}, \bibinfo {author}
  {\bibfnamefont {W.}~\bibnamefont {Tabis}}, \bibinfo {author} {\bibfnamefont
  {S.~W.}\ \bibnamefont {Huang}}, \bibinfo {author} {\bibfnamefont {V.~N.}\
  \bibnamefont {Strocov}}, \bibinfo {author} {\bibfnamefont {E.~M.}\
  \bibnamefont {Motoyama}}, \bibinfo {author} {\bibfnamefont {G.}~\bibnamefont
  {Yu}}, \bibinfo {author} {\bibfnamefont {B.}~\bibnamefont {Moritz}}, \bibinfo
  {author} {\bibfnamefont {H.~Y.}\ \bibnamefont {Huang}}, \bibinfo {author}
  {\bibfnamefont {R.~P.}\ \bibnamefont {Wang}}, \bibinfo {author}
  {\bibfnamefont {Y.~B.}\ \bibnamefont {Huang}}, \bibinfo {author}
  {\bibfnamefont {W.~B.}\ \bibnamefont {Wu}}, \bibinfo {author} {\bibfnamefont
  {C.~T.}\ \bibnamefont {Chen}}, \bibinfo {author} {\bibfnamefont {D.~J.}\
  \bibnamefont {Huang}}, \bibinfo {author} {\bibfnamefont {M.}~\bibnamefont
  {Greven}}, \bibinfo {author} {\bibfnamefont {T.}~\bibnamefont {Schmitt}},
  \bibinfo {author} {\bibfnamefont {Z.~X.}\ \bibnamefont {Shen}},\ and\
  \bibinfo {author} {\bibfnamefont {T.~P.}\ \bibnamefont {Devereaux}},\
  }\bibfield  {title} {\bibinfo {title} {Asymmetry of collective excitations in
  electron- and hole-doped cuprate superconductors},\ }\href
  {https://doi.org/10.1038/nphys3117} {\bibfield  {journal} {\bibinfo
  {journal} {Nature Physics}\ }\textbf {\bibinfo {volume} {10}},\ \bibinfo
  {pages} {883} (\bibinfo {year} {2014})}\BibitemShut {NoStop}%
\bibitem [{\citenamefont {Coldea}\ \emph {et~al.}(2001)\citenamefont {Coldea},
  \citenamefont {Hayden}, \citenamefont {Aeppli}, \citenamefont {Perring},
  \citenamefont {Frost}, \citenamefont {Mason}, \citenamefont {Cheong},\ and\
  \citenamefont {Fisk}}]{Coldea2001}%
  \BibitemOpen
  \bibfield  {author} {\bibinfo {author} {\bibfnamefont {R.}~\bibnamefont
  {Coldea}}, \bibinfo {author} {\bibfnamefont {S.~M.}\ \bibnamefont {Hayden}},
  \bibinfo {author} {\bibfnamefont {G.}~\bibnamefont {Aeppli}}, \bibinfo
  {author} {\bibfnamefont {T.~G.}\ \bibnamefont {Perring}}, \bibinfo {author}
  {\bibfnamefont {C.~D.}\ \bibnamefont {Frost}}, \bibinfo {author}
  {\bibfnamefont {T.~E.}\ \bibnamefont {Mason}}, \bibinfo {author}
  {\bibfnamefont {S.-W.}\ \bibnamefont {Cheong}},\ and\ \bibinfo {author}
  {\bibfnamefont {Z.}~\bibnamefont {Fisk}},\ }\bibfield  {title} {\bibinfo
  {title} {Spin waves and electronic interactions in
  ${\mathrm{la}}_{2}{\mathrm{cuo}}_{4}$},\ }\href
  {https://doi.org/10.1103/PhysRevLett.86.5377} {\bibfield  {journal} {\bibinfo
   {journal} {Phys. Rev. Lett.}\ }\textbf {\bibinfo {volume} {86}},\ \bibinfo
  {pages} {5377} (\bibinfo {year} {2001})}\BibitemShut {NoStop}%
\bibitem [{\citenamefont {Le~Tacon}\ \emph {et~al.}(2011)\citenamefont
  {Le~Tacon}, \citenamefont {Ghiringhelli}, \citenamefont {Chaloupka},
  \citenamefont {Sala}, \citenamefont {Hinkov}, \citenamefont {Haverkort},
  \citenamefont {Minola}, \citenamefont {Bakr}, \citenamefont {Zhou},
  \citenamefont {Blanco-Canosa}, \citenamefont {Monney}, \citenamefont {Song},
  \citenamefont {Sun}, \citenamefont {Lin}, \citenamefont {De~Luca},
  \citenamefont {Salluzzo}, \citenamefont {Khaliullin}, \citenamefont
  {Schmitt}, \citenamefont {Braicovich},\ and\ \citenamefont
  {Keimer}}]{LeTacon2011}%
  \BibitemOpen
  \bibfield  {author} {\bibinfo {author} {\bibfnamefont {M.}~\bibnamefont
  {Le~Tacon}}, \bibinfo {author} {\bibfnamefont {G.}~\bibnamefont
  {Ghiringhelli}}, \bibinfo {author} {\bibfnamefont {J.}~\bibnamefont
  {Chaloupka}}, \bibinfo {author} {\bibfnamefont {M.~M.}\ \bibnamefont {Sala}},
  \bibinfo {author} {\bibfnamefont {V.}~\bibnamefont {Hinkov}}, \bibinfo
  {author} {\bibfnamefont {M.~W.}\ \bibnamefont {Haverkort}}, \bibinfo {author}
  {\bibfnamefont {M.}~\bibnamefont {Minola}}, \bibinfo {author} {\bibfnamefont
  {M.}~\bibnamefont {Bakr}}, \bibinfo {author} {\bibfnamefont {K.~J.}\
  \bibnamefont {Zhou}}, \bibinfo {author} {\bibfnamefont {S.}~\bibnamefont
  {Blanco-Canosa}}, \bibinfo {author} {\bibfnamefont {C.}~\bibnamefont
  {Monney}}, \bibinfo {author} {\bibfnamefont {Y.~T.}\ \bibnamefont {Song}},
  \bibinfo {author} {\bibfnamefont {G.~L.}\ \bibnamefont {Sun}}, \bibinfo
  {author} {\bibfnamefont {C.~T.}\ \bibnamefont {Lin}}, \bibinfo {author}
  {\bibfnamefont {G.~M.}\ \bibnamefont {De~Luca}}, \bibinfo {author}
  {\bibfnamefont {M.}~\bibnamefont {Salluzzo}}, \bibinfo {author}
  {\bibfnamefont {G.}~\bibnamefont {Khaliullin}}, \bibinfo {author}
  {\bibfnamefont {T.}~\bibnamefont {Schmitt}}, \bibinfo {author} {\bibfnamefont
  {L.}~\bibnamefont {Braicovich}},\ and\ \bibinfo {author} {\bibfnamefont
  {B.}~\bibnamefont {Keimer}},\ }\bibfield  {title} {\bibinfo {title} {Intense
  paramagnon excitations in a large family of high-temperature
  superconductors},\ }\href {https://doi.org/10.1038/nphys2041} {\bibfield
  {journal} {\bibinfo  {journal} {Nature Physics}\ }\textbf {\bibinfo {volume}
  {7}},\ \bibinfo {pages} {725} (\bibinfo {year} {2011})}\BibitemShut {NoStop}%
\bibitem [{\citenamefont {Dean}\ \emph {et~al.}(2013)\citenamefont {Dean},
  \citenamefont {Dellea}, \citenamefont {Springell}, \citenamefont
  {Yakhou-Harris}, \citenamefont {Kummer}, \citenamefont {Brookes},
  \citenamefont {Liu}, \citenamefont {Sun}, \citenamefont {Strle},
  \citenamefont {Schmitt}, \citenamefont {Braicovich}, \citenamefont
  {Ghiringhelli}, \citenamefont {Bo{\v{z}}ovi{\'{c}}},\ and\ \citenamefont
  {Hill}}]{Dean2013}%
  \BibitemOpen
  \bibfield  {author} {\bibinfo {author} {\bibfnamefont {M.~P.~M.}\
  \bibnamefont {Dean}}, \bibinfo {author} {\bibfnamefont {G.}~\bibnamefont
  {Dellea}}, \bibinfo {author} {\bibfnamefont {R.~S.}\ \bibnamefont
  {Springell}}, \bibinfo {author} {\bibfnamefont {F.}~\bibnamefont
  {Yakhou-Harris}}, \bibinfo {author} {\bibfnamefont {K.}~\bibnamefont
  {Kummer}}, \bibinfo {author} {\bibfnamefont {N.~B.}\ \bibnamefont {Brookes}},
  \bibinfo {author} {\bibfnamefont {X.}~\bibnamefont {Liu}}, \bibinfo {author}
  {\bibfnamefont {Y.-J.}\ \bibnamefont {Sun}}, \bibinfo {author} {\bibfnamefont
  {J.}~\bibnamefont {Strle}}, \bibinfo {author} {\bibfnamefont
  {T.}~\bibnamefont {Schmitt}}, \bibinfo {author} {\bibfnamefont
  {L.}~\bibnamefont {Braicovich}}, \bibinfo {author} {\bibfnamefont
  {G.}~\bibnamefont {Ghiringhelli}}, \bibinfo {author} {\bibfnamefont
  {I.}~\bibnamefont {Bo{\v{z}}ovi{\'{c}}}},\ and\ \bibinfo {author}
  {\bibfnamefont {J.~P.}\ \bibnamefont {Hill}},\ }\bibfield  {title} {\bibinfo
  {title} {Persistence of magnetic excitations in {La}$_{2-x}${Sr}$_x${CuO}$_4$
  from the undoped insulator to the heavily overdoped non-superconducting
  metal},\ }\href {https://doi.org/10.1038/nmat3723} {\bibfield  {journal}
  {\bibinfo  {journal} {Nature Materials}\ }\textbf {\bibinfo {volume} {12}},\
  \bibinfo {pages} {1019} (\bibinfo {year} {2013})}\BibitemShut {NoStop}%
\bibitem [{\citenamefont {Kugel}\ and\ \citenamefont
  {Khomski{\u\i}}(1982)}]{kugel1982jahn}%
  \BibitemOpen
  \bibfield  {author} {\bibinfo {author} {\bibfnamefont {K.~I.}\ \bibnamefont
  {Kugel}}\ and\ \bibinfo {author} {\bibfnamefont {D.}~\bibnamefont
  {Khomski{\u\i}}},\ }\bibfield  {title} {\bibinfo {title} {The jahn-teller
  effect and magnetism: transition metal compounds},\ }\href
  {https://doi.org/10.1070/PU1982v025n04ABEH004537} {\bibfield  {journal}
  {\bibinfo  {journal} {Soviet Physics Uspekhi}\ }\textbf {\bibinfo {volume}
  {25}},\ \bibinfo {pages} {231} (\bibinfo {year} {1982})}\BibitemShut
  {NoStop}%
\bibitem [{\citenamefont {Tokura}\ and\ \citenamefont
  {Nagaosa}(2000)}]{tokura2000orbital}%
  \BibitemOpen
  \bibfield  {author} {\bibinfo {author} {\bibfnamefont {Y.}~\bibnamefont
  {Tokura}}\ and\ \bibinfo {author} {\bibfnamefont {N.}~\bibnamefont
  {Nagaosa}},\ }\bibfield  {title} {\bibinfo {title} {Orbital physics in
  transition-metal oxides},\ }\href
  {https://doi.org/10.1126/science.288.5465.462} {\bibfield  {journal}
  {\bibinfo  {journal} {Science}\ }\textbf {\bibinfo {volume} {288}},\ \bibinfo
  {pages} {462} (\bibinfo {year} {2000})}\BibitemShut {NoStop}%
\bibitem [{Sup()}]{SupplementaryMat}%
  \BibitemOpen
  \bibinfo {note} {Suppementary text, equations and figures}\BibitemShut
  {NoStop}%
\bibitem [{\citenamefont {Schlappa}\ \emph {et~al.}(2012)\citenamefont
  {Schlappa}, \citenamefont {Wohlfeld}, \citenamefont {Zhou}, \citenamefont
  {Mourigal}, \citenamefont {Haverkort}, \citenamefont {Strocov}, \citenamefont
  {Hozoi}, \citenamefont {Monney}, \citenamefont {Nishimoto}, \citenamefont
  {Singh} \emph {et~al.}}]{schlappa2012spin}%
  \BibitemOpen
  \bibfield  {author} {\bibinfo {author} {\bibfnamefont {J.}~\bibnamefont
  {Schlappa}}, \bibinfo {author} {\bibfnamefont {K.}~\bibnamefont {Wohlfeld}},
  \bibinfo {author} {\bibfnamefont {K.}~\bibnamefont {Zhou}}, \bibinfo {author}
  {\bibfnamefont {M.}~\bibnamefont {Mourigal}}, \bibinfo {author}
  {\bibfnamefont {M.}~\bibnamefont {Haverkort}}, \bibinfo {author}
  {\bibfnamefont {V.}~\bibnamefont {Strocov}}, \bibinfo {author} {\bibfnamefont
  {L.}~\bibnamefont {Hozoi}}, \bibinfo {author} {\bibfnamefont
  {C.}~\bibnamefont {Monney}}, \bibinfo {author} {\bibfnamefont
  {S.}~\bibnamefont {Nishimoto}}, \bibinfo {author} {\bibfnamefont
  {S.}~\bibnamefont {Singh}}, \emph {et~al.},\ }\bibfield  {title} {\bibinfo
  {title} {{Spin--orbital separation in the quasi-one-dimensional Mott
  insulator Sr$_2$CuO$_3$}},\ }\href {https://doi.org/10.1038/nature10974}
  {\bibfield  {journal} {\bibinfo  {journal} {Nature}\ }\textbf {\bibinfo
  {volume} {485}},\ \bibinfo {pages} {82} (\bibinfo {year} {2012})}\BibitemShut
  {NoStop}%
\bibitem [{\citenamefont {Bisogni}\ \emph {et~al.}(2015)\citenamefont
  {Bisogni}, \citenamefont {Wohlfeld}, \citenamefont {Nishimoto}, \citenamefont
  {Monney}, \citenamefont {Trinckauf}, \citenamefont {Zhou}, \citenamefont
  {Kraus}, \citenamefont {Koepernik}, \citenamefont {Sekar}, \citenamefont
  {Strocov}, \citenamefont {B\"uchner}, \citenamefont {Schmitt}, \citenamefont
  {van~den Brink},\ and\ \citenamefont {Geck}}]{bisogni2015orbital}%
  \BibitemOpen
  \bibfield  {author} {\bibinfo {author} {\bibfnamefont {V.}~\bibnamefont
  {Bisogni}}, \bibinfo {author} {\bibfnamefont {K.}~\bibnamefont {Wohlfeld}},
  \bibinfo {author} {\bibfnamefont {S.}~\bibnamefont {Nishimoto}}, \bibinfo
  {author} {\bibfnamefont {C.}~\bibnamefont {Monney}}, \bibinfo {author}
  {\bibfnamefont {J.}~\bibnamefont {Trinckauf}}, \bibinfo {author}
  {\bibfnamefont {K.}~\bibnamefont {Zhou}}, \bibinfo {author} {\bibfnamefont
  {R.}~\bibnamefont {Kraus}}, \bibinfo {author} {\bibfnamefont
  {K.}~\bibnamefont {Koepernik}}, \bibinfo {author} {\bibfnamefont
  {C.}~\bibnamefont {Sekar}}, \bibinfo {author} {\bibfnamefont
  {V.}~\bibnamefont {Strocov}}, \bibinfo {author} {\bibfnamefont
  {B.}~\bibnamefont {B\"uchner}}, \bibinfo {author} {\bibfnamefont
  {T.}~\bibnamefont {Schmitt}}, \bibinfo {author} {\bibfnamefont
  {J.}~\bibnamefont {van~den Brink}},\ and\ \bibinfo {author} {\bibfnamefont
  {J.}~\bibnamefont {Geck}},\ }\bibfield  {title} {\bibinfo {title} {{Orbital
  Control of Effective Dimensionality: From Spin-Orbital Fractionalization to
  Confinement in the Anisotropic Ladder System
  ${\mathrm{CaCu}}_{2}{\mathrm{O}}_{3}$}},\ }\href
  {https://doi.org/10.1103/PhysRevLett.114.096402} {\bibfield  {journal}
  {\bibinfo  {journal} {Phys. Rev. Lett.}\ }\textbf {\bibinfo {volume} {114}},\
  \bibinfo {pages} {096402} (\bibinfo {year} {2015})}\BibitemShut {NoStop}%
\bibitem [{\citenamefont {Fumagalli}\ \emph {et~al.}(2020)\citenamefont
  {Fumagalli}, \citenamefont {Heverhagen}, \citenamefont {Betto}, \citenamefont
  {Arpaia}, \citenamefont {Rossi}, \citenamefont {Di~Castro}, \citenamefont
  {Brookes}, \citenamefont {Moretti~Sala}, \citenamefont {Daghofer},
  \citenamefont {Braicovich}, \citenamefont {Wohlfeld},\ and\ \citenamefont
  {Ghiringhelli}}]{fumagalli2020mobile}%
  \BibitemOpen
  \bibfield  {author} {\bibinfo {author} {\bibfnamefont {R.}~\bibnamefont
  {Fumagalli}}, \bibinfo {author} {\bibfnamefont {J.}~\bibnamefont
  {Heverhagen}}, \bibinfo {author} {\bibfnamefont {D.}~\bibnamefont {Betto}},
  \bibinfo {author} {\bibfnamefont {R.}~\bibnamefont {Arpaia}}, \bibinfo
  {author} {\bibfnamefont {M.}~\bibnamefont {Rossi}}, \bibinfo {author}
  {\bibfnamefont {D.}~\bibnamefont {Di~Castro}}, \bibinfo {author}
  {\bibfnamefont {N.~B.}\ \bibnamefont {Brookes}}, \bibinfo {author}
  {\bibfnamefont {M.}~\bibnamefont {Moretti~Sala}}, \bibinfo {author}
  {\bibfnamefont {M.}~\bibnamefont {Daghofer}}, \bibinfo {author}
  {\bibfnamefont {L.}~\bibnamefont {Braicovich}}, \bibinfo {author}
  {\bibfnamefont {K.}~\bibnamefont {Wohlfeld}},\ and\ \bibinfo {author}
  {\bibfnamefont {G.}~\bibnamefont {Ghiringhelli}},\ }\bibfield  {title}
  {\bibinfo {title} {Mobile orbitons in {C}a$_2${C}u{O}$_3$: Crucial role of
  {H}und's exchange},\ }\href {https://doi.org/10.1103/PhysRevB.101.205117}
  {\bibfield  {journal} {\bibinfo  {journal} {Phys. Rev. B}\ }\textbf {\bibinfo
  {volume} {101}},\ \bibinfo {pages} {205117} (\bibinfo {year}
  {2020})}\BibitemShut {NoStop}%
\bibitem [{\citenamefont {Wohlfeld}\ \emph {et~al.}(2011)\citenamefont
  {Wohlfeld}, \citenamefont {Daghofer}, \citenamefont {Nishimoto},
  \citenamefont {Khaliullin},\ and\ \citenamefont {van~den
  Brink}}]{wohlfeld2011intrinsic}%
  \BibitemOpen
  \bibfield  {author} {\bibinfo {author} {\bibfnamefont {K.}~\bibnamefont
  {Wohlfeld}}, \bibinfo {author} {\bibfnamefont {M.}~\bibnamefont {Daghofer}},
  \bibinfo {author} {\bibfnamefont {S.}~\bibnamefont {Nishimoto}}, \bibinfo
  {author} {\bibfnamefont {G.}~\bibnamefont {Khaliullin}},\ and\ \bibinfo
  {author} {\bibfnamefont {J.}~\bibnamefont {van~den Brink}},\ }\bibfield
  {title} {\bibinfo {title} {Intrinsic coupling of orbital excitations to spin
  fluctuations in mott insulators},\ }\href
  {https://doi.org/10.1103/PhysRevLett.107.147201} {\bibfield  {journal}
  {\bibinfo  {journal} {Phys. Rev. Lett.}\ }\textbf {\bibinfo {volume} {107}},\
  \bibinfo {pages} {147201} (\bibinfo {year} {2011})}\BibitemShut {NoStop}%
\bibitem [{\citenamefont {Wohlfeld}\ \emph {et~al.}(2013)\citenamefont
  {Wohlfeld}, \citenamefont {Nishimoto}, \citenamefont {Haverkort},\ and\
  \citenamefont {van~den Brink}}]{Wohlfeld2013}%
  \BibitemOpen
  \bibfield  {author} {\bibinfo {author} {\bibfnamefont {K.}~\bibnamefont
  {Wohlfeld}}, \bibinfo {author} {\bibfnamefont {S.}~\bibnamefont {Nishimoto}},
  \bibinfo {author} {\bibfnamefont {M.~W.}\ \bibnamefont {Haverkort}},\ and\
  \bibinfo {author} {\bibfnamefont {J.}~\bibnamefont {van~den Brink}},\
  }\bibfield  {title} {\bibinfo {title} {{Microscopic origin of spin-orbital
  separation in Sr$_{2}$CuO$_{3}$}},\ }\href
  {https://doi.org/10.1103/PhysRevB.88.195138} {\bibfield  {journal} {\bibinfo
  {journal} {Phys. Rev. B}\ }\textbf {\bibinfo {volume} {88}},\ \bibinfo
  {pages} {195138} (\bibinfo {year} {2013})}\BibitemShut {NoStop}%
\bibitem [{\citenamefont {Heverhagen}\ and\ \citenamefont
  {Daghofer}(2018)}]{heverhagen2018spinon}%
  \BibitemOpen
  \bibfield  {author} {\bibinfo {author} {\bibfnamefont {J.}~\bibnamefont
  {Heverhagen}}\ and\ \bibinfo {author} {\bibfnamefont {M.}~\bibnamefont
  {Daghofer}},\ }\bibfield  {title} {\bibinfo {title} {Spinon-orbiton repulsion
  and attraction mediated by hund's rule},\ }\href
  {https://doi.org/10.1103/PhysRevB.98.085120} {\bibfield  {journal} {\bibinfo
  {journal} {Phys. Rev. B}\ }\textbf {\bibinfo {volume} {98}},\ \bibinfo
  {pages} {085120} (\bibinfo {year} {2018})}\BibitemShut {NoStop}%
\bibitem [{\citenamefont {Kim}\ \emph {et~al.}(1996)\citenamefont {Kim},
  \citenamefont {Matsuura}, \citenamefont {Shen}, \citenamefont {Motoyama},
  \citenamefont {Eisaki}, \citenamefont {Uchida}, \citenamefont {Tohyama},\
  and\ \citenamefont {Maekawa}}]{Kim1996}%
  \BibitemOpen
  \bibfield  {author} {\bibinfo {author} {\bibfnamefont {C.}~\bibnamefont
  {Kim}}, \bibinfo {author} {\bibfnamefont {A.~Y.}\ \bibnamefont {Matsuura}},
  \bibinfo {author} {\bibfnamefont {Z.-X.}\ \bibnamefont {Shen}}, \bibinfo
  {author} {\bibfnamefont {N.}~\bibnamefont {Motoyama}}, \bibinfo {author}
  {\bibfnamefont {H.}~\bibnamefont {Eisaki}}, \bibinfo {author} {\bibfnamefont
  {S.}~\bibnamefont {Uchida}}, \bibinfo {author} {\bibfnamefont
  {T.}~\bibnamefont {Tohyama}},\ and\ \bibinfo {author} {\bibfnamefont
  {S.}~\bibnamefont {Maekawa}},\ }\bibfield  {title} {\bibinfo {title}
  {Observation of spin-charge separation in one-dimensional
  srcu${\mathrm{o}}_{2}$},\ }\href
  {https://doi.org/10.1103/PhysRevLett.77.4054} {\bibfield  {journal} {\bibinfo
   {journal} {Phys. Rev. Lett.}\ }\textbf {\bibinfo {volume} {77}},\ \bibinfo
  {pages} {4054} (\bibinfo {year} {1996})}\BibitemShut {NoStop}%
\bibitem [{\citenamefont {{Kim}}\ \emph {et~al.}(2006)\citenamefont {{Kim}},
  \citenamefont {{Koh}}, \citenamefont {{Rotenberg}}, \citenamefont {{Oh}},
  \citenamefont {{Eisaki}}, \citenamefont {{Motoyama}}, \citenamefont
  {{Uchida}}, \citenamefont {{Tohyama}}, \citenamefont {{Maekawa}},
  \citenamefont {{Shen}},\ and\ \citenamefont {{Kim}}}]{Kim2006}%
  \BibitemOpen
  \bibfield  {author} {\bibinfo {author} {\bibfnamefont {B.~J.}\ \bibnamefont
  {{Kim}}}, \bibinfo {author} {\bibfnamefont {H.}~\bibnamefont {{Koh}}},
  \bibinfo {author} {\bibfnamefont {E.}~\bibnamefont {{Rotenberg}}}, \bibinfo
  {author} {\bibfnamefont {S.~J.}\ \bibnamefont {{Oh}}}, \bibinfo {author}
  {\bibfnamefont {H.}~\bibnamefont {{Eisaki}}}, \bibinfo {author}
  {\bibfnamefont {N.}~\bibnamefont {{Motoyama}}}, \bibinfo {author}
  {\bibfnamefont {S.}~\bibnamefont {{Uchida}}}, \bibinfo {author}
  {\bibfnamefont {T.}~\bibnamefont {{Tohyama}}}, \bibinfo {author}
  {\bibfnamefont {S.}~\bibnamefont {{Maekawa}}}, \bibinfo {author}
  {\bibfnamefont {Z.~X.}\ \bibnamefont {{Shen}}},\ and\ \bibinfo {author}
  {\bibfnamefont {C.}~\bibnamefont {{Kim}}},\ }\bibfield  {title} {\bibinfo
  {title} {{Distinct spinon and holon dispersions in photoemission spectral
  functions from one-dimensional SrCuO$_{2}$}},\ }\href
  {https://doi.org/10.1038/nphys316} {\bibfield  {journal} {\bibinfo  {journal}
  {Nature Physics}\ }\textbf {\bibinfo {volume} {2}},\ \bibinfo {pages} {397}
  (\bibinfo {year} {2006})}\BibitemShut {NoStop}%
\bibitem [{\citenamefont {Wohlfeld}\ \emph {et~al.}(2012)\citenamefont
  {Wohlfeld}, \citenamefont {Daghofer}, \citenamefont {Khaliullin},\ and\
  \citenamefont {van~den Brink}}]{Wohlfeld2012dispersion}%
  \BibitemOpen
  \bibfield  {author} {\bibinfo {author} {\bibfnamefont {K.}~\bibnamefont
  {Wohlfeld}}, \bibinfo {author} {\bibfnamefont {M.}~\bibnamefont {Daghofer}},
  \bibinfo {author} {\bibfnamefont {G.}~\bibnamefont {Khaliullin}},\ and\
  \bibinfo {author} {\bibfnamefont {J.}~\bibnamefont {van~den Brink}},\
  }\bibfield  {title} {\bibinfo {title} {Dispersion of orbital excitations in
  2d quantum antiferromagnets},\ }\href
  {https://doi.org/10.1088/1742-6596/391/1/012168} {\bibfield  {journal}
  {\bibinfo  {journal} {Journal of Physics: Conference Series}\ }\textbf
  {\bibinfo {volume} {391}},\ \bibinfo {pages} {012168} (\bibinfo {year}
  {2012})}\BibitemShut {NoStop}%
\bibitem [{\citenamefont {Kim}\ \emph {et~al.}(2014)\citenamefont {Kim},
  \citenamefont {Daghofer}, \citenamefont {Said}, \citenamefont {Gog},
  \citenamefont {van~den Brink}, \citenamefont {Khaliullin},\ and\
  \citenamefont {Kim}}]{kim2014excitonic}%
  \BibitemOpen
  \bibfield  {author} {\bibinfo {author} {\bibfnamefont {J.}~\bibnamefont
  {Kim}}, \bibinfo {author} {\bibfnamefont {M.}~\bibnamefont {Daghofer}},
  \bibinfo {author} {\bibfnamefont {A.~H.}\ \bibnamefont {Said}}, \bibinfo
  {author} {\bibfnamefont {T.}~\bibnamefont {Gog}}, \bibinfo {author}
  {\bibfnamefont {J.}~\bibnamefont {van~den Brink}}, \bibinfo {author}
  {\bibfnamefont {G.}~\bibnamefont {Khaliullin}},\ and\ \bibinfo {author}
  {\bibfnamefont {B.~J.}\ \bibnamefont {Kim}},\ }\bibfield  {title} {\bibinfo
  {title} {Excitonic quasiparticles in a spin–orbit {Mott} insulator},\
  }\href {https://doi.org/10.1038/ncomms5453} {\bibfield  {journal} {\bibinfo
  {journal} {Nature Communications}\ }\textbf {\bibinfo {volume} {5}},\
  \bibinfo {pages} {4453} (\bibinfo {year} {2014})}\BibitemShut {NoStop}%
\bibitem [{\citenamefont {van~den Brink}\ \emph {et~al.}(1999)\citenamefont
  {van~den Brink}, \citenamefont {Horsch}, \citenamefont {Mack},\ and\
  \citenamefont {Ole\ifmmode~\acute{s}\else
  \'{s}\fi{}}}]{vandenbrink1999orbital}%
  \BibitemOpen
  \bibfield  {author} {\bibinfo {author} {\bibfnamefont {J.}~\bibnamefont
  {van~den Brink}}, \bibinfo {author} {\bibfnamefont {P.}~\bibnamefont
  {Horsch}}, \bibinfo {author} {\bibfnamefont {F.}~\bibnamefont {Mack}},\ and\
  \bibinfo {author} {\bibfnamefont {A.~M.}\ \bibnamefont
  {Ole\ifmmode~\acute{s}\else \'{s}\fi{}}},\ }\bibfield  {title} {\bibinfo
  {title} {Orbital dynamics in ferromagnetic transition-metal oxides},\ }\href
  {https://doi.org/10.1103/PhysRevB.59.6795} {\bibfield  {journal} {\bibinfo
  {journal} {Phys. Rev. B}\ }\textbf {\bibinfo {volume} {59}},\ \bibinfo
  {pages} {6795} (\bibinfo {year} {1999})}\BibitemShut {NoStop}%
\bibitem [{\citenamefont {Nasu}\ and\ \citenamefont
  {Ishihara}(2013)}]{Nasu2013vibronic}%
  \BibitemOpen
  \bibfield  {author} {\bibinfo {author} {\bibfnamefont {J.}~\bibnamefont
  {Nasu}}\ and\ \bibinfo {author} {\bibfnamefont {S.}~\bibnamefont
  {Ishihara}},\ }\bibfield  {title} {\bibinfo {title} {Vibronic excitation
  dynamics in orbitally degenerate correlated electron system},\ }\href
  {https://doi.org/10.1103/PhysRevB.88.205110} {\bibfield  {journal} {\bibinfo
  {journal} {Phys. Rev. B}\ }\textbf {\bibinfo {volume} {88}},\ \bibinfo
  {pages} {205110} (\bibinfo {year} {2013})}\BibitemShut {NoStop}%
\bibitem [{\citenamefont {Tanaka}\ \emph {et~al.}(2004)\citenamefont {Tanaka},
  \citenamefont {Baron}, \citenamefont {Kim}, \citenamefont {Thomas},
  \citenamefont {Hill}, \citenamefont {Honda}, \citenamefont {Iga},
  \citenamefont {Tsutsui}, \citenamefont {Ishikawa},\ and\ \citenamefont
  {Nelson}}]{tanaka2004search}%
  \BibitemOpen
  \bibfield  {author} {\bibinfo {author} {\bibfnamefont {Y.}~\bibnamefont
  {Tanaka}}, \bibinfo {author} {\bibfnamefont {A.}~\bibnamefont {Baron}},
  \bibinfo {author} {\bibfnamefont {Y.-J.}\ \bibnamefont {Kim}}, \bibinfo
  {author} {\bibfnamefont {K.}~\bibnamefont {Thomas}}, \bibinfo {author}
  {\bibfnamefont {J.}~\bibnamefont {Hill}}, \bibinfo {author} {\bibfnamefont
  {Z.}~\bibnamefont {Honda}}, \bibinfo {author} {\bibfnamefont
  {F.}~\bibnamefont {Iga}}, \bibinfo {author} {\bibfnamefont {S.}~\bibnamefont
  {Tsutsui}}, \bibinfo {author} {\bibfnamefont {D.}~\bibnamefont {Ishikawa}},\
  and\ \bibinfo {author} {\bibfnamefont {C.}~\bibnamefont {Nelson}},\
  }\bibfield  {title} {\bibinfo {title} {{Search for orbitons in LaMnO$_3$,
  YTiO$_3$ and KCuF$_3$ using high-resolution inelastic x-ray scattering}},\
  }\href {https://doi.org/https://doi.org/10.1088/1367-2630/6/1/161} {\bibfield
   {journal} {\bibinfo  {journal} {New Journal of Physics}\ }\textbf {\bibinfo
  {volume} {6}},\ \bibinfo {pages} {161} (\bibinfo {year} {2004})}\BibitemShut
  {NoStop}%
\bibitem [{\citenamefont {Li}\ \emph {et~al.}(2021)\citenamefont {Li},
  \citenamefont {Xu}, \citenamefont {Garcia-Fernandez}, \citenamefont {Nag},
  \citenamefont {Robarts}, \citenamefont {Walters}, \citenamefont {Liu},
  \citenamefont {Zhou}, \citenamefont {Wohlfeld}, \citenamefont {van~den
  Brink}, \citenamefont {Ding},\ and\ \citenamefont {Zhou}}]{li2021unraveling}%
  \BibitemOpen
  \bibfield  {author} {\bibinfo {author} {\bibfnamefont {J.}~\bibnamefont
  {Li}}, \bibinfo {author} {\bibfnamefont {L.}~\bibnamefont {Xu}}, \bibinfo
  {author} {\bibfnamefont {M.}~\bibnamefont {Garcia-Fernandez}}, \bibinfo
  {author} {\bibfnamefont {A.}~\bibnamefont {Nag}}, \bibinfo {author}
  {\bibfnamefont {H.~C.}\ \bibnamefont {Robarts}}, \bibinfo {author}
  {\bibfnamefont {A.~C.}\ \bibnamefont {Walters}}, \bibinfo {author}
  {\bibfnamefont {X.}~\bibnamefont {Liu}}, \bibinfo {author} {\bibfnamefont
  {J.}~\bibnamefont {Zhou}}, \bibinfo {author} {\bibfnamefont {K.}~\bibnamefont
  {Wohlfeld}}, \bibinfo {author} {\bibfnamefont {J.}~\bibnamefont {van~den
  Brink}}, \bibinfo {author} {\bibfnamefont {H.}~\bibnamefont {Ding}},\ and\
  \bibinfo {author} {\bibfnamefont {K.-J.}\ \bibnamefont {Zhou}},\ }\bibfield
  {title} {\bibinfo {title} {Unraveling the orbital physics in a canonical
  orbital system ${\mathrm{kcuf}}_{3}$},\ }\href
  {https://doi.org/10.1103/PhysRevLett.126.106401} {\bibfield  {journal}
  {\bibinfo  {journal} {Phys. Rev. Lett.}\ }\textbf {\bibinfo {volume} {126}},\
  \bibinfo {pages} {106401} (\bibinfo {year} {2021})}\BibitemShut {NoStop}%
\bibitem [{\citenamefont {Saitoh}\ \emph {et~al.}(2001)\citenamefont {Saitoh},
  \citenamefont {Okamoto}, \citenamefont {Takahashi}, \citenamefont {Tobe},
  \citenamefont {Yamamoto}, \citenamefont {Kimura}, \citenamefont {Ishihara},
  \citenamefont {Maekawa},\ and\ \citenamefont
  {Tokura}}]{saitoh2001observation}%
  \BibitemOpen
  \bibfield  {author} {\bibinfo {author} {\bibfnamefont {E.}~\bibnamefont
  {Saitoh}}, \bibinfo {author} {\bibfnamefont {S.}~\bibnamefont {Okamoto}},
  \bibinfo {author} {\bibfnamefont {K.~T.}\ \bibnamefont {Takahashi}}, \bibinfo
  {author} {\bibfnamefont {K.}~\bibnamefont {Tobe}}, \bibinfo {author}
  {\bibfnamefont {K.}~\bibnamefont {Yamamoto}}, \bibinfo {author}
  {\bibfnamefont {T.}~\bibnamefont {Kimura}}, \bibinfo {author} {\bibfnamefont
  {S.}~\bibnamefont {Ishihara}}, \bibinfo {author} {\bibfnamefont
  {S.}~\bibnamefont {Maekawa}},\ and\ \bibinfo {author} {\bibfnamefont
  {Y.}~\bibnamefont {Tokura}},\ }\bibfield  {title} {\bibinfo {title}
  {{Observation of Orbital Waves as Elementary Excitations in a Solid}},\
  }\href {https://doi.org/10.1038/35065547} {\bibfield  {journal} {\bibinfo
  {journal} {Nature}\ }\textbf {\bibinfo {volume} {410}},\ \bibinfo {pages}
  {180} (\bibinfo {year} {2001})}\BibitemShut {NoStop}%
\bibitem [{\citenamefont {Gr{\"u}ninger}\ \emph {et~al.}(2002)\citenamefont
  {Gr{\"u}ninger}, \citenamefont {R{\"u}ckamp}, \citenamefont {Windt},
  \citenamefont {Reutler}, \citenamefont {Zobel}, \citenamefont {Lorenz},
  \citenamefont {Freimuth},\ and\ \citenamefont
  {Revcolevschi}}]{gruninger2002experimental}%
  \BibitemOpen
  \bibfield  {author} {\bibinfo {author} {\bibfnamefont {M.}~\bibnamefont
  {Gr{\"u}ninger}}, \bibinfo {author} {\bibfnamefont {R.}~\bibnamefont
  {R{\"u}ckamp}}, \bibinfo {author} {\bibfnamefont {M.}~\bibnamefont {Windt}},
  \bibinfo {author} {\bibfnamefont {P.}~\bibnamefont {Reutler}}, \bibinfo
  {author} {\bibfnamefont {C.}~\bibnamefont {Zobel}}, \bibinfo {author}
  {\bibfnamefont {T.}~\bibnamefont {Lorenz}}, \bibinfo {author} {\bibfnamefont
  {A.}~\bibnamefont {Freimuth}},\ and\ \bibinfo {author} {\bibfnamefont
  {A.}~\bibnamefont {Revcolevschi}},\ }\bibfield  {title} {\bibinfo {title}
  {Experimental quest for orbital waves},\ }\href
  {https://doi.org/10.1038/418039a} {\bibfield  {journal} {\bibinfo  {journal}
  {Nature}\ }\textbf {\bibinfo {volume} {418}},\ \bibinfo {pages} {39}
  (\bibinfo {year} {2002})}\BibitemShut {NoStop}%
\bibitem [{\citenamefont {Ulrich}\ \emph {et~al.}(2009)\citenamefont {Ulrich},
  \citenamefont {Ament}, \citenamefont {Ghiringhelli}, \citenamefont
  {Braicovich}, \citenamefont {Moretti~Sala}, \citenamefont {Pezzotta},
  \citenamefont {Schmitt}, \citenamefont {Khaliullin}, \citenamefont {van~den
  Brink}, \citenamefont {Roth}, \citenamefont {Lorenz},\ and\ \citenamefont
  {Keimer}}]{ulrich2009momentum}%
  \BibitemOpen
  \bibfield  {author} {\bibinfo {author} {\bibfnamefont {C.}~\bibnamefont
  {Ulrich}}, \bibinfo {author} {\bibfnamefont {L.~J.~P.}\ \bibnamefont
  {Ament}}, \bibinfo {author} {\bibfnamefont {G.}~\bibnamefont {Ghiringhelli}},
  \bibinfo {author} {\bibfnamefont {L.}~\bibnamefont {Braicovich}}, \bibinfo
  {author} {\bibfnamefont {M.}~\bibnamefont {Moretti~Sala}}, \bibinfo {author}
  {\bibfnamefont {N.}~\bibnamefont {Pezzotta}}, \bibinfo {author}
  {\bibfnamefont {T.}~\bibnamefont {Schmitt}}, \bibinfo {author} {\bibfnamefont
  {G.}~\bibnamefont {Khaliullin}}, \bibinfo {author} {\bibfnamefont
  {J.}~\bibnamefont {van~den Brink}}, \bibinfo {author} {\bibfnamefont
  {H.}~\bibnamefont {Roth}}, \bibinfo {author} {\bibfnamefont {T.}~\bibnamefont
  {Lorenz}},\ and\ \bibinfo {author} {\bibfnamefont {B.}~\bibnamefont
  {Keimer}},\ }\bibfield  {title} {\bibinfo {title} {Momentum dependence of
  orbital excitations in mott-insulating titanates},\ }\href
  {https://doi.org/10.1103/PhysRevLett.103.107205} {\bibfield  {journal}
  {\bibinfo  {journal} {Phys. Rev. Lett.}\ }\textbf {\bibinfo {volume} {103}},\
  \bibinfo {pages} {107205} (\bibinfo {year} {2009})}\BibitemShut {NoStop}%
\bibitem [{\citenamefont {Fumagalli}\ \emph {et~al.}(2019)\citenamefont
  {Fumagalli}, \citenamefont {Braicovich}, \citenamefont {Minola},
  \citenamefont {Peng}, \citenamefont {Kummer}, \citenamefont {Betto},
  \citenamefont {Rossi}, \citenamefont {Lefran\ifmmode~\mbox{\c{c}}\else
  \c{c}\fi{}ois}, \citenamefont {Morawe}, \citenamefont {Salluzzo},
  \citenamefont {Suzuki}, \citenamefont {Yakhou}, \citenamefont {Le~Tacon},
  \citenamefont {Keimer}, \citenamefont {Brookes}, \citenamefont {Sala},\ and\
  \citenamefont {Ghiringhelli}}]{fumagalli2019polarization}%
  \BibitemOpen
  \bibfield  {author} {\bibinfo {author} {\bibfnamefont {R.}~\bibnamefont
  {Fumagalli}}, \bibinfo {author} {\bibfnamefont {L.}~\bibnamefont
  {Braicovich}}, \bibinfo {author} {\bibfnamefont {M.}~\bibnamefont {Minola}},
  \bibinfo {author} {\bibfnamefont {Y.~Y.}\ \bibnamefont {Peng}}, \bibinfo
  {author} {\bibfnamefont {K.}~\bibnamefont {Kummer}}, \bibinfo {author}
  {\bibfnamefont {D.}~\bibnamefont {Betto}}, \bibinfo {author} {\bibfnamefont
  {M.}~\bibnamefont {Rossi}}, \bibinfo {author} {\bibfnamefont
  {E.}~\bibnamefont {Lefran\ifmmode~\mbox{\c{c}}\else \c{c}\fi{}ois}}, \bibinfo
  {author} {\bibfnamefont {C.}~\bibnamefont {Morawe}}, \bibinfo {author}
  {\bibfnamefont {M.}~\bibnamefont {Salluzzo}}, \bibinfo {author}
  {\bibfnamefont {H.}~\bibnamefont {Suzuki}}, \bibinfo {author} {\bibfnamefont
  {F.}~\bibnamefont {Yakhou}}, \bibinfo {author} {\bibfnamefont
  {M.}~\bibnamefont {Le~Tacon}}, \bibinfo {author} {\bibfnamefont
  {B.}~\bibnamefont {Keimer}}, \bibinfo {author} {\bibfnamefont {N.~B.}\
  \bibnamefont {Brookes}}, \bibinfo {author} {\bibfnamefont {M.~M.}\
  \bibnamefont {Sala}},\ and\ \bibinfo {author} {\bibfnamefont
  {G.}~\bibnamefont {Ghiringhelli}},\ }\bibfield  {title} {\bibinfo {title}
  {Polarization-resolved {C}u l3-edge resonant inelastic x-ray scattering of
  orbital and spin excitations in {N}d{B}a$_2${C}u$_3${O}$_{7-\delta}$},\
  }\href {https://doi.org/10.1103/PhysRevB.99.134517} {\bibfield  {journal}
  {\bibinfo  {journal} {Phys. Rev. B}\ }\textbf {\bibinfo {volume} {99}},\
  \bibinfo {pages} {134517} (\bibinfo {year} {2019})}\BibitemShut {NoStop}%
\bibitem [{Note1()}]{Note1}%
  \BibitemOpen
  \bibinfo {note} {The observed orbiton dispersion in the {\protect \it doped}
  cuprates~\cite {Ellis2015} has a distinct nature than the one reported here
  and probably can be attributed to the large itinerancy of the doped
  systems.}\BibitemShut {Stop}%
\bibitem [{Note2()}]{Note2}%
  \BibitemOpen
  \bibinfo {note} {Dispersing spin-orbitons have been found in $5d$ iridates
  \cite {kim2012magnetic, kim2014excitonic}, where however the orbital quantum
  number is no more a good one because of the spin-orbit coupling of
  iridium.}\BibitemShut {Stop}%
\bibitem [{\citenamefont {Slater}\ and\ \citenamefont
  {Koster}(1954)}]{Slater1954}%
  \BibitemOpen
  \bibfield  {author} {\bibinfo {author} {\bibfnamefont {J.~C.}\ \bibnamefont
  {Slater}}\ and\ \bibinfo {author} {\bibfnamefont {G.~F.}\ \bibnamefont
  {Koster}},\ }\bibfield  {title} {\bibinfo {title} {Simplified lcao method for
  the periodic potential problem},\ }\href
  {https://doi.org/10.1103/PhysRev.94.1498} {\bibfield  {journal} {\bibinfo
  {journal} {Phys. Rev.}\ }\textbf {\bibinfo {volume} {94}},\ \bibinfo {pages}
  {1498} (\bibinfo {year} {1954})}\BibitemShut {NoStop}%
\bibitem [{\citenamefont {Jefferson}\ \emph {et~al.}(1992)\citenamefont
  {Jefferson}, \citenamefont {Eskes},\ and\ \citenamefont
  {Feiner}}]{Jefferson1992}%
  \BibitemOpen
  \bibfield  {author} {\bibinfo {author} {\bibfnamefont {J.~H.}\ \bibnamefont
  {Jefferson}}, \bibinfo {author} {\bibfnamefont {H.}~\bibnamefont {Eskes}},\
  and\ \bibinfo {author} {\bibfnamefont {L.~F.}\ \bibnamefont {Feiner}},\
  }\bibfield  {title} {\bibinfo {title} {Derivation of a single-band model for
  ${\mathrm{cuo}}_{2}$ planes by a cell-perturbation method},\ }\href
  {https://doi.org/10.1103/PhysRevB.45.7959} {\bibfield  {journal} {\bibinfo
  {journal} {Phys. Rev. B}\ }\textbf {\bibinfo {volume} {45}},\ \bibinfo
  {pages} {7959} (\bibinfo {year} {1992})}\BibitemShut {NoStop}%
\bibitem [{\citenamefont {Bozovic}\ \emph {et~al.}(2002)\citenamefont
  {Bozovic}, \citenamefont {Logvenov}, \citenamefont {Belca}, \citenamefont
  {Narimbetov},\ and\ \citenamefont {Sveklo}}]{bozovic2002epitaxial}%
  \BibitemOpen
  \bibfield  {author} {\bibinfo {author} {\bibfnamefont {I.}~\bibnamefont
  {Bozovic}}, \bibinfo {author} {\bibfnamefont {G.}~\bibnamefont {Logvenov}},
  \bibinfo {author} {\bibfnamefont {I.}~\bibnamefont {Belca}}, \bibinfo
  {author} {\bibfnamefont {B.}~\bibnamefont {Narimbetov}},\ and\ \bibinfo
  {author} {\bibfnamefont {I.}~\bibnamefont {Sveklo}},\ }\bibfield  {title}
  {\bibinfo {title} {{Epitaxial Strain and Superconductivity in
  La$_{2-x}$Sr$_x$CuO$_4$ Thin Films}},\ }\href
  {https://doi.org/10.1103/PhysRevLett.89.107001} {\bibfield  {journal}
  {\bibinfo  {journal} {Phys. Rev. Lett.}\ }\textbf {\bibinfo {volume} {89}},\
  \bibinfo {pages} {107001} (\bibinfo {year} {2002})}\BibitemShut {NoStop}%
\bibitem [{\citenamefont {Castro}\ \emph {et~al.}(2014)\citenamefont {Castro},
  \citenamefont {Aruta}, \citenamefont {Tebano}, \citenamefont {Innocenti},
  \citenamefont {Minola}, \citenamefont {Sala}, \citenamefont {Prellier},
  \citenamefont {Lebedev},\ and\ \citenamefont
  {Balestrino}}]{Di_Castro_2014superconductivity}%
  \BibitemOpen
  \bibfield  {author} {\bibinfo {author} {\bibfnamefont {D.~D.}\ \bibnamefont
  {Castro}}, \bibinfo {author} {\bibfnamefont {C.}~\bibnamefont {Aruta}},
  \bibinfo {author} {\bibfnamefont {A.}~\bibnamefont {Tebano}}, \bibinfo
  {author} {\bibfnamefont {D.}~\bibnamefont {Innocenti}}, \bibinfo {author}
  {\bibfnamefont {M.}~\bibnamefont {Minola}}, \bibinfo {author} {\bibfnamefont
  {M.~M.}\ \bibnamefont {Sala}}, \bibinfo {author} {\bibfnamefont
  {W.}~\bibnamefont {Prellier}}, \bibinfo {author} {\bibfnamefont
  {O.}~\bibnamefont {Lebedev}},\ and\ \bibinfo {author} {\bibfnamefont
  {G.}~\bibnamefont {Balestrino}},\ }\bibfield  {title} {\bibinfo {title}
  {{T$_{\text{c}}$ up to 50 {K} in superlattices of insulating oxides}},\
  }\href {https://doi.org/10.1088/0953-2048/27/4/044016} {\bibfield  {journal}
  {\bibinfo  {journal} {Superconductor Science and Technology}\ }\textbf
  {\bibinfo {volume} {27}},\ \bibinfo {pages} {044016} (\bibinfo {year}
  {2014})}\BibitemShut {NoStop}%
\bibitem [{\citenamefont {Di~Castro}\ \emph {et~al.}(2015)\citenamefont
  {Di~Castro}, \citenamefont {Cantoni}, \citenamefont {Ridolfi}, \citenamefont
  {Aruta}, \citenamefont {Tebano}, \citenamefont {Yang},\ and\ \citenamefont
  {Balestrino}}]{dicastro2015high}%
  \BibitemOpen
  \bibfield  {author} {\bibinfo {author} {\bibfnamefont {D.}~\bibnamefont
  {Di~Castro}}, \bibinfo {author} {\bibfnamefont {C.}~\bibnamefont {Cantoni}},
  \bibinfo {author} {\bibfnamefont {F.}~\bibnamefont {Ridolfi}}, \bibinfo
  {author} {\bibfnamefont {C.}~\bibnamefont {Aruta}}, \bibinfo {author}
  {\bibfnamefont {A.}~\bibnamefont {Tebano}}, \bibinfo {author} {\bibfnamefont
  {N.}~\bibnamefont {Yang}},\ and\ \bibinfo {author} {\bibfnamefont
  {G.}~\bibnamefont {Balestrino}},\ }\bibfield  {title} {\bibinfo {title}
  {{High-${T}_{c}$ Superconductivity at the Interface between the
  ${\mathrm{CaCuO}}_{2}$ and ${\mathrm{SrTiO}}_{3}$ Insulating Oxides}},\
  }\href {https://doi.org/10.1103/PhysRevLett.115.147001} {\bibfield  {journal}
  {\bibinfo  {journal} {Phys. Rev. Lett.}\ }\textbf {\bibinfo {volume} {115}},\
  \bibinfo {pages} {147001} (\bibinfo {year} {2015})}\BibitemShut {NoStop}%
\bibitem [{\citenamefont {Krockenberger}\ \emph {et~al.}(2013)\citenamefont
  {Krockenberger}, \citenamefont {Irie}, \citenamefont {Matsumoto},
  \citenamefont {Yamagami}, \citenamefont {Mitsuhashi}, \citenamefont
  {Tsukada}, \citenamefont {Naito},\ and\ \citenamefont
  {Yamamoto}}]{krockenberger2013emerging}%
  \BibitemOpen
  \bibfield  {author} {\bibinfo {author} {\bibfnamefont {Y.}~\bibnamefont
  {Krockenberger}}, \bibinfo {author} {\bibfnamefont {H.}~\bibnamefont {Irie}},
  \bibinfo {author} {\bibfnamefont {O.}~\bibnamefont {Matsumoto}}, \bibinfo
  {author} {\bibfnamefont {K.}~\bibnamefont {Yamagami}}, \bibinfo {author}
  {\bibfnamefont {M.}~\bibnamefont {Mitsuhashi}}, \bibinfo {author}
  {\bibfnamefont {A.}~\bibnamefont {Tsukada}}, \bibinfo {author} {\bibfnamefont
  {M.}~\bibnamefont {Naito}},\ and\ \bibinfo {author} {\bibfnamefont
  {H.}~\bibnamefont {Yamamoto}},\ }\bibfield  {title} {\bibinfo {title}
  {Emerging superconductivity hidden beneath charge-transfer insulators},\
  }\href {https://doi.org/https://doi.org/10.1038/srep02235} {\bibfield
  {journal} {\bibinfo  {journal} {Scientific Reports}\ }\textbf {\bibinfo
  {volume} {3}},\ \bibinfo {pages} {2235} (\bibinfo {year} {2013})}\BibitemShut
  {NoStop}%
\bibitem [{\citenamefont {Krockenberger}\ \emph {et~al.}(2012)\citenamefont
  {Krockenberger}, \citenamefont {Yamamoto}, \citenamefont {Tsukada},
  \citenamefont {Mitsuhashi},\ and\ \citenamefont
  {Naito}}]{krockenberger2012unconventional}%
  \BibitemOpen
  \bibfield  {author} {\bibinfo {author} {\bibfnamefont {Y.}~\bibnamefont
  {Krockenberger}}, \bibinfo {author} {\bibfnamefont {H.}~\bibnamefont
  {Yamamoto}}, \bibinfo {author} {\bibfnamefont {A.}~\bibnamefont {Tsukada}},
  \bibinfo {author} {\bibfnamefont {M.}~\bibnamefont {Mitsuhashi}},\ and\
  \bibinfo {author} {\bibfnamefont {M.}~\bibnamefont {Naito}},\ }\bibfield
  {title} {\bibinfo {title} {Unconventional transport and superconducting
  properties in electron-doped cuprates},\ }\href
  {https://doi.org/10.1103/PhysRevB.85.184502} {\bibfield  {journal} {\bibinfo
  {journal} {Phys. Rev. B}\ }\textbf {\bibinfo {volume} {85}},\ \bibinfo
  {pages} {184502} (\bibinfo {year} {2012})}\BibitemShut {NoStop}%
\bibitem [{\citenamefont {Brookes}\ \emph {et~al.}(2018)\citenamefont
  {Brookes}, \citenamefont {{Yakhou-Harris}}, \citenamefont {Kummer},
  \citenamefont {Fondacaro}, \citenamefont {Cezar}, \citenamefont {Betto},
  \citenamefont {{Velez-Fort}}, \citenamefont {Amorese}, \citenamefont
  {Ghiringhelli}, \citenamefont {Braicovich}, \citenamefont {Barrett},
  \citenamefont {Berruyer}, \citenamefont {Cianciosi}, \citenamefont {Eybert},
  \citenamefont {Marion}, \citenamefont {{van der Linden}},\ and\ \citenamefont
  {Zhang}}]{brookes2018beamline}%
  \BibitemOpen
  \bibfield  {author} {\bibinfo {author} {\bibfnamefont {N.~B.}\ \bibnamefont
  {Brookes}}, \bibinfo {author} {\bibfnamefont {F.}~\bibnamefont
  {{Yakhou-Harris}}}, \bibinfo {author} {\bibfnamefont {K.}~\bibnamefont
  {Kummer}}, \bibinfo {author} {\bibfnamefont {A.}~\bibnamefont {Fondacaro}},
  \bibinfo {author} {\bibfnamefont {J.~C.}\ \bibnamefont {Cezar}}, \bibinfo
  {author} {\bibfnamefont {D.}~\bibnamefont {Betto}}, \bibinfo {author}
  {\bibfnamefont {E.}~\bibnamefont {{Velez-Fort}}}, \bibinfo {author}
  {\bibfnamefont {A.}~\bibnamefont {Amorese}}, \bibinfo {author} {\bibfnamefont
  {G.}~\bibnamefont {Ghiringhelli}}, \bibinfo {author} {\bibfnamefont
  {L.}~\bibnamefont {Braicovich}}, \bibinfo {author} {\bibfnamefont
  {R.}~\bibnamefont {Barrett}}, \bibinfo {author} {\bibfnamefont
  {G.}~\bibnamefont {Berruyer}}, \bibinfo {author} {\bibfnamefont
  {F.}~\bibnamefont {Cianciosi}}, \bibinfo {author} {\bibfnamefont
  {L.}~\bibnamefont {Eybert}}, \bibinfo {author} {\bibfnamefont
  {P.}~\bibnamefont {Marion}}, \bibinfo {author} {\bibfnamefont
  {P.}~\bibnamefont {{van der Linden}}},\ and\ \bibinfo {author} {\bibfnamefont
  {L.}~\bibnamefont {Zhang}},\ }\bibfield  {title} {\bibinfo {title} {The
  beamline {{ID32}} at the {{ESRF}} for soft {{X}}-ray high energy resolution
  resonant inelastic {{X}}-ray scattering and polarisation dependent {{X}}-ray
  absorption spectroscopy},\ }\href
  {https://doi.org/10.1016/j.nima.2018.07.001} {\bibfield  {journal} {\bibinfo
  {journal} {Nucl. Instrum. Methods A}\ }\textbf {\bibinfo {volume} {903}},\
  \bibinfo {pages} {175} (\bibinfo {year} {2018})}\BibitemShut {NoStop}%
\bibitem [{\citenamefont {Zhou}\ \emph {et~al.}(2022)\citenamefont {Zhou},
  \citenamefont {Walters}, \citenamefont {{Garcia-Fernandez}}, \citenamefont
  {Rice}, \citenamefont {Hand}, \citenamefont {Nag}, \citenamefont {Li},
  \citenamefont {Agrestini}, \citenamefont {Garland}, \citenamefont {Wang},
  \citenamefont {Alcock}, \citenamefont {Nistea}, \citenamefont {Nutter},
  \citenamefont {Rubies}, \citenamefont {Knap}, \citenamefont {Gaughran},
  \citenamefont {Yuan}, \citenamefont {Chang}, \citenamefont {Emmins},\ and\
  \citenamefont {Howell}}]{zhou2022I21}%
  \BibitemOpen
  \bibfield  {author} {\bibinfo {author} {\bibfnamefont {K.-J.}\ \bibnamefont
  {Zhou}}, \bibinfo {author} {\bibfnamefont {A.}~\bibnamefont {Walters}},
  \bibinfo {author} {\bibfnamefont {M.}~\bibnamefont {{Garcia-Fernandez}}},
  \bibinfo {author} {\bibfnamefont {T.}~\bibnamefont {Rice}}, \bibinfo {author}
  {\bibfnamefont {M.}~\bibnamefont {Hand}}, \bibinfo {author} {\bibfnamefont
  {A.}~\bibnamefont {Nag}}, \bibinfo {author} {\bibfnamefont {J.}~\bibnamefont
  {Li}}, \bibinfo {author} {\bibfnamefont {S.}~\bibnamefont {Agrestini}},
  \bibinfo {author} {\bibfnamefont {P.}~\bibnamefont {Garland}}, \bibinfo
  {author} {\bibfnamefont {H.}~\bibnamefont {Wang}}, \bibinfo {author}
  {\bibfnamefont {S.}~\bibnamefont {Alcock}}, \bibinfo {author} {\bibfnamefont
  {I.}~\bibnamefont {Nistea}}, \bibinfo {author} {\bibfnamefont
  {B.}~\bibnamefont {Nutter}}, \bibinfo {author} {\bibfnamefont
  {N.}~\bibnamefont {Rubies}}, \bibinfo {author} {\bibfnamefont
  {G.}~\bibnamefont {Knap}}, \bibinfo {author} {\bibfnamefont {M.}~\bibnamefont
  {Gaughran}}, \bibinfo {author} {\bibfnamefont {F.}~\bibnamefont {Yuan}},
  \bibinfo {author} {\bibfnamefont {P.}~\bibnamefont {Chang}}, \bibinfo
  {author} {\bibfnamefont {J.}~\bibnamefont {Emmins}},\ and\ \bibinfo {author}
  {\bibfnamefont {G.}~\bibnamefont {Howell}},\ }\bibfield  {title} {\bibinfo
  {title} {I21: An advanced high-resolution resonant inelastic {{X-ray}}
  scattering beamline at {{Diamond Light Source}}},\ }\href
  {https://doi.org/10.1107/S1600577522000601} {\bibfield  {journal} {\bibinfo
  {journal} {Journal of Synchrotron Radiation}\ }\textbf {\bibinfo {volume}
  {29}},\ \bibinfo {pages} {563} (\bibinfo {year} {2022})}\BibitemShut
  {NoStop}%
\bibitem [{\citenamefont {Peng}\ \emph {et~al.}(2017)\citenamefont {Peng},
  \citenamefont {Dellea}, \citenamefont {Minola}, \citenamefont {Conni},
  \citenamefont {Amorese}, \citenamefont {Di~Castro}, \citenamefont {De~Luca},
  \citenamefont {Kummer}, \citenamefont {Salluzzo}, \citenamefont {Sun},
  \citenamefont {Zhou}, \citenamefont {Balestrino}, \citenamefont {Le~Tacon},
  \citenamefont {Keimer}, \citenamefont {Braicovich}, \citenamefont {Brookes},\
  and\ \citenamefont {Ghiringhelli}}]{peng2017influence}%
  \BibitemOpen
  \bibfield  {author} {\bibinfo {author} {\bibfnamefont {Y.~Y.}\ \bibnamefont
  {Peng}}, \bibinfo {author} {\bibfnamefont {G.}~\bibnamefont {Dellea}},
  \bibinfo {author} {\bibfnamefont {M.}~\bibnamefont {Minola}}, \bibinfo
  {author} {\bibfnamefont {M.}~\bibnamefont {Conni}}, \bibinfo {author}
  {\bibfnamefont {A.}~\bibnamefont {Amorese}}, \bibinfo {author} {\bibfnamefont
  {D.}~\bibnamefont {Di~Castro}}, \bibinfo {author} {\bibfnamefont {G.~M.}\
  \bibnamefont {De~Luca}}, \bibinfo {author} {\bibfnamefont {K.}~\bibnamefont
  {Kummer}}, \bibinfo {author} {\bibfnamefont {M.}~\bibnamefont {Salluzzo}},
  \bibinfo {author} {\bibfnamefont {X.}~\bibnamefont {Sun}}, \bibinfo {author}
  {\bibfnamefont {X.~J.}\ \bibnamefont {Zhou}}, \bibinfo {author}
  {\bibfnamefont {G.}~\bibnamefont {Balestrino}}, \bibinfo {author}
  {\bibfnamefont {M.}~\bibnamefont {Le~Tacon}}, \bibinfo {author}
  {\bibfnamefont {B.}~\bibnamefont {Keimer}}, \bibinfo {author} {\bibfnamefont
  {L.}~\bibnamefont {Braicovich}}, \bibinfo {author} {\bibfnamefont {N.~B.}\
  \bibnamefont {Brookes}},\ and\ \bibinfo {author} {\bibfnamefont
  {G.}~\bibnamefont {Ghiringhelli}},\ }\bibfield  {title} {\bibinfo {title}
  {Influence of apical oxygen on the extent of in-plane exchange interaction in
  cuprate superconductors},\ }\href
  {https://doi.org/https://doi.org/10.1038/nphys4248} {\bibfield  {journal}
  {\bibinfo  {journal} {Nature Physics}\ }\textbf {\bibinfo {volume} {13}},\
  \bibinfo {pages} {1201–1206} (\bibinfo {year} {2017})}\BibitemShut
  {NoStop}%
\bibitem [{\citenamefont {Martinelli}\ \emph {et~al.}(2022)\citenamefont
  {Martinelli}, \citenamefont {Betto}, \citenamefont {Kummer}, \citenamefont
  {Arpaia}, \citenamefont {Braicovich}, \citenamefont {Di~Castro},
  \citenamefont {Brookes}, \citenamefont {Moretti~Sala},\ and\ \citenamefont
  {Ghiringhelli}}]{martinelli2022fractional}%
  \BibitemOpen
  \bibfield  {author} {\bibinfo {author} {\bibfnamefont {L.}~\bibnamefont
  {Martinelli}}, \bibinfo {author} {\bibfnamefont {D.}~\bibnamefont {Betto}},
  \bibinfo {author} {\bibfnamefont {K.}~\bibnamefont {Kummer}}, \bibinfo
  {author} {\bibfnamefont {R.}~\bibnamefont {Arpaia}}, \bibinfo {author}
  {\bibfnamefont {L.}~\bibnamefont {Braicovich}}, \bibinfo {author}
  {\bibfnamefont {D.}~\bibnamefont {Di~Castro}}, \bibinfo {author}
  {\bibfnamefont {N.~B.}\ \bibnamefont {Brookes}}, \bibinfo {author}
  {\bibfnamefont {M.}~\bibnamefont {Moretti~Sala}},\ and\ \bibinfo {author}
  {\bibfnamefont {G.}~\bibnamefont {Ghiringhelli}},\ }\bibfield  {title}
  {\bibinfo {title} {Fractional spin excitations in the infinite-layer cuprate
  ${\mathrm{cacuo}}_{2}$},\ }\href {https://doi.org/10.1103/PhysRevX.12.021041}
  {\bibfield  {journal} {\bibinfo  {journal} {Phys. Rev. X}\ }\textbf {\bibinfo
  {volume} {12}},\ \bibinfo {pages} {021041} (\bibinfo {year}
  {2022})}\BibitemShut {NoStop}%
\bibitem [{\citenamefont {Ohta}\ \emph {et~al.}(1991)\citenamefont {Ohta},
  \citenamefont {Tohyama},\ and\ \citenamefont {Maekawa}}]{ohta1991charge}%
  \BibitemOpen
  \bibfield  {author} {\bibinfo {author} {\bibfnamefont {Y.}~\bibnamefont
  {Ohta}}, \bibinfo {author} {\bibfnamefont {T.}~\bibnamefont {Tohyama}},\ and\
  \bibinfo {author} {\bibfnamefont {S.}~\bibnamefont {Maekawa}},\ }\bibfield
  {title} {\bibinfo {title} {Charge-transfer gap and superexchange interaction
  in insulating cuprates},\ }\href
  {https://doi.org/10.1103/PhysRevLett.66.1228} {\bibfield  {journal} {\bibinfo
   {journal} {Phys. Rev. Lett.}\ }\textbf {\bibinfo {volume} {66}},\ \bibinfo
  {pages} {1228} (\bibinfo {year} {1991})}\BibitemShut {NoStop}%
\bibitem [{\citenamefont {Weber}\ \emph {et~al.}(2012)\citenamefont {Weber},
  \citenamefont {Yee}, \citenamefont {Haule},\ and\ \citenamefont
  {Kotliar}}]{Weber2012}%
  \BibitemOpen
  \bibfield  {author} {\bibinfo {author} {\bibfnamefont {C.}~\bibnamefont
  {Weber}}, \bibinfo {author} {\bibfnamefont {C.}~\bibnamefont {Yee}}, \bibinfo
  {author} {\bibfnamefont {K.}~\bibnamefont {Haule}},\ and\ \bibinfo {author}
  {\bibfnamefont {G.}~\bibnamefont {Kotliar}},\ }\bibfield  {title} {\bibinfo
  {title} {Scaling of the transition temperature of hole-doped cuprate
  superconductors with the charge-transfer energy},\ }\href
  {https://doi.org/10.1209/0295-5075/100/37001} {\bibfield  {journal} {\bibinfo
   {journal} {Europhysics Letters}\ }\textbf {\bibinfo {volume} {100}},\
  \bibinfo {pages} {37001} (\bibinfo {year} {2012})}\BibitemShut {NoStop}%
\bibitem [{\citenamefont {Pavarini}\ \emph {et~al.}(2001)\citenamefont
  {Pavarini}, \citenamefont {Dasgupta}, \citenamefont {Saha-Dasgupta},
  \citenamefont {Jepsen},\ and\ \citenamefont {Andersen}}]{pavarini2001band}%
  \BibitemOpen
  \bibfield  {author} {\bibinfo {author} {\bibfnamefont {E.}~\bibnamefont
  {Pavarini}}, \bibinfo {author} {\bibfnamefont {I.}~\bibnamefont {Dasgupta}},
  \bibinfo {author} {\bibfnamefont {T.}~\bibnamefont {Saha-Dasgupta}}, \bibinfo
  {author} {\bibfnamefont {O.}~\bibnamefont {Jepsen}},\ and\ \bibinfo {author}
  {\bibfnamefont {O.~K.}\ \bibnamefont {Andersen}},\ }\bibfield  {title}
  {\bibinfo {title} {Band-structure trend in hole-doped cuprates and
  correlation with ${\mathit{t}}_{\mathit{c}\mathrm{max}}$},\ }\href
  {https://doi.org/10.1103/PhysRevLett.87.047003} {\bibfield  {journal}
  {\bibinfo  {journal} {Phys. Rev. Lett.}\ }\textbf {\bibinfo {volume} {87}},\
  \bibinfo {pages} {047003} (\bibinfo {year} {2001})}\BibitemShut {NoStop}%
\bibitem [{Note3()}]{Note3}%
  \BibitemOpen
  \bibinfo {note} {For simplicity, we focus below on the $xy$ excitation
  leaving an even more complex~\cite {SupplementaryMat} $xz/yz$ case for
  further studies.}\BibitemShut {Stop}%
\bibitem [{\citenamefont {Bari{\v{s}}i{\'c}}\ and\ \citenamefont
  {Sunko}(2022)}]{barivsic2022high}%
  \BibitemOpen
  \bibfield  {author} {\bibinfo {author} {\bibfnamefont {N.}~\bibnamefont
  {Bari{\v{s}}i{\'c}}}\ and\ \bibinfo {author} {\bibfnamefont {D.~K.}\
  \bibnamefont {Sunko}},\ }\bibfield  {title} {\bibinfo {title} {{High-T$_c$
  Cuprates: a Story of Two Electronic Subsystems}},\ }\href
  {https://doi.org/10.1007/s10948-022-06183-y} {\bibfield  {journal} {\bibinfo
  {journal} {Journal of Superconductivity and Novel Magnetism}\ ,\ \bibinfo
  {pages} {1}} (\bibinfo {year} {2022})}\BibitemShut {NoStop}%
\bibitem [{\citenamefont {Perucchi}\ \emph {et~al.}(2018)\citenamefont
  {Perucchi}, \citenamefont {Di~Pietro}, \citenamefont {Lupi}, \citenamefont
  {Sopracase}, \citenamefont {Tebano}, \citenamefont {Giovannetti},
  \citenamefont {Petocchi}, \citenamefont {Capone},\ and\ \citenamefont
  {Di~Castro}}]{perucchi2018electrodynamic}%
  \BibitemOpen
  \bibfield  {author} {\bibinfo {author} {\bibfnamefont {A.}~\bibnamefont
  {Perucchi}}, \bibinfo {author} {\bibfnamefont {P.}~\bibnamefont {Di~Pietro}},
  \bibinfo {author} {\bibfnamefont {S.}~\bibnamefont {Lupi}}, \bibinfo {author}
  {\bibfnamefont {R.}~\bibnamefont {Sopracase}}, \bibinfo {author}
  {\bibfnamefont {A.}~\bibnamefont {Tebano}}, \bibinfo {author} {\bibfnamefont
  {G.}~\bibnamefont {Giovannetti}}, \bibinfo {author} {\bibfnamefont
  {F.}~\bibnamefont {Petocchi}}, \bibinfo {author} {\bibfnamefont
  {M.}~\bibnamefont {Capone}},\ and\ \bibinfo {author} {\bibfnamefont
  {D.}~\bibnamefont {Di~Castro}},\ }\bibfield  {title} {\bibinfo {title}
  {Electrodynamic properties of an artificial heterostructured superconducting
  cuprate},\ }\href {https://doi.org/10.1103/PhysRevB.97.045114} {\bibfield
  {journal} {\bibinfo  {journal} {Phys. Rev. B}\ }\textbf {\bibinfo {volume}
  {97}},\ \bibinfo {pages} {045114} (\bibinfo {year} {2018})}\BibitemShut
  {NoStop}%
\bibitem [{\citenamefont {Kan}\ \emph {et~al.}(2004)\citenamefont {Kan},
  \citenamefont {Yamanaka}, \citenamefont {Terashima},\ and\ \citenamefont
  {Takano}}]{kan2004preparation}%
  \BibitemOpen
  \bibfield  {author} {\bibinfo {author} {\bibfnamefont {D.}~\bibnamefont
  {Kan}}, \bibinfo {author} {\bibfnamefont {A.}~\bibnamefont {Yamanaka}},
  \bibinfo {author} {\bibfnamefont {T.}~\bibnamefont {Terashima}},\ and\
  \bibinfo {author} {\bibfnamefont {M.}~\bibnamefont {Takano}},\ }\bibfield
  {title} {\bibinfo {title} {{Preparation and optical properties of
  single-crystalline CaCuO2 thin films with infinite layer structure}},\ }\href
  {https://doi.org/https://doi.org/10.1016/j.physc.2003.12.040} {\bibfield
  {journal} {\bibinfo  {journal} {Physica C: Superconductivity}\ }\textbf
  {\bibinfo {volume} {412-414}},\ \bibinfo {pages} {298} (\bibinfo {year}
  {2004})},\ \bibinfo {note} {proceedings of the 16th International Symposium
  on Superconductivity (ISS 2003). Advances in Superconductivity XVI. Part
  I}\BibitemShut {NoStop}%
\bibitem [{\citenamefont {Yoshida}\ \emph {et~al.}(1992)\citenamefont
  {Yoshida}, \citenamefont {Tajima}, \citenamefont {Koshizuka}, \citenamefont
  {Tanaka}, \citenamefont {Uchida},\ and\ \citenamefont
  {Itoh}}]{yoshida1992two}%
  \BibitemOpen
  \bibfield  {author} {\bibinfo {author} {\bibfnamefont {M.}~\bibnamefont
  {Yoshida}}, \bibinfo {author} {\bibfnamefont {S.}~\bibnamefont {Tajima}},
  \bibinfo {author} {\bibfnamefont {N.}~\bibnamefont {Koshizuka}}, \bibinfo
  {author} {\bibfnamefont {S.}~\bibnamefont {Tanaka}}, \bibinfo {author}
  {\bibfnamefont {S.}~\bibnamefont {Uchida}},\ and\ \bibinfo {author}
  {\bibfnamefont {T.}~\bibnamefont {Itoh}},\ }\bibfield  {title} {\bibinfo
  {title} {Two-magnon and two-phonon excitations in some parent insulating
  compounds of the high-${\mathit{t}}_{\mathit{c}}$ cuprates},\ }\href
  {https://doi.org/10.1103/PhysRevB.46.6505} {\bibfield  {journal} {\bibinfo
  {journal} {Phys. Rev. B}\ }\textbf {\bibinfo {volume} {46}},\ \bibinfo
  {pages} {6505} (\bibinfo {year} {1992})}\BibitemShut {NoStop}%
\bibitem [{\citenamefont {Botana}\ and\ \citenamefont
  {Norman}(2020)}]{Botana2020}%
  \BibitemOpen
  \bibfield  {author} {\bibinfo {author} {\bibfnamefont {A.~S.}\ \bibnamefont
  {Botana}}\ and\ \bibinfo {author} {\bibfnamefont {M.~R.}\ \bibnamefont
  {Norman}},\ }\bibfield  {title} {\bibinfo {title} {Similarities and
  differences between ${\mathrm{lanio}}_{2}$ and ${\mathrm{cacuo}}_{2}$ and
  implications for superconductivity},\ }\href
  {https://doi.org/10.1103/PhysRevX.10.011024} {\bibfield  {journal} {\bibinfo
  {journal} {Phys. Rev. X}\ }\textbf {\bibinfo {volume} {10}},\ \bibinfo
  {pages} {011024} (\bibinfo {year} {2020})}\BibitemShut {NoStop}%
\bibitem [{\citenamefont {Neudert}\ \emph {et~al.}(2000)\citenamefont
  {Neudert}, \citenamefont {Drechsler}, \citenamefont {M\'alek}, \citenamefont
  {Rosner}, \citenamefont {Kielwein}, \citenamefont {Hu}, \citenamefont
  {Knupfer}, \citenamefont {Golden}, \citenamefont {Fink}, \citenamefont
  {N\"ucker}, \citenamefont {Merz}, \citenamefont {Schuppler}, \citenamefont
  {Motoyama}, \citenamefont {Eisaki}, \citenamefont {Uchida}, \citenamefont
  {Domke},\ and\ \citenamefont {Kaindl}}]{neudert2000four}%
  \BibitemOpen
  \bibfield  {author} {\bibinfo {author} {\bibfnamefont {R.}~\bibnamefont
  {Neudert}}, \bibinfo {author} {\bibfnamefont {S.-L.}\ \bibnamefont
  {Drechsler}}, \bibinfo {author} {\bibfnamefont {J.}~\bibnamefont {M\'alek}},
  \bibinfo {author} {\bibfnamefont {H.}~\bibnamefont {Rosner}}, \bibinfo
  {author} {\bibfnamefont {M.}~\bibnamefont {Kielwein}}, \bibinfo {author}
  {\bibfnamefont {Z.}~\bibnamefont {Hu}}, \bibinfo {author} {\bibfnamefont
  {M.}~\bibnamefont {Knupfer}}, \bibinfo {author} {\bibfnamefont {M.~S.}\
  \bibnamefont {Golden}}, \bibinfo {author} {\bibfnamefont {J.}~\bibnamefont
  {Fink}}, \bibinfo {author} {\bibfnamefont {N.}~\bibnamefont {N\"ucker}},
  \bibinfo {author} {\bibfnamefont {M.}~\bibnamefont {Merz}}, \bibinfo {author}
  {\bibfnamefont {S.}~\bibnamefont {Schuppler}}, \bibinfo {author}
  {\bibfnamefont {N.}~\bibnamefont {Motoyama}}, \bibinfo {author}
  {\bibfnamefont {H.}~\bibnamefont {Eisaki}}, \bibinfo {author} {\bibfnamefont
  {S.}~\bibnamefont {Uchida}}, \bibinfo {author} {\bibfnamefont
  {M.}~\bibnamefont {Domke}},\ and\ \bibinfo {author} {\bibfnamefont
  {G.}~\bibnamefont {Kaindl}},\ }\bibfield  {title} {\bibinfo {title}
  {Four-band extended hubbard hamiltonian for the one-dimensional cuprate
  ${\mathrm{sr}}_{2}{\mathrm{cuo}}_{3}:$ distribution of oxygen holes and its
  relation to strong intersite coulomb interaction},\ }\href
  {https://doi.org/10.1103/PhysRevB.62.10752} {\bibfield  {journal} {\bibinfo
  {journal} {Phys. Rev. B}\ }\textbf {\bibinfo {volume} {62}},\ \bibinfo
  {pages} {10752} (\bibinfo {year} {2000})}\BibitemShut {NoStop}%
\bibitem [{\citenamefont {Kim}\ \emph {et~al.}(2003)\citenamefont {Kim},
  \citenamefont {Rosner}, \citenamefont {Drechsler}, \citenamefont {Hu},
  \citenamefont {Sekar}, \citenamefont {Krabbes}, \citenamefont {M\'alek},
  \citenamefont {Knupfer}, \citenamefont {Fink},\ and\ \citenamefont
  {Eschrig}}]{kim2003unusual}%
  \BibitemOpen
  \bibfield  {author} {\bibinfo {author} {\bibfnamefont {T.~K.}\ \bibnamefont
  {Kim}}, \bibinfo {author} {\bibfnamefont {H.}~\bibnamefont {Rosner}},
  \bibinfo {author} {\bibfnamefont {S.-L.}\ \bibnamefont {Drechsler}}, \bibinfo
  {author} {\bibfnamefont {Z.}~\bibnamefont {Hu}}, \bibinfo {author}
  {\bibfnamefont {C.}~\bibnamefont {Sekar}}, \bibinfo {author} {\bibfnamefont
  {G.}~\bibnamefont {Krabbes}}, \bibinfo {author} {\bibfnamefont
  {J.}~\bibnamefont {M\'alek}}, \bibinfo {author} {\bibfnamefont
  {M.}~\bibnamefont {Knupfer}}, \bibinfo {author} {\bibfnamefont
  {J.}~\bibnamefont {Fink}},\ and\ \bibinfo {author} {\bibfnamefont
  {H.}~\bibnamefont {Eschrig}},\ }\bibfield  {title} {\bibinfo {title} {Unusual
  electronic structure of the pseudoladder compound
  ${\mathrm{cacu}}_{2}{\mathrm{o}}_{3}$},\ }\href
  {https://doi.org/10.1103/PhysRevB.67.024516} {\bibfield  {journal} {\bibinfo
  {journal} {Phys. Rev. B}\ }\textbf {\bibinfo {volume} {67}},\ \bibinfo
  {pages} {024516} (\bibinfo {year} {2003})}\BibitemShut {NoStop}%
\bibitem [{\citenamefont {Watanabe}\ \emph {et~al.}(2021)\citenamefont
  {Watanabe}, \citenamefont {Shirakawa}, \citenamefont {Seki}, \citenamefont
  {Sakakibara}, \citenamefont {Kotani}, \citenamefont {Ikeda},\ and\
  \citenamefont {Yunoki}}]{watanabe2021unified}%
  \BibitemOpen
  \bibfield  {author} {\bibinfo {author} {\bibfnamefont {H.}~\bibnamefont
  {Watanabe}}, \bibinfo {author} {\bibfnamefont {T.}~\bibnamefont {Shirakawa}},
  \bibinfo {author} {\bibfnamefont {K.}~\bibnamefont {Seki}}, \bibinfo {author}
  {\bibfnamefont {H.}~\bibnamefont {Sakakibara}}, \bibinfo {author}
  {\bibfnamefont {T.}~\bibnamefont {Kotani}}, \bibinfo {author} {\bibfnamefont
  {H.}~\bibnamefont {Ikeda}},\ and\ \bibinfo {author} {\bibfnamefont
  {S.}~\bibnamefont {Yunoki}},\ }\bibfield  {title} {\bibinfo {title} {Unified
  description of cuprate superconductors using a four-band
  $d\text{\ensuremath{-}}p$ model},\ }\href
  {https://doi.org/10.1103/PhysRevResearch.3.033157} {\bibfield  {journal}
  {\bibinfo  {journal} {Phys. Rev. Res.}\ }\textbf {\bibinfo {volume} {3}},\
  \bibinfo {pages} {033157} (\bibinfo {year} {2021})}\BibitemShut {NoStop}%
\bibitem [{\citenamefont {Kim}\ \emph {et~al.}(2012)\citenamefont {Kim},
  \citenamefont {Casa}, \citenamefont {Upton}, \citenamefont {Gog},
  \citenamefont {Kim}, \citenamefont {Mitchell}, \citenamefont {van
  Veenendaal}, \citenamefont {Daghofer}, \citenamefont {van~den Brink},
  \citenamefont {Khaliullin},\ and\ \citenamefont {Kim}}]{kim2012magnetic}%
  \BibitemOpen
  \bibfield  {author} {\bibinfo {author} {\bibfnamefont {J.}~\bibnamefont
  {Kim}}, \bibinfo {author} {\bibfnamefont {D.}~\bibnamefont {Casa}}, \bibinfo
  {author} {\bibfnamefont {M.~H.}\ \bibnamefont {Upton}}, \bibinfo {author}
  {\bibfnamefont {T.}~\bibnamefont {Gog}}, \bibinfo {author} {\bibfnamefont
  {Y.-J.}\ \bibnamefont {Kim}}, \bibinfo {author} {\bibfnamefont {J.~F.}\
  \bibnamefont {Mitchell}}, \bibinfo {author} {\bibfnamefont {M.}~\bibnamefont
  {van Veenendaal}}, \bibinfo {author} {\bibfnamefont {M.}~\bibnamefont
  {Daghofer}}, \bibinfo {author} {\bibfnamefont {J.}~\bibnamefont {van~den
  Brink}}, \bibinfo {author} {\bibfnamefont {G.}~\bibnamefont {Khaliullin}},\
  and\ \bibinfo {author} {\bibfnamefont {B.~J.}\ \bibnamefont {Kim}},\
  }\bibfield  {title} {\bibinfo {title} {{Magnetic Excitation Spectra of
  ${\mathrm{Sr}}_{2}{\mathrm{IrO}}_{4}$ Probed by Resonant Inelastic X-Ray
  Scattering: Establishing Links to Cuprate Superconductors}},\ }\href
  {https://doi.org/10.1103/PhysRevLett.108.177003} {\bibfield  {journal}
  {\bibinfo  {journal} {Phys. Rev. Lett.}\ }\textbf {\bibinfo {volume} {108}},\
  \bibinfo {pages} {177003} (\bibinfo {year} {2012})}\BibitemShut {NoStop}%
\bibitem [{\citenamefont {Marciniak}\ \emph {et~al.}(2021)\citenamefont
  {Marciniak}, \citenamefont {Marcantoni}, \citenamefont {Giusti},
  \citenamefont {Glerean}, \citenamefont {Sparapassi}, \citenamefont {Nova},
  \citenamefont {Cartella}, \citenamefont {Latini}, \citenamefont {Valiera},
  \citenamefont {Rubio}, \citenamefont {{van den Brink}}, \citenamefont
  {Benatti},\ and\ \citenamefont
  {Fausti}}]{marciniakVibrationalCoherentControl2021}%
  \BibitemOpen
  \bibfield  {author} {\bibinfo {author} {\bibfnamefont {A.}~\bibnamefont
  {Marciniak}}, \bibinfo {author} {\bibfnamefont {S.}~\bibnamefont
  {Marcantoni}}, \bibinfo {author} {\bibfnamefont {F.}~\bibnamefont {Giusti}},
  \bibinfo {author} {\bibfnamefont {F.}~\bibnamefont {Glerean}}, \bibinfo
  {author} {\bibfnamefont {G.}~\bibnamefont {Sparapassi}}, \bibinfo {author}
  {\bibfnamefont {T.}~\bibnamefont {Nova}}, \bibinfo {author} {\bibfnamefont
  {A.}~\bibnamefont {Cartella}}, \bibinfo {author} {\bibfnamefont
  {S.}~\bibnamefont {Latini}}, \bibinfo {author} {\bibfnamefont
  {F.}~\bibnamefont {Valiera}}, \bibinfo {author} {\bibfnamefont
  {A.}~\bibnamefont {Rubio}}, \bibinfo {author} {\bibfnamefont
  {J.}~\bibnamefont {{van den Brink}}}, \bibinfo {author} {\bibfnamefont
  {F.}~\bibnamefont {Benatti}},\ and\ \bibinfo {author} {\bibfnamefont
  {D.}~\bibnamefont {Fausti}},\ }\bibfield  {title} {\bibinfo {title}
  {Vibrational coherent control of localized d\textendash d electronic
  excitation},\ }\href {https://doi.org/10.1038/s41567-020-01098-8} {\bibfield
  {journal} {\bibinfo  {journal} {Nature Physics}\ }\textbf {\bibinfo {volume}
  {17}},\ \bibinfo {pages} {368} (\bibinfo {year} {2021})}\BibitemShut
  {NoStop}%
\bibitem [{\citenamefont {Andersen}\ \emph {et~al.}(1996)\citenamefont
  {Andersen}, \citenamefont {Savrasov}, \citenamefont {Jepsen},\ and\
  \citenamefont {Liechtenstein}}]{andersen1996out}%
  \BibitemOpen
  \bibfield  {author} {\bibinfo {author} {\bibfnamefont {O.~K.}\ \bibnamefont
  {Andersen}}, \bibinfo {author} {\bibfnamefont {S.~Y.}\ \bibnamefont
  {Savrasov}}, \bibinfo {author} {\bibfnamefont {O.}~\bibnamefont {Jepsen}},\
  and\ \bibinfo {author} {\bibfnamefont {A.~I.}\ \bibnamefont
  {Liechtenstein}},\ }\bibfield  {title} {\bibinfo {title} {Out-of-plane
  instability and electron-phonon contribution tos- andd-wave pairing in
  high-temperature superconductors; {{LDA}} linear-response calculation for
  doped {{CaCuO2}} and a generic tight-binding model},\ }\href
  {https://doi.org/10.1007/BF00768402} {\bibfield  {journal} {\bibinfo
  {journal} {Journal of Low Temperature Physics}\ }\textbf {\bibinfo {volume}
  {105}},\ \bibinfo {pages} {285} (\bibinfo {year} {1996})}\BibitemShut
  {NoStop}%
\bibitem [{\citenamefont {Bieniasz}\ \emph {et~al.}(2022)\citenamefont
  {Bieniasz}, \citenamefont {Johnston},\ and\ \citenamefont
  {Berciu}}]{Bieniasz2022}%
  \BibitemOpen
  \bibfield  {author} {\bibinfo {author} {\bibfnamefont {K.}~\bibnamefont
  {Bieniasz}}, \bibinfo {author} {\bibfnamefont {S.}~\bibnamefont {Johnston}},\
  and\ \bibinfo {author} {\bibfnamefont {M.}~\bibnamefont {Berciu}},\
  }\bibfield  {title} {\bibinfo {title} {Theory of dispersive optical phonons
  in resonant inelastic x-ray scattering experiments},\ }\href
  {https://doi.org/10.1103/PhysRevB.105.L180302} {\bibfield  {journal}
  {\bibinfo  {journal} {Phys. Rev. B}\ }\textbf {\bibinfo {volume} {105}},\
  \bibinfo {pages} {L180302} (\bibinfo {year} {2022})}\BibitemShut {NoStop}%
\bibitem [{\citenamefont {Bulaevskii}\ \emph {et~al.}(1968)\citenamefont
  {Bulaevskii}, \citenamefont {Nagaev},\ and\ \citenamefont
  {Khomskii}}]{Bulaevskii1968}%
  \BibitemOpen
  \bibfield  {author} {\bibinfo {author} {\bibfnamefont {L.~N.}\ \bibnamefont
  {Bulaevskii}}, \bibinfo {author} {\bibfnamefont {E.~L.}\ \bibnamefont
  {Nagaev}},\ and\ \bibinfo {author} {\bibfnamefont {D.~I.}\ \bibnamefont
  {Khomskii}},\ }\bibfield  {title} {\bibinfo {title} {{A New Type of
  Auto-localized State of a Conduction Electron in an Antiferromagnetic
  Semiconductor}},\ }\href
  {http://www.jetp.ac.ru/cgi-bin/e/index/e/27/5/p836?a=list} {\bibfield
  {journal} {\bibinfo  {journal} {JETP}\ }\textbf {\bibinfo {volume} {27}},\
  \bibinfo {pages} {836} (\bibinfo {year} {1968})}\BibitemShut {NoStop}%
\bibitem [{\citenamefont {Mart\'inez}\ and\ \citenamefont
  {Horsch}(1991)}]{Martinez1991}%
  \BibitemOpen
  \bibfield  {author} {\bibinfo {author} {\bibfnamefont {G.}~\bibnamefont
  {Mart\'inez}}\ and\ \bibinfo {author} {\bibfnamefont {P.}~\bibnamefont
  {Horsch}},\ }\bibfield  {title} {\bibinfo {title} {{Spin polarons in the
  $t$-$J$ model}},\ }\href {https://doi.org/10.1103/PhysRevB.44.317} {\bibfield
   {journal} {\bibinfo  {journal} {Phys. Rev. B}\ }\textbf {\bibinfo {volume}
  {44}},\ \bibinfo {pages} {317} (\bibinfo {year} {1991})}\BibitemShut
  {NoStop}%
\bibitem [{\citenamefont {Bohrdt}\ \emph {et~al.}(2020)\citenamefont {Bohrdt},
  \citenamefont {Demler}, \citenamefont {Pollmann}, \citenamefont {Knap},\ and\
  \citenamefont {Grusdt}}]{Bohrdt2020}%
  \BibitemOpen
  \bibfield  {author} {\bibinfo {author} {\bibfnamefont {A.}~\bibnamefont
  {Bohrdt}}, \bibinfo {author} {\bibfnamefont {E.}~\bibnamefont {Demler}},
  \bibinfo {author} {\bibfnamefont {F.}~\bibnamefont {Pollmann}}, \bibinfo
  {author} {\bibfnamefont {M.}~\bibnamefont {Knap}},\ and\ \bibinfo {author}
  {\bibfnamefont {F.}~\bibnamefont {Grusdt}},\ }\bibfield  {title} {\bibinfo
  {title} {Parton theory of angle-resolved photoemission spectroscopy spectra
  in antiferromagnetic mott insulators},\ }\href
  {https://doi.org/10.1103/PhysRevB.102.035139} {\bibfield  {journal} {\bibinfo
   {journal} {Phys. Rev. B}\ }\textbf {\bibinfo {volume} {102}},\ \bibinfo
  {pages} {035139} (\bibinfo {year} {2020})}\BibitemShut {NoStop}%
\bibitem [{\citenamefont {Devereaux}\ \emph {et~al.}(2016)\citenamefont
  {Devereaux}, \citenamefont {Shvaika}, \citenamefont {Wu}, \citenamefont
  {Wohlfeld}, \citenamefont {Jia}, \citenamefont {Wang}, \citenamefont
  {Moritz}, \citenamefont {Chaix}, \citenamefont {Lee}, \citenamefont {Shen},
  \citenamefont {Ghiringhelli},\ and\ \citenamefont
  {Braicovich}}]{Devereaux_Shvaika2016}%
  \BibitemOpen
  \bibfield  {author} {\bibinfo {author} {\bibfnamefont {T.~P.}\ \bibnamefont
  {Devereaux}}, \bibinfo {author} {\bibfnamefont {A.~M.}\ \bibnamefont
  {Shvaika}}, \bibinfo {author} {\bibfnamefont {K.}~\bibnamefont {Wu}},
  \bibinfo {author} {\bibfnamefont {K.}~\bibnamefont {Wohlfeld}}, \bibinfo
  {author} {\bibfnamefont {C.~J.}\ \bibnamefont {Jia}}, \bibinfo {author}
  {\bibfnamefont {Y.}~\bibnamefont {Wang}}, \bibinfo {author} {\bibfnamefont
  {B.}~\bibnamefont {Moritz}}, \bibinfo {author} {\bibfnamefont
  {L.}~\bibnamefont {Chaix}}, \bibinfo {author} {\bibfnamefont {W.-S.}\
  \bibnamefont {Lee}}, \bibinfo {author} {\bibfnamefont {Z.-X.}\ \bibnamefont
  {Shen}}, \bibinfo {author} {\bibfnamefont {G.}~\bibnamefont {Ghiringhelli}},\
  and\ \bibinfo {author} {\bibfnamefont {L.}~\bibnamefont {Braicovich}},\
  }\bibfield  {title} {\bibinfo {title} {Directly characterizing the relative
  strength and momentum dependence of electron-phonon coupling using resonant
  inelastic x-ray scattering},\ }\href
  {https://doi.org/10.1103/PhysRevX.6.041019} {\bibfield  {journal} {\bibinfo
  {journal} {Phys. Rev. X}\ }\textbf {\bibinfo {volume} {6}},\ \bibinfo {pages}
  {041019} (\bibinfo {year} {2016})}\BibitemShut {NoStop}%
\bibitem [{\citenamefont {Lee}\ \emph {et~al.}(2013)\citenamefont {Lee},
  \citenamefont {Johnston}, \citenamefont {Moritz}, \citenamefont {Lee},
  \citenamefont {Yi}, \citenamefont {Zhou}, \citenamefont {Schmitt},
  \citenamefont {Patthey}, \citenamefont {Strocov}, \citenamefont {Kudo},
  \citenamefont {Koike}, \citenamefont {van~den Brink}, \citenamefont
  {Devereaux},\ and\ \citenamefont {Shen}}]{Lee2013}%
  \BibitemOpen
  \bibfield  {author} {\bibinfo {author} {\bibfnamefont {W.~S.}\ \bibnamefont
  {Lee}}, \bibinfo {author} {\bibfnamefont {S.}~\bibnamefont {Johnston}},
  \bibinfo {author} {\bibfnamefont {B.}~\bibnamefont {Moritz}}, \bibinfo
  {author} {\bibfnamefont {J.}~\bibnamefont {Lee}}, \bibinfo {author}
  {\bibfnamefont {M.}~\bibnamefont {Yi}}, \bibinfo {author} {\bibfnamefont
  {K.~J.}\ \bibnamefont {Zhou}}, \bibinfo {author} {\bibfnamefont
  {T.}~\bibnamefont {Schmitt}}, \bibinfo {author} {\bibfnamefont
  {L.}~\bibnamefont {Patthey}}, \bibinfo {author} {\bibfnamefont
  {V.}~\bibnamefont {Strocov}}, \bibinfo {author} {\bibfnamefont
  {K.}~\bibnamefont {Kudo}}, \bibinfo {author} {\bibfnamefont {Y.}~\bibnamefont
  {Koike}}, \bibinfo {author} {\bibfnamefont {J.}~\bibnamefont {van~den
  Brink}}, \bibinfo {author} {\bibfnamefont {T.~P.}\ \bibnamefont
  {Devereaux}},\ and\ \bibinfo {author} {\bibfnamefont {Z.~X.}\ \bibnamefont
  {Shen}},\ }\bibfield  {title} {\bibinfo {title} {Role of lattice coupling in
  establishing electronic and magnetic properties in quasi-one-dimensional
  cuprates},\ }\href {https://doi.org/10.1103/PhysRevLett.110.265502}
  {\bibfield  {journal} {\bibinfo  {journal} {Phys. Rev. Lett.}\ }\textbf
  {\bibinfo {volume} {110}},\ \bibinfo {pages} {265502} (\bibinfo {year}
  {2013})}\BibitemShut {NoStop}%
\bibitem [{\citenamefont {Walters}\ \emph {et~al.}(2009)\citenamefont
  {Walters}, \citenamefont {Perring}, \citenamefont {Caux}, \citenamefont
  {Savici}, \citenamefont {Gu}, \citenamefont {Lee}, \citenamefont {Ku},\ and\
  \citenamefont {Zaliznyak}}]{Walters2009}%
  \BibitemOpen
  \bibfield  {author} {\bibinfo {author} {\bibfnamefont {A.~C.}\ \bibnamefont
  {Walters}}, \bibinfo {author} {\bibfnamefont {T.~G.}\ \bibnamefont
  {Perring}}, \bibinfo {author} {\bibfnamefont {J.-S.}\ \bibnamefont {Caux}},
  \bibinfo {author} {\bibfnamefont {A.~T.}\ \bibnamefont {Savici}}, \bibinfo
  {author} {\bibfnamefont {G.~D.}\ \bibnamefont {Gu}}, \bibinfo {author}
  {\bibfnamefont {C.-C.}\ \bibnamefont {Lee}}, \bibinfo {author} {\bibfnamefont
  {W.}~\bibnamefont {Ku}},\ and\ \bibinfo {author} {\bibfnamefont {I.~A.}\
  \bibnamefont {Zaliznyak}},\ }\bibfield  {title} {\bibinfo {title} {Effect of
  covalent bonding on magnetism and the missing neutron intensity in copper
  oxide compounds},\ }\href {https://doi.org/10.1038/nphys1405} {\bibfield
  {journal} {\bibinfo  {journal} {Nature Physics}\ }\textbf {\bibinfo {volume}
  {5}},\ \bibinfo {pages} {867} (\bibinfo {year} {2009})}\BibitemShut {NoStop}%
\bibitem [{\citenamefont {Maiti}\ \emph {et~al.}(1998)\citenamefont {Maiti},
  \citenamefont {Sarma}, \citenamefont {Mizokawa},\ and\ \citenamefont
  {Fujimori}}]{Maiti1998}%
  \BibitemOpen
  \bibfield  {author} {\bibinfo {author} {\bibfnamefont {K.}~\bibnamefont
  {Maiti}}, \bibinfo {author} {\bibfnamefont {D.~D.}\ \bibnamefont {Sarma}},
  \bibinfo {author} {\bibfnamefont {T.}~\bibnamefont {Mizokawa}},\ and\
  \bibinfo {author} {\bibfnamefont {A.}~\bibnamefont {Fujimori}},\ }\bibfield
  {title} {\bibinfo {title} {Electronic structure of one-dimensional
  cuprates},\ }\href {https://doi.org/10.1103/PhysRevB.57.1572} {\bibfield
  {journal} {\bibinfo  {journal} {Phys. Rev. B}\ }\textbf {\bibinfo {volume}
  {57}},\ \bibinfo {pages} {1572} (\bibinfo {year} {1998})}\BibitemShut
  {NoStop}%
\bibitem [{\citenamefont {Barantani}\ \emph {et~al.}(2022)\citenamefont
  {Barantani}, \citenamefont {Tran}, \citenamefont {Madan}, \citenamefont
  {Kapon}, \citenamefont {Bachar}, \citenamefont {Asmara}, \citenamefont
  {Paris}, \citenamefont {Tseng}, \citenamefont {Zhang}, \citenamefont {Hu},
  \citenamefont {Giannini}, \citenamefont {Gu}, \citenamefont {Devereaux},
  \citenamefont {Berthod}, \citenamefont {Carbone}, \citenamefont {Schmitt},\
  and\ \citenamefont {van~der Marel}}]{barantani2022resonant}%
  \BibitemOpen
  \bibfield  {author} {\bibinfo {author} {\bibfnamefont {F.}~\bibnamefont
  {Barantani}}, \bibinfo {author} {\bibfnamefont {M.~K.}\ \bibnamefont {Tran}},
  \bibinfo {author} {\bibfnamefont {I.}~\bibnamefont {Madan}}, \bibinfo
  {author} {\bibfnamefont {I.}~\bibnamefont {Kapon}}, \bibinfo {author}
  {\bibfnamefont {N.}~\bibnamefont {Bachar}}, \bibinfo {author} {\bibfnamefont
  {T.~C.}\ \bibnamefont {Asmara}}, \bibinfo {author} {\bibfnamefont
  {E.}~\bibnamefont {Paris}}, \bibinfo {author} {\bibfnamefont
  {Y.}~\bibnamefont {Tseng}}, \bibinfo {author} {\bibfnamefont
  {W.}~\bibnamefont {Zhang}}, \bibinfo {author} {\bibfnamefont
  {Y.}~\bibnamefont {Hu}}, \bibinfo {author} {\bibfnamefont {E.}~\bibnamefont
  {Giannini}}, \bibinfo {author} {\bibfnamefont {G.}~\bibnamefont {Gu}},
  \bibinfo {author} {\bibfnamefont {T.~P.}\ \bibnamefont {Devereaux}}, \bibinfo
  {author} {\bibfnamefont {C.}~\bibnamefont {Berthod}}, \bibinfo {author}
  {\bibfnamefont {F.}~\bibnamefont {Carbone}}, \bibinfo {author} {\bibfnamefont
  {T.}~\bibnamefont {Schmitt}},\ and\ \bibinfo {author} {\bibfnamefont
  {D.}~\bibnamefont {van~der Marel}},\ }\bibfield  {title} {\bibinfo {title}
  {Resonant inelastic x-ray scattering study of electron-exciton coupling in
  high-${T}_{c}$ cuprates},\ }\href
  {https://doi.org/10.1103/PhysRevX.12.021068} {\bibfield  {journal} {\bibinfo
  {journal} {Phys. Rev. X}\ }\textbf {\bibinfo {volume} {12}},\ \bibinfo
  {pages} {021068} (\bibinfo {year} {2022})}\BibitemShut {NoStop}%
\bibitem [{\citenamefont {Barantani}\ \emph {et~al.}(2023)\citenamefont
  {Barantani}, \citenamefont {Berthod},\ and\ \citenamefont {van {der
  Marel}}}]{barantani2023can}%
  \BibitemOpen
  \bibfield  {author} {\bibinfo {author} {\bibfnamefont {F.}~\bibnamefont
  {Barantani}}, \bibinfo {author} {\bibfnamefont {C.}~\bibnamefont {Berthod}},\
  and\ \bibinfo {author} {\bibfnamefont {D.}~\bibnamefont {van {der Marel}}},\
  }\bibfield  {title} {\bibinfo {title} {Can dd excitations mediate pairing?},\
  }\href {https://doi.org/10.1016/j.physc.2023.1354321} {\bibfield  {journal}
  {\bibinfo  {journal} {Physica C: Superconductivity and its Applications}\
  }\textbf {\bibinfo {volume} {613}},\ \bibinfo {pages} {1354321} (\bibinfo
  {year} {2023})}\BibitemShut {NoStop}%
\bibitem [{\citenamefont {Ellis}\ \emph {et~al.}(2015)\citenamefont {Ellis},
  \citenamefont {Huang}, \citenamefont {Olalde-Velasco}, \citenamefont {Dantz},
  \citenamefont {Pelliciari}, \citenamefont {Drachuck}, \citenamefont {Ofer},
  \citenamefont {Bazalitsky}, \citenamefont {Berger}, \citenamefont {Schmitt},\
  and\ \citenamefont {Keren}}]{Ellis2015}%
  \BibitemOpen
  \bibfield  {author} {\bibinfo {author} {\bibfnamefont {D.~S.}\ \bibnamefont
  {Ellis}}, \bibinfo {author} {\bibfnamefont {Y.-B.}\ \bibnamefont {Huang}},
  \bibinfo {author} {\bibfnamefont {P.}~\bibnamefont {Olalde-Velasco}},
  \bibinfo {author} {\bibfnamefont {M.}~\bibnamefont {Dantz}}, \bibinfo
  {author} {\bibfnamefont {J.}~\bibnamefont {Pelliciari}}, \bibinfo {author}
  {\bibfnamefont {G.}~\bibnamefont {Drachuck}}, \bibinfo {author}
  {\bibfnamefont {R.}~\bibnamefont {Ofer}}, \bibinfo {author} {\bibfnamefont
  {G.}~\bibnamefont {Bazalitsky}}, \bibinfo {author} {\bibfnamefont
  {J.}~\bibnamefont {Berger}}, \bibinfo {author} {\bibfnamefont
  {T.}~\bibnamefont {Schmitt}},\ and\ \bibinfo {author} {\bibfnamefont
  {A.}~\bibnamefont {Keren}},\ }\bibfield  {title} {\bibinfo {title}
  {{Correlation of the superconducting critical temperature with spin and
  orbital excitations in
  $({\mathrm{Ca}}_{x}{\mathrm{La}}_{1\ensuremath{-}x})({\mathrm{Ba}}_{1.75\ensuremath{-}x}\text{La}$
  ${}_{0.25+x}){\mathrm{Cu}}_{3}{\mathrm{O}}_{y}$ as measured by resonant
  inelastic x-ray scattering}},\ }\href
  {https://doi.org/10.1103/PhysRevB.92.104507} {\bibfield  {journal} {\bibinfo
  {journal} {Phys. Rev. B}\ }\textbf {\bibinfo {volume} {92}},\ \bibinfo
  {pages} {104507} (\bibinfo {year} {2015})}\BibitemShut {NoStop}%
\end{thebibliography}%

\newpage
\clearpage



\section*{Supplementary material}

# Supplementary material

## S1 Further experimental details

*Methods* – The $La_2CuO_4$ (LCO) film (30 nm thick) was grown by pulsed laser deposition (KrF excimer laser, $\lambda$ = 248 nm) on $LaSrAlO_4$ (001) (LSAO) substrate. The substrate holder was at a distance of 2.5 cm from the LCO target, which was prepared by the standard solid-state reaction method. Before the growth, the substrate was pre-annealed in situ at 700 °C under 1 mbar oxygen for 10 min. During the growth, the substrate temperature was around 700 °C and the oxygen pressure around 0.8 mbar. After the growth, to obtain a more insulating LCO, the film was post-annealed in vacuum at 200 °C for 30 minutes [1]. The lattice constants of LCO are $a = b = 3.77$ Å and $c = 13.1$ Å. The resistivity of the LCO film is reported in Fig. S1, demonstrating its insulating behaviour.

The $CaCuO_2$ (CCO) films (30 nm thick) were grown by pulsed laser deposition (KrF excimer laser, $\lambda$ = 248 nm) at a temperature around 600 °C and an oxygen pressure of 0.1 mbar, on $NdGaO_3$ (NGO) (1 1 0) substrate. The substrate holder was at a distance of 2.5 cm from the CCO target, which was prepared by standard solid-state reaction [2, 3]. The lattice constants for CCO, as determined from XRD, are $a = b = 3.86$ Å and $c = 3.18$ Å. The measured sheet resistance of CCO at room temperature (300 K) was 2 MΩ, corresponding to a resistivity of ~ 6000 mΩcm. Below 250 K, the resistivity of the film increased too much and the measurements became unreliable.

The thin films of T'-$Nd_2CuO_4$ used in this work were grown by molecular beam epitaxy under ultrahigh vacuum using Nd and Cu metal sources and atomic oxygen generated in situ from a radiofrequency oxygen source [4]. Reflection high energy electron diffraction and electron impact emission spectroscopy were used to monitor and control the growth of the NCO films on (001) SrTiO3 substrates in real time. High-resolution reciprocal space mapping data show that films with a thickness of 100 nm are grown fully relaxed. The films were subjected to a two-step annealing process [5]. The resistivity of the NCO film is reported in Fig. S1, demonstrating its insulating behaviour.

The RIXS experiments on LCO and CCO samples were conducted at the ID32 beamline of the ESRF – The European Synchrotron, France. The scattering angle was kept fixed at 149.5°, and the momentum scans along the (1,0) and (1,1) directions were acquired rotating the incident angle $\theta$. The energy was fixed at the Cu $L_3$ resonance (~ 931 eV) and we employed incident linear-vertical ($\sigma$) and linear-horizontal ($\pi$) polarizations. Total resolution was estimated to be ~ 43 meV both for normal and polarimetric scans. Temperature was kept at 20 K. Although we are not analyzing the intensity of the orbital excitations, all of the spectra have been corrected for self-absorption as described in [6]. In the case of CCO and LCO, self-absorption correction was taking the finite thickness of the sample into consideration.

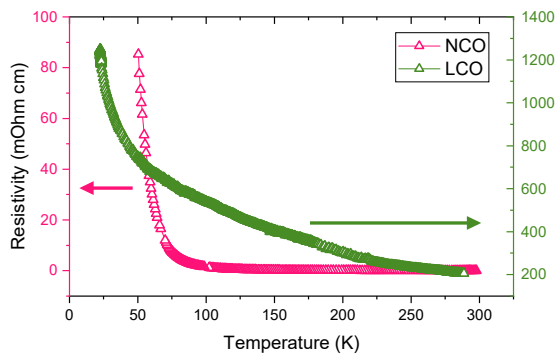

Figure S1: Resistivity of the LCO and NCO film.

## S2 Extraction of the $xy$-orbital dispersion

In order to evaluate the dispersion of the $xy$ orbital, we have extracted its energy in two ways:

- as the first relative maximum of the (inverted) second derivative, smoothed using adjacent-averaging over 5 points. Since the second derivative curves are noisy, after having identified the first relative maximum, we have calculated "a centre of mass" with the four adjacent points, using the intensities of the second derivative in those points as a weight. The extracted dispersion at energy loss $\omega_i$ therefore is:

$$\omega_i = \frac{\sum_{j=-2}^{+2} f''(\omega_{i+j}) \cdot \omega_{i+j}}{\sum_{j=-2}^{+2} f''(\omega_{i+j})} \tag{S1}$$

  where $-f''(\omega_{i+j})$ is the second derivative of the RIXS spectrum calculated at energy loss $\omega_{i+j}$. Close to the maximum, the inverted second derivative is always well above zero so that the denominator is clearly positive. An example of the second derivative is shown Fig. S3(f) for RIXS spectra collected at $(0,0)$, $(1/4,1/4)$, and $(1/2,0)$, with $\sigma$ incident polarization. Points are the raw curves, while the solid lines are smoothed over 5 points. In the case of CCO, the error bars have been taken, for the last 5 points along the $(1,0)$ direction close to $(1/2,0)$, as two times the energy separation between experimental points ($\sim 9$ meV). This is because the intensity of the dispersing feature decreases and its width increases. In Fig. S2 we report five representative RIXS spectra along the $(1,0)$ and $(1,1)$ directions, along with the corresponding smoothed second derivatives.

- fitting the spectra with functions representing the different features. The $xy$ excitation is fitted with two different peaks, named $xy_1$ and $xy_2$. $xy_1$ is fitted with a Voigt function. Motivated by the asymmetry of the peaks, and by theoretical consideration of the nature of the excitation (see discussion in the next seections), we fit $xy_2$ using a skewed Gaussian profile. The asymmetry parameter $\alpha$ is kept fixed at 4. All other excitations ($xz/yz$, $4^{th}$ peak and $z^2$ are fitted using symmetric Gaussians. The intensity, energy and width of the different features are left as free parameters with the exception of the Gaussian width of $xy_1$, which is fixed at the experimental resolution. The energy of the $xy_2$ peak has been taken as the maximum of the curve. Bottom figures refer to CCO, where the peaks are particularly sharp and this estimation is more reliable. Examples of the fitting results at different momenta are given in panels (a)-(e) of Fig. S3.

As seen in panel (c) of Fig. S3, there is a very good agreement between the two approaches along the diagonal and close to the $(0,0)$ point; close to $(\pi,0)$, the intensity drops and the peak broadens, so that the uncertainty increases. The estimation of the dispersion of orbital excitations is complicated by the fact that two or more peaks might be close in energy, and their intensity might be strongly dependent on the scattering geometry. Therefore, a "dispersion" might just be the result of two or more peaks with fixed energy, but with intensities that depend differently on the momentum $\mathbf{q}$. In order to rule out this scenario, we take advantage of the fact that the atomic cross-sections of the dd excitations (which are still a good approximation of the real ones) change drastically when one changes polarization and experimental geometry. Therefore, we have extracted the dispersion of the $xy$ orbital using both $\sigma$ and $\pi$ incident polarization, and by changing between grazing-in and grazing-out geometries. Figure S3(g) shows a RIXS map as a function of momentum collected along the path depicted as a red line the inset, with $\sigma$ polarization. Blue (red) dots report the local maxima of the inverted second derivative extracted from RIXS spectra collected with $\pi$ ($\sigma$) polarization, along the blue (red) path in the inset corresponding to grazing-out (grazing-incident) geometry. Evidently, the dispersions of the $xy$ excitation extracted from the two datasets are in very good agreement with each other.

## S3 Momentum dependence of the $dd$ intensities

Fitting the spectra as described in the previous section allows us to extract the spectral weight of $dd$ excitations as a function of momentum transfer. This is useful to compare the measured behaviour of

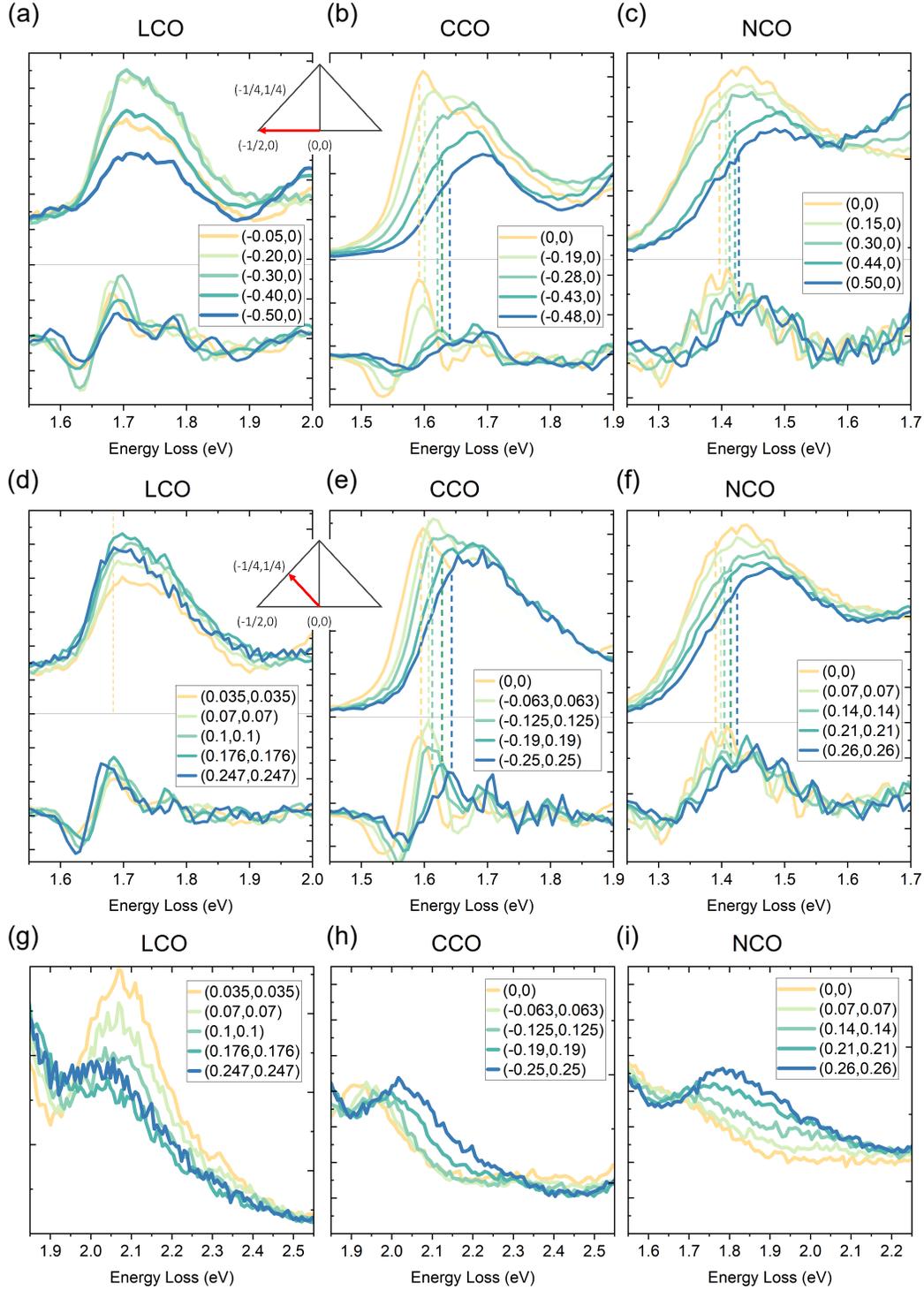

Figure S2: Representative RIXS spectra of the $xy$ orbital excitations acquired along the (1,0) (panels (a)-(c)) and (1,1) directions (panels (d)-(f)) using $\sigma$ incident polarization. In each panel, the smoothed second derivative curves are plotted below the raw RIXS spectra. Panels (g)-(i) report the RIXS spectra of $xz/yz$ excitation along (1,1) direction.

3

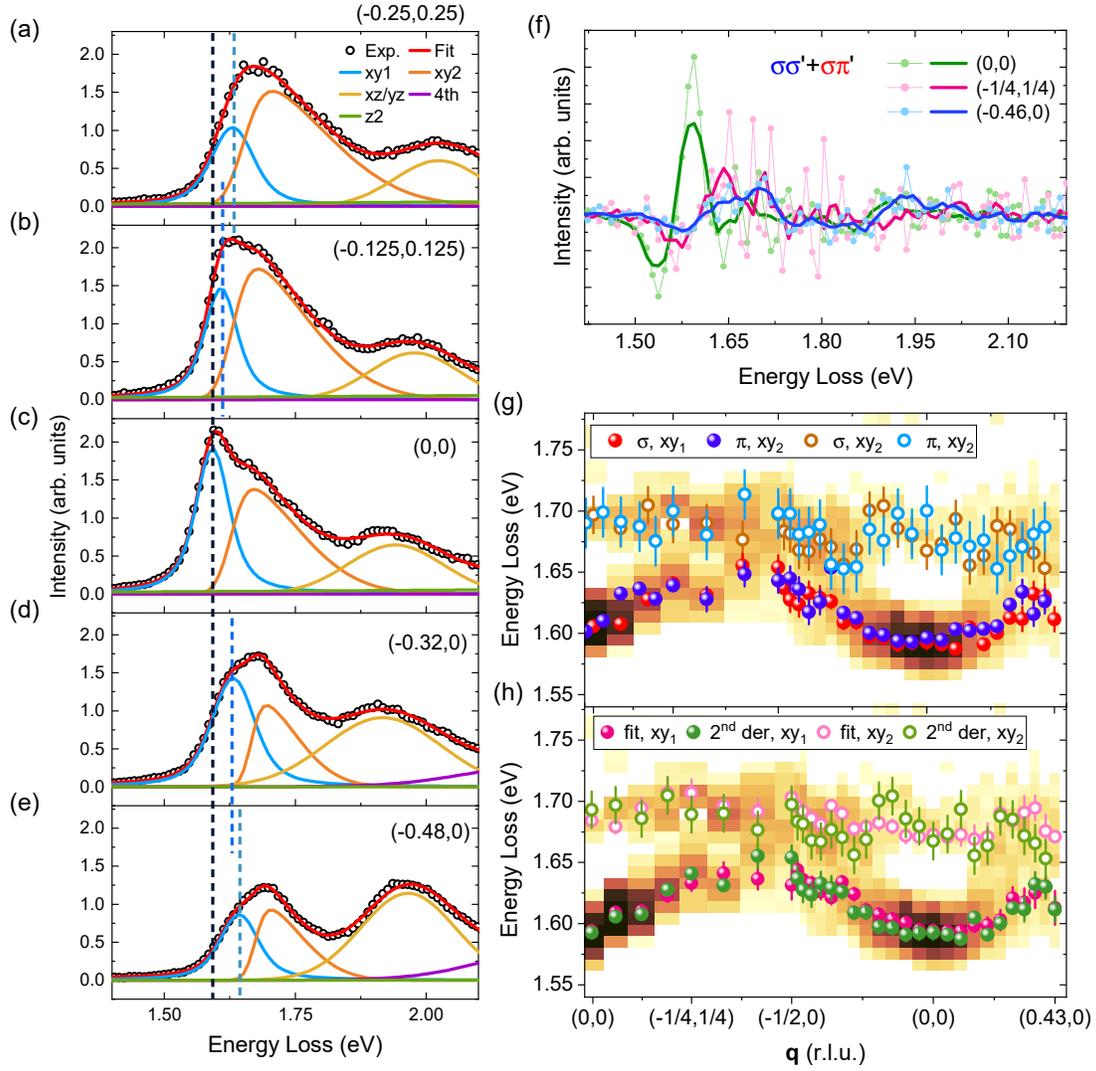

Figure S3: (a)-(e) Examples of fitting procedure (zoomed close to $xy$ excitation) on the RIXS spectra of CCO. All the spectra were acquired with linear vertical polarization. Vertical dashed lines report the energy of the dispersing $xy$ peak. (f) examples of the (inverted) second derivative of the RIXS spectra collected with $\sigma$ polarization at $(0,0)$ (green dots and line) and $(1/4,1/4)$ (violet dots and line). Dots (line) refers to the raw (smoothed with 5 points moving average) curve. (g) Second derivative map of the RIXS spectra acquired with $\sigma$ and $\pi$ polarizations. Blue (red) dots correspond to energies extracted with the second-derivative method on spectra collected with $\pi$ polarization and grazing-out geometry ($\sigma$ polarization and grazing-in geometry), as depicted in the inset above the panel. (h) Second derivative map of the RIXS spectra acquired with $\sigma$ polarization. Green (violet) dots represent energy of the $xy$ excitation extracted. Red line corresponds to $\sigma$ polarization, blue to $\pi$ polarization.

4

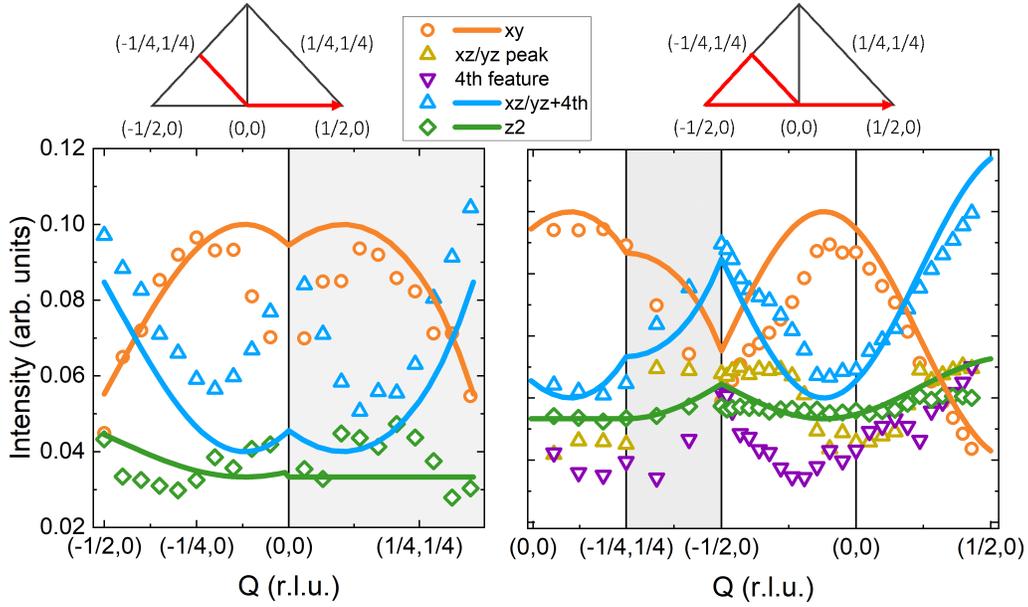

Figure S4: spectral weight of *dd* excitations as a function of momentum for LCO (left panel) and CCO (right panel). Symbols report intensities extracted from experimental data, solid lines are curves calculated with the single-ion model.

the intensity with the one expected from single-ion calculations. It must be noted, however, that in a simple ionic picture the *dd* intensity just depends on the experimental geometry (incident angle $\theta$ and azimuthal angle $\phi$) and only indirectly on momentum.

The measured and calculated intensities are shown in Fig. S4. In LCO, the single ion picture gives a good description of the experimental geometry, as was demonstrated before [7]. Evidently, there is only a small discrepancy close to the $\Gamma$ point for the $xy$ and $xz/yz$ peaks. This might be due to the asymmetric lineshape of the peaks, not completely captured by the fitting and resulting in an imperfect assignment of the spectral weight. Constraining the intensity of $xy$ peak would result in a better assignment of the spectral weight.

In CCO, the weight of the $xy$ excitation (which is calculated as the sum of the two peaks with which it was fitted) is consistently underestimated. This might be due again to the strong asymmetric lineshape of the 1.7 eV peak, not captured by our fitting. We note that the small spin-orbit coupling of cuprates might also redistribute part of the spectral weight. The more interesting fact comes from the observation of the $xz/yz$ and 4$^{\text{th}}$ feature. Evidently, the single-ion calculation only agrees with experimental data when the two spectral weights are summed together. Taken separately (yellow and purple points in the right panel of Fig. S4), neither of them agrees with the single-ion prediction even qualitatively.

## S4 The orbital superexchange model: derivation, comparison with experiment and its validity

A dispersion of *dd* excitations of the right order of the magnitude can be obtained considering a simple 2-site 2-orbital Hubbard model, solved with an exact diagonalization calculation and assuming the strong Hubbard $U$ limit. There are then two excited states with momenta 0 and $\pi$. The expected dispersion originates in the spin-orbital superexchange and becomes $E(\pi) - E(0) = -4(t_a t_b)/(U - 3J_\text{H})$, where $t_a, t_b$ are the nearest neighbor (NN) hopping integrals for the two orbitals, and $U$ and $J_\text{H}$ are the on-site repulsion and exchange integral respectively. However, the considered dimer does not capture the effect of the underlying two-dimensional AFM lattice. The strong interaction between the orbital excitations and the magnetic background strongly renormalizes the dispersion originating from such a NN orbital exchange, cf. [8] and below. Therefore, as the starting point we consider a charge transfer model and from that we derive a realistic orbital (Kugel-Khomskii type) superexchange model. It

5

turns out that the latter model contains a (surprisingly) large next nearest neighbor (NNN) orbital superexchange which gives rise to the free orbiton hopping on the same antiferromagnetic (AFM) sublattice and, therefore, yields a large orbiton dispersion (see below for details).

### S4.1 Charge transfer model in the 'standard' basis and its parameters for CCO

As the starting point of the calculation we use a charge transfer model with two $d$ orbitals ($d_{x^2-y^2} \equiv x^2-y^2$ and $d_{xy} \equiv xy$) and two $p$ orbitals ($p_x$ and $p_y$) per each copper and each oxygen atom (respectively). Thus, in the CuO$_2$ unit cell we have in total six orbitals. The model contains eight parameters. Six charge transfer model parameters are defined in the usual manner (in the hole language): (i) the Cu-O hopping $t_{pd\sigma}$, (ii) the Cu-O hopping $t_{pd\pi}$, (iii) the O-O hoppings $t_{pp}$ (nearest neighbor between distinct oxygen orbitals), (iv) the charge transfer energy for all oxygen orbitals $\Delta$, (v) the on-site copper repulsion $U$, (vi) the Hund's exchange $J_H$. We also consider two crystal field energies: (i) the energy difference between the $\pi$ and the $\sigma$ bonding oxygen orbitals $\varepsilon_{\pi\sigma}$ (originates in the *higher* energies for holes in the $\sigma$ bonding, i.e. 'pointing-towards copper', oxygen orbitals), (ii) the crystal field energy of the $xy$ orbital $\varepsilon_{xy}$. Note that we skip the Coulomb repulsion $U_p$ on oxygen, since below we consider bonding or antibonding combinations of oxygen orbitals and the effective repulsion between oxygen holes is strongly reduced (due to the charge delocalisation). Altogether this leads to the following charge transfer Hamiltonian in the hole notation:

$$\mathcal{H} = \mathcal{H}_{\text{kin}} + \mathcal{H}_{\text{coul}}, \tag{S2}$$

where

$$\begin{aligned}
\mathcal{H}_{\text{kin}} = \sum_{i,s \in \{\uparrow\downarrow\}} &\Big\{ \varepsilon_{xy} n_{i,xy,s} + \Delta(n_{i+L,p_x,s} + n_{i+R,p_x,s} + n_{i+B,p_y,s} + n_{i+T,p_y,s}) \\
&+ (\Delta + \varepsilon_{\pi\sigma})(n_{i+L,p_y,s} + n_{i+R,p_y,s} + n_{i+B,p_x,s} + n_{i+T,p_x,s}) \\
&+ t_{pd\sigma}[c^\dagger_{i,x^2-y^2,s}(c_{i+L,p_x,s} - c_{i+R,p_x,s} - c_{i+B,p_y,s} + c_{i+T,p_y,s}) + \text{h.c.}] \\
&+ t_{pd\pi}[c^\dagger_{i,xy,s}(-c_{i+L,p_y,s} + c_{i+R,p_y,s} - c_{i+B,p_x,s} + c_{i+T,p_x,s}) + \text{h.c.}] \\
&+ t_{pp}(c^\dagger_{i+B,p_y,s} c_{i+L,p_x,s} - c^\dagger_{i+L,p_x,s} c_{i+T,p_y,s} + c^\dagger_{i+T,p_y,s} c_{i+R,p_x,s} - c^\dagger_{i+R,p_x,s} c_{i+B,p_y,s} + \text{h.c.}) \\
&+ t_{pp}(c^\dagger_{i+B,p_x,s} c_{i+L,p_y,s} - c^\dagger_{i+L,p_y,s} c_{i+T,p_x,s} + c^\dagger_{i+T,p_x,s} c_{i+R,p_y,s} - c^\dagger_{i+R,p_y,s} c_{i+B,p_x,s} + \text{h.c.}) \Big\},
\end{aligned} \tag{S3}$$

and

$$\mathcal{H}_{\text{coul}} = \sum_{i,s \in \{\uparrow\downarrow\}} \left[ (U - 3J_H) n_{i,xy,s} n_{i,x^2-y^2,s} + (U - 2J_H) n_{i,xy,s} n_{i,x^2-y^2,\bar{s}} \right]. \tag{S4}$$

Here $c_{i,f,s}$ ($c^\dagger_{i,f,s}$) are annihilation (creation) operators at site $i$, orbital $f$ and with spin $s$ ($n_{i,f,s}$ are the corresponding number operators). The sum goes over the copper sites and the indices $i+L, i+R, i+B, i+T$ describe the nearest neighbor oxygen sites situated to the left, right, bottom and top of the copper site $i$.

The chosen charge transfer parameters for CCO, which correctly yield the observed value of the orbiton dispersion (see main text) as well as the on-site $xy$ orbital energy within 20% error (see below) can be found in Tab. 1. While in general the estimates of the charge transfer parameters given in Tab. 1 are rather standard, a few of them require a more detailed discussion:

First, the charge transfer energy $\Delta$ is relatively low for CCO. This can be explained by the lack of apical oxygens, since the charge transfer energy significantly decreases with the distance to apical oxygens, cf. Fig. 2 or Table 1 in the appendix of [9]; this value is also consistent with the one extracted from optical absorption measurements [10, 11].

Second, the O-O hoppings $t_{pp}$ are a bit larger than sometimes assumed. The considered here estimate follows, for instance, from [9] or the XPS on quasi 1D cuprates (see first column of Table I of [12]). Besides, a recent review [13] strongly advocates that $t_{pp}$ should be even of the order 1 eV for the $\sigma$ bonding oxygens (cf. [14]). Note, however, that this value is larger than e.g. the one estimated by DFT for CCO [15].

| $t_{pd\sigma}$ | $t_{pd\pi}$ | $t_{pp}$ | $\Delta$ | $\varepsilon_{\pi\sigma}$ | $\varepsilon_{xy}$ | $U$ | $J_H$ |
|---|---|---|---|---|---|---|---|
| 1.3 | 0.7 | 0.7 | 1.8 | $-1.6$ | 1.0 | 8.0 | 1.0 |

Table 1: Parameters of the charge transfer model (S2) that are considered to be realistic for CCO (in eV; see text for more details).

Third, the estimates of the (relative) on-site energies of the $\pi$ bonding oxygen orbitals and of the $xy$ copper orbital follow from Ref. [15]. These are *pure* crystal field parameters, which should not include any interactions (or even covalency effects) and therefore DFT is the correct approach here. Note that the crystal field energy of the $xy$ orbital is *not* (and should not be) equal to the $xy$ orbital excitation energy measured by RIXS (which includes the covalency effects and, in fact, can be reproduced by the approach used here, see below).

### S4.2 Charge transfer model in the bonding-antibonding basis

Inspired by [16] we express the charge transfer model in the bonding-antibonding basis:
*First*, we consider two copper orbitals

1. $|x^2 - y^2\rangle$ at zero energy,
2. $|xy\rangle$ orbital with energy $\varepsilon_{xy}$,

and construct the bonding and antibonding combinations of four oxygen $\pi$ and four oxygen $\sigma$ orbitals on a single CuO$_4$ plaquette. To this end, we move from the $p_x$ and $p_y$ oxygen basis to the one formed by the bonding and antibonding oxygen orbitals (note that the nonbonding orbitals do not contribute to the exchange processese considered below and can be skipped and that the bonding and antibonding terminology refers to the electron language):

1. the $\sigma$ bonding oxygen orbital $|B\sigma\rangle = \frac{1}{2}(|B\rangle - |R\rangle - |T\rangle + |R\rangle)$ with energy $\varepsilon_{B\sigma} = \Delta + 2t_{pp}$ (this orbital corresponds to $-|\beta\rangle$ in [16]),
2. the $\sigma$ antibonding oxygen orbital $|A\sigma\rangle = \frac{1}{2}(-|B\rangle - |R\rangle + |T\rangle + |L\rangle)$ with energy $\varepsilon_{A\sigma} = \Delta - 2t_{pp}$ (this orbital corresponds to $-|\alpha\rangle$ in [16]),
3. the $\pi$ bonding oxygen orbital $|B\pi\rangle = \frac{1}{2}(|B\rangle - |R\rangle - |T\rangle + |L\rangle)$ with energy $\varepsilon_{B\pi} = \Delta + 2t_{pp} + \varepsilon_{\pi\sigma}$,
4. the $\pi$ antibonding oxygen orbitals $|A\pi\rangle = \frac{1}{2}(-|B\rangle - |R\rangle + |T\rangle + |L\rangle)$ with energy $\varepsilon_{A\pi} = \Delta - 2t_{pp} + \varepsilon_{\pi\sigma}$.

Note that we skip the $t_{pp}$ hopping which mixes the $|B\sigma\rangle$ and $|A\pi\rangle$ orbitals. On one hand, including the hybridisation between these states hugely complicates matters and is also neglected in [16]. On the other hand, such a hopping does not lead to a coherent orbiton propagation (for it leads to the double orbiton creation and annihilation terms).
*Second*, due to the overlaps between the nearest neighbor CuO$_4$ plaquettes, we itroduce the orthogonalisation factors in the definition of the bonding and antibonding oxygen orbitals. This is done exactly in the same manner as in [16]: *inter alia* we use the following values for the relevant orthogonalisation factors (notation as in [16])

$$\mu(0) = 0.958, \ \nu(0) = 0.727, \quad \mu(1) = -0.14, \ \nu(1) = -0.273, \quad \mu(\sqrt{2}) = -0.02, \ \nu(\sqrt{2}) = 0.122. \qquad \text{(S5)}$$

Note that introducing these factors leads to changes in the energies of the bonding and antibonding states (cf. [16]). In particular, we obtain for the relevant (see below) $\sigma$ antibonding and $\pi$ bonding states

$$\varepsilon_{A\sigma} = \Delta - 2\nu(0)t_{pp}, \quad \varepsilon_{B\pi} = \Delta + 2\nu(0)t_{pp} + \varepsilon_{\pi\sigma}. \qquad \text{(S6)}$$

*Third*, we include hybridisation between all four oxygen and two copper orbitals. Then, it turns out that only the $|A\sigma\rangle$ and $|B\pi\rangle$ orbitals hybridise with copper $x^2-y^2$ and $xy$ orbitals (respectively). We obtain that the relevant charge transfer model in such a bonding-antibodning basis reads:

$$\mathcal{H} \simeq \mathcal{H}_{\text{kin}}^{(0)} + \mathcal{H}_{\text{kin}}^{(1)} + \mathcal{H}_{\text{kin}}^{(\sqrt{2})} + \mathcal{H}_{\text{coul}}, \tag{S7}$$

with

$$\mathcal{H}_{\text{kin}}^{(0)} = \sum_{i,s\in\{\uparrow\downarrow\}} \Big[ \varepsilon_{xy} n_{i,xy,s} + \varepsilon_{A\sigma} n_{i,A\sigma,s} + \varepsilon_{B\pi} n_{i,B\pi,s}$$
$$+ T_{pd\sigma}^{(0)} \left( c_{i,x^2-y^2,s}^\dagger c_{i,A\sigma,s} + \text{h.c.} \right) + T_{pd\pi}^{(0)} \left( c_{i,xy,s}^\dagger c_{i,B\pi,s} + \text{h.c.} \right) \Big], \tag{S8}$$

$$\mathcal{H}_{\text{kin}}^{(1)} = \sum_{\langle i,j\rangle,s\in\{\uparrow\downarrow\}} \Big[ T_{pd\sigma}^{(1)} \left( c_{i,x^2-y^2,s}^\dagger c_{j,A\sigma,s} + \text{h.c.} \right) + T_{pd\pi}^{(1)} \left( c_{i,xy,s}^\dagger c_{j,B\pi,s} + \text{h.c.} \right)$$
$$+ T_{A\sigma}^{(1)} \left( c_{i,A\sigma,s}^\dagger c_{j,A\sigma,s} + \text{h.c.} \right) + T_{B\pi}^{(1)} \left( c_{i,B\pi,s}^\dagger c_{j,B\pi,s} + \text{h.c.} \right) \Big], \tag{S9}$$

$$\mathcal{H}_{\text{kin}}^{(\sqrt{2})} = \sum_{\langle\langle i,j\rangle\rangle,s\in\{\uparrow\downarrow\}} \Big[ T_{A\sigma}^{(\sqrt{2})} \left( c_{i,A\sigma,s}^\dagger c_{j,A\sigma,s} + \text{h.c.} \right) + T_{B\pi}^{(\sqrt{2})} \left( c_{i,B\pi,s}^\dagger c_{j,B\pi,s} + \text{h.c.} \right) \Big], \tag{S10}$$

where $\mathcal{H}_{\text{coul}}$ is the same as in the previous subsection and the relevant hopping elements are (after including the orthogonalisation factors):

$$T_{pd\sigma}^{(0)} = 2\mu(0) t_{pd\sigma}, \quad T_{pd\pi}^{(0)} = -2\mu(0) t_{pd\pi}, \tag{S11}$$

$$T_{pd\sigma}^{(1)} = 2\mu(1) t_{pd\sigma}, \quad T_{pd\pi}^{(1)} = -2\mu(1) t_{pd\pi}, \quad T_{A\sigma}^{(1)} = -2\nu(1) t_{pp}, \quad T_{B\pi}^{(1)} = 2\nu(1) t_{pp}, \tag{S12}$$

$$T_{A\sigma}^{(\sqrt{2})} = -2\nu(\sqrt{2}) t_{pp}, \quad T_{B\pi}^{(\sqrt{2})} = 2\nu(\sqrt{2}) t_{pp}. \tag{S13}$$

Note that $\mathcal{H}_{\text{kin}}^{(0)}$ depicts hybridisation between states within the same (orthogonalised) plaquette, $\mathcal{H}_{\text{kin}}^{(1)}$ stands for the hybridisation between states on the nearest neigbor (NN) plaquettes, $\mathcal{H}_{\text{kin}}^{(\sqrt{2})}$ stands for the nonnegligible hopping between the next-nearest neighbor (NNN) plaquettes (note that for the NNN the hopping between copper and oxygen orbitals can be neglected due to very small $\mu(\sqrt{2})$ [16]). As in the previous subsection $c_{i,f,s}$ ($c_{i,f,s}^\dagger$) are annihilation (creation) operators at site $i$, orbital $f$ and with spin $s$ ($n_{i,f,s}$ are the corresponding number operators).

### S4.3 Basis states of the cell perturbation theory

We construct the *single-particle* cell perturbation theory basis by diagonalising the single cell Hamiltonian $\mathcal{H}_{\text{kin}}^{(0)}$. We obtain four single-particle eigenstates on the single CuO$_4$ plaquette:

1. $|\psi_-\rangle = \cos\psi |x^2-y^2\rangle + \sin\psi |A\sigma\rangle$ with energy $E_{\psi_-} = \frac{1}{2}\Big[\varepsilon_{A\sigma} - \sqrt{\varepsilon_{A\sigma}^2 + 4(T_{pd\sigma}^{(0)})^2}\Big]$ —this is the CCO ground on the single plaquette,

2. $|\psi_+\rangle = -\sin\psi |x^2-y^2\rangle + \cos\psi |A\sigma\rangle$ with energy $E_{\psi_+} = \frac{1}{2}\Big[\varepsilon_{A\sigma} + \sqrt{\varepsilon_{A\sigma}^2 + 4(T_{pd\sigma}^{(0)})^2}\Big]$ —this is the charge transfer–like excitation on the single plaquette,

3. $|\phi_-\rangle = \cos\phi |xy\rangle + \sin\phi |B\pi\rangle$ with energy $E_{\phi_-} = \frac{1}{2}\Big[\varepsilon_{xy} + \varepsilon_{B\pi} - \sqrt{(\varepsilon_{B\pi} - \varepsilon_{xy})^2 + 4(T_{pd\pi}^{(0)})^2}\Big]$ —this is the $xy$ orbital excitation on the single plaquette (its energy is the one that is observed by RIXS),

4. $|\phi_+\rangle = -\sin\phi |xy\rangle + \cos\phi |B\pi\rangle$ with energy $E_{\phi_+} = \frac{1}{2}\Big[\varepsilon_{xy} + \varepsilon_{B\pi} + \sqrt{(\varepsilon_{B\pi} - \varepsilon_{xy})^2 + 4(T_{pd\pi}^{(0)})^2}\Big]$ —this is the charge transfer–like excitation of the $xy$ character on the single plaquette.

The angles $\phi$ and $\psi$ can be calculated from the following formulae:

$$\tan(2\psi) = -\frac{2T_{pd\sigma}^{(0)}}{\varepsilon_{A\sigma}}, \quad \tan(2\phi) = -\frac{2T_{pd\phi}^{(0)}}{\varepsilon_{B\pi} - \varepsilon_{xy}}. \tag{S14}$$

We assume that the *two-particle* cell perturbation theory basis consists of just one state [the other states are not of relevant for the considered (in the next subsection) exchange of the $xy$ orbital excitation, i.e. $|\phi_-\rangle$]:

$$|\text{INT}\rangle \equiv |\phi_-\rangle|\psi_-\rangle, \tag{S15}$$

since all other states will have a much higher excitation energy. Its energy can be estimated by considering contributions from $\mathcal{H}_{\text{kin}}^{(0)}$ and $\mathcal{H}_{\text{coul}}$:

$$E_{\text{INT}} = E_{\psi_-} + E_{\phi_-} + U_{\text{eff}} \approx E_{\psi_-} + E_{\phi_-} + \cos^2\psi \cos^2\phi(U_{\text{eff}}), \tag{S16}$$

where the last (approximate) equation follows from the fact that Coulomb repulsion is only paid for the charge located in the copper orbitals ($\cos^2\psi \cos^2\phi$) and that we define the Coulomb repulsion $U_{\text{eff}}$ to be:

$$U_{\text{eff}} = U - 3J_H \tag{S17}$$

for the NNN orbital exchange (the NNN spins are parallel in the AFM ordered state) and

$$U_{\text{eff}} = U - 2J_H \tag{S18}$$

for the NN orbital exchange (the NN spins are antiparaller in the AFM ordered state).

### S4.4 Next-nearest neighbor orbital superexchange in the cell perturbation theory

Having obtained the proper cell perturbation theory basis, we are now ready to derive the next-nearest neighbor (NNN) orbital superexchange. This is done using second order perturbation theory and assuming that the initial state is

$$|\text{I}\rangle = |\phi_{-,I}\rangle|\psi_{-,II}\rangle, \tag{S19}$$

the intermediate is

$$|\text{INT}\rangle = |\text{VAC}_I\rangle|\phi_{-,II}\rangle|\psi_{-,II}\rangle, \tag{S20}$$

and the final state is

$$|\text{F}\rangle = |\psi_{-,I}\rangle|\phi_{-,II}\rangle, \tag{S21}$$

where the indices $I$ and $II$ denote the two distinct NNN sites. We obtain:

$$\begin{aligned}
J_{\text{NNN}}^{\text{orb}} &= -4\frac{\langle \text{F}|\mathcal{H}_{\text{kin}}^{(\sqrt{2})}|\text{INT}\rangle\langle \text{INT}|\mathcal{H}_{\text{kin}}^{(\sqrt{2})}|\text{I}\rangle}{E_{\text{INT}} - E_I} = -4\frac{\langle \phi_{-,II}|\langle\psi_{-,I}|\mathcal{H}_{\text{kin}}^{(\sqrt{2})}|\text{INT}\rangle\langle \text{INT}|\mathcal{H}_{\text{kin}}^{(\sqrt{2})}|\phi_{-,I}\rangle|\psi_{-,II}\rangle}{\cos^2\psi\cos^2\phi(U - 3J_H)} \\
&= 4\frac{\langle\psi_{-,I}|\mathcal{H}_{\text{kin}}^{(\sqrt{2})}|\psi_{-,II}\rangle\langle\phi_{-,II}|\mathcal{H}_{\text{kin}}^{(\sqrt{2})}|\phi_{-,I}\rangle}{\cos^2\psi\cos^2\phi(U - 3J_H)} = 4\frac{\sin^2\psi T_{A\sigma}^{(\sqrt{2})}\sin^2\phi T_{B\pi}^{(\sqrt{2})}}{\cos^2\psi\cos^2\phi(U - 3J_H)} \\
&= -\frac{16\nu(\sqrt{2})^2 t_{pp}^2 \sin^2\psi\sin^2\phi}{\cos^2\psi\cos^2\phi(U - 3J_H)}.
\end{aligned} \tag{S22}$$

Note that this exchange 'around the corner' is allowed due to the finite $t_{pp}$, which leads to the formation of the bonding (B) and antibonding (A) states of the $\pi$ and $\sigma$ orbitals and then their hybridisation on the NNN plaquettes. Besides, the factor of four in the formula above stems from the fact that: (i) there is another superexchange process with the doubly occupied site $I$ in the intermediate state ($|\text{INT}\rangle = |\phi_{-,I}\rangle|\psi_{-,I}\rangle|\text{VAC}_{II}\rangle$) which gives rise to one factor of two and (ii) the superexchange constant is implicitly multiplying the orbital pseudospin raising and lowering with a prefactor of $1/2$ (hence another factor of two).

The NNN orbital superexchange is always negative and, for the realistic parameters of the charge transfer model for CCO of Table 1, we obtain

$$J_{\text{NNN}}^{\text{orb}} \approx -0.015 \text{ eV}. \tag{S23}$$

.

Note that from the above expression for the NNN orbital exchange we can easily extract the NNN hoppings of the ground state $|\psi_-\rangle$

$$t'_a = -2\nu(\sqrt{2})\sin(\psi)^2 t_{pp} \approx -0.072 \text{ eV}, \tag{S24}$$

and excited orbital $|\phi_-\rangle$

$$t'_b = 2\nu(\sqrt{2})\sin(\phi)^2 t_{pp} \approx 0.078 \text{ eV}. \tag{S25}$$

Note that in the simplified approach presented in the main text of the paper the $t'_a$ and $t'_b$ hoppings are depicted in the cartoon Fig. 3.

### S4.5 Nearest neighbor orbital superexchange in the cell perturbation theory

A similar (though a bit more lengthy) derivation as in the subsection above leads to the NN orbital superexchange:

$$J_{\text{NN}}^{\text{orb}} = -4 \frac{\langle F|\mathcal{H}_{\text{kin}}^{(1)}|\text{INT}\rangle\langle\text{INT}|\mathcal{H}_{\text{kin}}^{(1)}|I\rangle}{E_{\text{INT}} - E_I} =$$
$$= \frac{16[-\nu(1)\sin^2\psi\ t_{pp} + 2\mu(1)\sin\psi\cos\psi\ t_{pd\sigma}][\nu(1)\sin^2\phi\ t_{pp} - 2\mu(1)\sin\phi\cos\phi\ t_{pd\pi}]}{\cos^2\psi\cos^2\phi(U - 2J_H)}. \tag{S26}$$

Interestingly, this exchange is *positive* for realistic parameters of the charge transfer model of CCO in Table 1—in this case we obtain

$$J_{\text{NN}}^{\text{orb}} \approx 0.022 \text{ eV}. \tag{S27}$$

The fact that it is positive is important, since this means that the small, but effectively free orbiton motion originating from the NN orbital exchange *and* finite Hund's exchange, can lead to the dispersion relation along the 'downwards' direction (i.e. opposite as the fully free NNN orbiton exchange).

Note that (just as in the previous subsection) from the above expression for the NN orbital exchange we can easily extract the NN hoppings of the ground state $|\psi_-\rangle$

$$t_a = 2[-\nu(1)\sin(\psi)^2\ t_{pp} + 2\mu(1)\sin\psi\cos\psi\ t_{pd\sigma}] \approx 0.52 \text{ eV}, \tag{S28}$$

and excited orbital $|\phi_-\rangle$

$$t_b = 2[\nu(1)\sin(\phi)^2\ t_{pp} - 2\mu(1)\sin\phi\cos\phi\ t_{pd\pi}] \approx 0.020 \text{ eV}. \tag{S29}$$

### S4.6 Orbiton dispersion and the magnetic string effect

The NNN orbital superexchange leads to the free orbiton hopping, not frustrating the AFM order, see below. This leads to the following orbiton dispersion

$$\varepsilon_{\mathbf{k}} = 2J_{\text{NNN}}^{\text{orb}}\cos k_x \cos k_y. \tag{S30}$$

Such a dispersion gives rise to the bandwidth, defined as the energy difference between $(\pi, 0)$ and $(0, 0)$ momenta, $W_{\text{orb}} = 4J_{\text{NNN}}^{\text{orb}}$.

Naturally, besides the NNN orbital superexchange, one should consider the contribution of the NN orbital superexchange to the orbiton propagation. However, the latter one frustrates the AFM order (due to the AFM spin exchange $J_{\text{NN}}^{\text{spin}} \approx 160$ meV [17]) and in the first order of approximation can be neglected, see below.

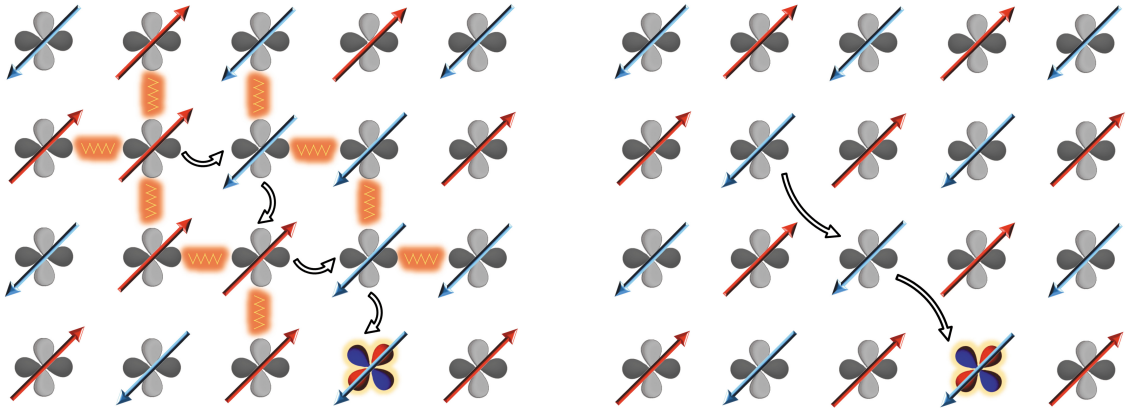

Figure S5: Orbiton propagation in the 2D AFM via the orbital superexchange: the nearest neighbor (NN) exchange leading to the onset of defects in the AFM background and the magnetic string effect (left panel); the next nearest neighbor (NNN) exchange leading to a lack of magnetic defects and the free orbiton propagation (right panel). See text for further details.

| $|\sin\psi|^2$ | $|\sin\phi|^2$ | $E_{\phi_-} - E_{\psi_-}$ | $W_{\text{orb}}$ |
|---|---|---|---|
| 0.42 | 0.46 | 1.89 | 0.058 |

Table 2: The numerical results of the calculations, assuming the charge transfer parameters from Tab. 1, that can be compared with their experimental counterparts: the oxygen content in the CCO ground state $|\sin\psi|^2$, the oxygen content in the $xy$ orbital excitation $|\sin\phi|^2$, the energy of the $xy$ orbital excitations measured by RIXS $E_{\phi_-} - E_{\psi_-}$ in eV, the bandwidth of the orbiton excitation, measured as the difference between energy at $(\pi, 0)$ and $(0, 0)$ momenta $W_{\text{orb}}$, in eV.

There is a very simple and intuitive way to understand the qualitative differences between orbiton propagation with NN and NNN superexchanges, see Fig. S5. When one considers the NN orbital exchange, subsequent hoppings of the $xy$ excitation leaves a string of misaligned spins. They are represented in the figure as glowing orange lines. Since each orbiton hop costs magnetic energy, the orbital excitation tends to be localized by the magnon-orbiton interaction. With the NNN orbital exchange, instead, the orbiton motion costs no magnetic energy since the superexchange happens between atoms belonging to the same magnetic sublattice. This effect is characteristic of 2D and 3D AFM ordered systems: it is absent in 1D AFM compounds [18].

### S4.7 Comparison with the experimental results for CCO (and NCO)

The primary result of the calculation is that the orbiton dispersion, which goes 'upwards' from the $\Gamma$ point both along the nodal and antinodal directions, is in qualitative agreement with the experiment. Note that at the quantitative level the size of the numerical value of the orbiton bandwidth (for the realistic charge transfer model parameters of Tab. 1) agrees pretty well with the experimentally observed value of 50-60 meV (in CCO and NCO), cf. $W_{\text{orb}}$ Tab. 2. Besides, we also observe that the on-site energy cost of the $xy$ orbital excitation in RIXS (which amounts to about 1.6 eV in CCO) is off by about 20% w.r.t. the obtained here numerical value $E_{\phi_-} - E_{\psi_-}$, cf. Tab. 2. This is a surprisingly good estimate, given the relative simplicity of the model (note that the quantum chemistry calculations typically give similar agreement with the experimental result [19]). Finally, the obtained oxygen content of the ground and excited state ($|\sin\psi|^2$ and $|\sin\phi|^2$) favourably compares against e.g. the values typically observed for the cuprates in XAS, cf. the 40% 'covalency' aspect ratio suggested by [20].

## S4.8 Effect of the apical oxygens and lack od orbiton dispersion in LCO

Finally, one has to check whether the model is capable of rationalizing the very different behaviour measured in LCO and in the infinite-layer compounds. The key lies in understanding what happens to the physics of the system when the apical oxygens are included. Then, first of all the oxygen content both in the $\sigma$ and $\pi$ bonding orbitals *in the plane* is reduced in LCO more than in CCO. This happens primarily because the charge-transfer energy increases in LCO. Ref. [9] clearly shows that the charge transfer energy $\Delta$ decreases with an increase in the apical oxygen distance (see Fig. 2 of [9])—and is, thus, even further reduced once the apical oxygens are absent. Altogether, this leads to overall less charge participating in the orbital superexchange and thus lowers the value of the exchange constants in the presence of apical oxygens. Moreover, the work of [13] suggests that the dominant contribution to the $t_{pp}$ hopping comes from a second-order process which involves virtual occupancy of the copper $4s$ orbital at energy $\Delta_{ps}$ (thus $t_{pp} \sim t_{ps}^2/\Delta_{ps}$ where $t_{ps}$ is the direct oxygen 2p and copper 4s orbital). Since the hole on-site energy in the $4s$ orbital goes up when the apical oxygens are present, we can speculate that the oxygen-oxygen hopping $t_{pp}$ *should* also be reduced once the apical oxygens are included (albeit [9] does not seem to suggest that this indeed happens). This would reduce even more the orbiton bandwidth $W_{\text{orb}}$.

Since small changes in any of the above parameters hugely influence the NNN orbiton superexchange, the orbiton dispersion may get significantly reduced in LCO. In fact, just changing the charge transfer energy $\Delta$ from 1.8 eV (the value suggested for CCO or NCO, see above) to 2.6 eV (which is the value commonly accepted for LCO [9, 21]) reduces the orbiton bandwidth in LCO to $W_{\text{orb}} \approx 24$ meV, i.e. it is almost two-and-a-half times smaller as the one calculated for CCO. Furthermore, this number can be even further reduced once the expected reduction of $t_{pp}$ in LCO is taken into account. In particular, assuming $t_{pp} = 0.5$ eV (so 30% smaller than the assumed value for CCO) gives $W_{\text{orb}} \sim 13$ meV for LCO, three times smaller than the experimental resolution and well below the level of sensitivity of our current instrumentation.

## S4.9 Validity of the orbital superexchange model: exact diagonalisation of an effective two-orbital Hubbard model

We verified the validity of the superexchange dispersion by comparing it to numerical spectra obtained for an effective two-orbital model modelling the processes illustrated in Fig. 2 of the main text. The models kinetic energy is

$$H_{\text{kin}} = t_a \sum_{\langle i,j \rangle, \sigma} c^\dagger_{i,\sigma,a} c_{j,\sigma,a} + t_b \sum_{\langle i,j \rangle, \sigma} c^\dagger_{i,\sigma,b} c_{j,\sigma,b} + \\ + t'_a \sum_{\langle\langle i,j \rangle\rangle, \sigma} c^\dagger_{i,\sigma,A} c_{j,\sigma,A} + t'_b \sum_{\langle\langle i,j \rangle\rangle, \sigma} c^\dagger_{i,\sigma,B} c_{j,\sigma,B} \tag{S31}$$

where $c^\dagger_{i,\sigma,\alpha}$ ($c_{i,\sigma,\alpha}$) creates (annhiliates) a hole on site $i$, with spin $\sigma$ and in orbital $\alpha = a,b$. Orbital $a$ has $x^2 - y^2$ symmetry and corresponds to the $|\psi_-\rangle$ introduced above, orbital $b$ represents the $|\phi_-\rangle$ state with $xy$ symmetry discussed above. Hopping parameters for nearest-neighbor pairs $\langle i,j \rangle$ and next-nearest-neighbor pairs $\langle\langle i,j \rangle\rangle$ follow from the effective values derived with cell-perturbation theory above [cf. (S24-S25) and (S28-S29)]: $t_{\text{NN},a} = -0.51$ eV and $t_{\text{NN},b} = 0.03$ eV, $t'_a = -0.069$ eV, and $t'_b = 0.072$ eV. The two NNN hoppings are necessarily of opposite sign, see main text.

In the ground state, the $B$ state (i.e. $|\phi_-\rangle$) is empty, as its energy is raised by the crystal field

$$H_{\text{CF}} = \Delta_b \sum_{i,\sigma} n_{i,\sigma,b} = \Delta_b \sum_{i,\sigma} c^\dagger_{i,\sigma,b} c_{i,\sigma,b}, \tag{S32}$$

with $\Delta_b = 1.9$ eV. The $a$ (i.e. $|\psi_-\rangle$) state is half-filled. Onsite Coulomb interactions are

$$H_{\text{int}} = U \sum_{i,\alpha=a,b} n_{i\alpha\uparrow} n_{i\alpha\downarrow} + \frac{U'}{2} \sum_{i,\sigma} \sum_{\alpha \neq \beta} n_{i\alpha\sigma} n_{i\beta\bar{\sigma}} + \frac{1}{2}(U' - J_H) \sum_{i,\sigma} \sum_{\alpha \neq \beta} n_{i\alpha\sigma} n_{i\beta\sigma} \\ - J_H \sum_{i,\alpha \neq \beta} (c^\dagger_{i\alpha\uparrow} c_{i\alpha\downarrow} c^\dagger_{i,\beta\downarrow} c_{i\beta\uparrow}) + J_H \sum_{i,\alpha \neq \beta} (c^\dagger_{i\alpha\uparrow} c^\dagger_{i\alpha\downarrow} c_{i\beta\downarrow} c_{i\beta\uparrow}) \tag{S33}$$

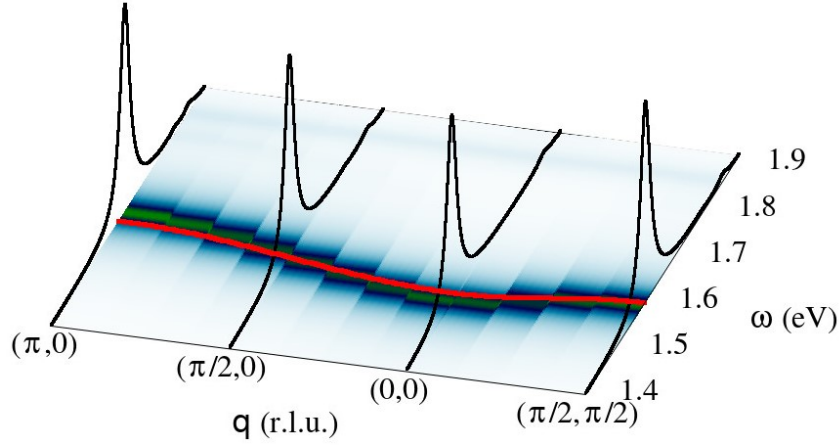

Figure S6: Spectral function obtained by the ED calculations at different momenta. The profile is reported in black solid curves. The asymmetric lineshape is barely visible, though a small almost non-dispersive satellite peak at energy higher by ~200 meV w.r.t. the main peak is relatively well-pronounced. Also shown is the SE model dispersion, reported with the solid red line.

where $U' = U - 2J_H$ and effective $U = 2.4$ eV and Hund's-rule coupling $J_H = 0.34$ eV are relatively strong for two electrons (with opposite spin) in the same orbital. Such states are thus treated in second-order perturbation theory, i.e., the ground state corresponds to that of a $t$-$J$ model, where coupling $J_{NN} = 160$ meV of the effective spin-spin coupling

$$H_{\text{Heis}} = J_{NN} \sum_{\langle i,j \rangle, \sigma} \mathbf{S}_{i,a} \mathbf{S}_{j,a} \tag{S34}$$

was fitted to experiment.

An orbital excitation moves one particle from the lower into the upper orbital. Such states where two holes are on the same site, but in different orbitals, are not projected out, but are kept in the model's Hilbert space. The model can thus in principle go beyond the superexchange process discussed in the main text and can e.g. include the impact of Hund's-rule coupling in more detail. The good agreement between numerical results and superexchange indicate, however, that the latter provides an excellent approximation. The orbital excitation can additionally move via analogous NN processes (and these are automatically included in the model). However, their impact onto orbiton motion turns out to be small, which is explained above by the fact that they disturb the spin background.

The orbital Green's function

$$\langle GS | T^\dagger(\mathbf{k}) \frac{1}{\omega - (H - E_0)} T(\mathbf{k}) | GS \rangle \tag{S35}$$

is obtained using the Lanczos algorithm for a cluster with $4 \times 4$ sites. Operator

$$T(\mathbf{k}) = \sum_j \exp[-i\mathbf{k}\mathbf{r}_j] c^\dagger_{j,\sigma,b} c_{j,\sigma',a} \tag{S36}$$

with $c^\dagger_{j,\sigma,\alpha}$ ($c_{j,\sigma,\alpha}$) being the creation (annihilation) operator for a particle in orbital $\alpha$ and spin $\sigma$ on site $j$. $\mathbf{k}$ are the momenta accessible to a $4 \times 4$ cluster. The excitation can arise without a spin flip ($\sigma = \sigma'$) or with a spin flip ($\sigma = -\sigma'$). While the associated energies slightly differ due to Hund's-rule coupling, the splitting is small, so that only the excitation without flip is plotted in the main text.

## S5 Discussion on non-dispersing $xy$ feature

To explain the presence of the non-dispersing $xy$ feature, we propose two possible scenarios (with one more probable than the other one, with the future experimental studies with better resolution possibly

being able to discriminate between the two): the $xy$ satellite arises because either i) a coupling of orbitons to magnetic excitations, or ii) coupling of orbitons to phonon modes.

Before we have a closer look at these two possible scenarios, let us make a small 'detour' and make the point that, disregarding of the type of excitation to which the orbitons couples (either magnons or phonons), the resulting satellite peak would *not* show any appreciable dispersion. A first guess here would be that the satellite peak, which is composed from an orbiton coupled to phonon or magnon, would 'inherit' a dispersion from the first peak (i.e. the orbiton). Nevertheless, the existing research as well as calculations that we have performed (see below) show the opposite – the inherited dispersion should be strongly reduced or nonexistent at all. The reason for this is as follows:

- First, if the bosons to which the quasiparticle (=orbiton) couples to are dispersionless, then the satellite peak would not show any dispersion— even if the (main) quasiparticle peak is strongly dispersive. This is best visible in the recent RIXS studies on the phonon response, see Fig. 1(a) of [22]. Such a result can be rationalised by noting that the quasiparticle becomes immobile when coupled to a massive local excitation (this is the 'string effect' which gives rise to the onset of a ladder spectrum, cf. [23, 24]).

- Second, if a realistic phonon or magnon dispersion is included, then still the acquired dispersion of the satellite peak should be very small. For instance, in the case of phonons, this is visible in the realistic calculations of the case when itinerant (so highly mobile) electrons couple to several distinct phonon modes in the cuprates [25]. Ultimately, coupling to each phonon introduces an extra peak in the spectrum which has some dispersion—however, its bandwidth is (for realistic cuprate parameters) of the order of 10 meV (see Fig. 10 of [25]) and hence below the experimental resolution of the RIXS experiment. We stress that we have verified that a similar situation happens if we assume that the satellite peak is of magnetic origin, as shown in Fig.S6.

A first explanation of the satellite peak, which perhaps would seem more natural considering the nature of the orbiton and its strong coupling to magnon, would be that it is of magnetic origin. In other words, the satellite peak arises due to the coupling between orbiton and magnons which in principle always leads to the onset of the whole set of satellite peaks, the so-called ladder spectrum [23, 24]. However, as the spin quantum fluctuations are strong in the quasi-2D quantum antiferromagnet, in the end of the day only the first satellite peak should be visible in such a case, see Fig. 1(a) of [26]. Indeed, as our ED calculations show (Fig.S6) also in our model there is just one satellite peak visible in the broad incoherent orbiton-magnon spectrum. We also note that the peak is located about 80 meV above the main orbiton peak which well agrees with the experimental observation. Moreover, from the theory point of view such a peak should also be visible even for the non-dispersive orbiton, for the its origin comes from a different type of spin-orbital exchange than the free dispersion. The latter statement also agrees with the experiment, since LCO also shows a satellite peak (as evident e.g. from Fig. S2). Nevertheless, the theoretically obtained ratio of intensities of the satellite and main peaks do not match the experimental ones: in the former case this ratio is equal to 10:1, while in the latter is closer to 1:1. Such a strong discrepancy may partially be explained by the use of small cluster in the calculations (which from our experience always substantially overestimate the intensity of quasiparticle peaks and suppress the continuum), yet it is unclear whether this could explain this large mismatch between theoretical and experimental intensities.

The explanation that we propose as more likely is that the satellite peak arises from coupling to phonons. This is highly probable, since both theory [25, 27] and experimental data show a phonon mode at this energy range in RIXS (see e.g. Fig. S7). The strong coupling between *dd* excitations and phonons in copper oxides is well-known from optical spectroscopy studies, that reveal the clear presence of phonon-assisted orbital transitions in the experimental spectra (see e.g. [28]). Besides, typically such a phonon satellite would show weak dispersion (below the RIXS resolution, see discussion in the previous answer), has rather strong intensity w.r.t. the electronic-only part [25], and is mostly dominated by the first phonon peak (with the coupling to more phonons contributing to peaks of smaller intensity) [27]. That is why we tend to believe that the satellite peak is of phonon origin. Nevertheless, further experimental studies with better resolution should confirm this hypothesis.

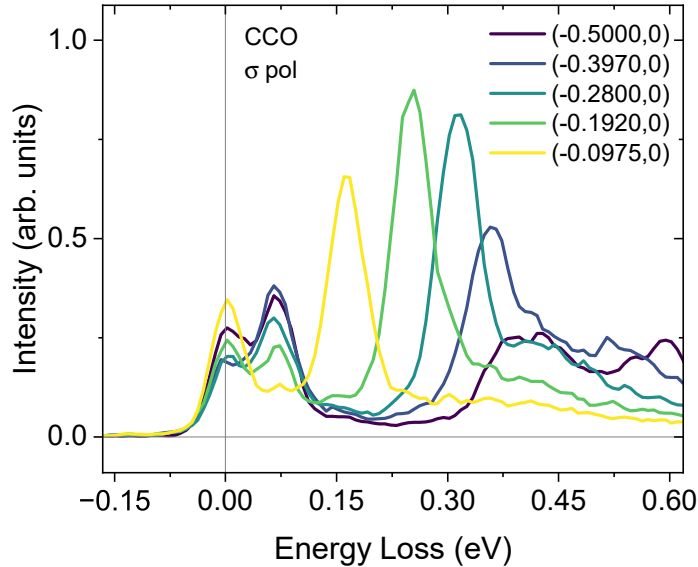

Figure S7: Low-energy RIXS spectra of CCO, measured with incident $\sigma$ polarization and grazing-incidence geometry. Evident is a strong phonon mode around 75 meV.

# References

[1] I. Bozovic, G. Logvenov, et al., "Epitaxial Strain and Superconductivity in $La_{2-x}Sr_xCuO_4$ Thin Films", Phys. Rev. Lett. **89**, 107001 (2002).

[2] D. D. Castro, C. Aruta, et al., "$T_c$ up to 50 K in superlattices of insulating oxides", Superconductor Science and Technology **27**, 044016 (2014).

[3] D. Di Castro, C. Cantoni, et al., "High-$T_c$ Superconductivity at the Interface between the $CaCuO_2$ and $SrTiO_3$ Insulating Oxides", Phys. Rev. Lett. **115**, 147001 (2015).

[4] Y. Krockenberger, H. Irie, et al., "Emerging superconductivity hidden beneath charge-transfer insulators", Scientific Reports **3**, 2235 (2013).

[5] Y. Krockenberger, H. Yamamoto, et al., "Unconventional transport and superconducting properties in electron-doped cuprates", Phys. Rev. B **85**, 184502 (2012).

[6] R. Fumagalli, L. Braicovich, et al., "Polarization-resolved Cu l3-edge resonant inelastic x-ray scattering of orbital and spin excitations in $NdBa_2Cu_3O_{7-\delta}$", Phys. Rev. B **99**, 134517 (2019).

[7] M. M. Sala, V. Bisogni, et al., "Energy and symmetry of dd excitations in undoped layered cuprates measured by Cu L3 resonant inelastic x-ray scattering", New Journal of Physics **13**, 043026 (2011).

[8] K. Wohlfeld, M. Daghofer, et al., "Dispersion of orbital excitations in 2d quantum antiferromagnets", Journal of Physics: Conference Series **391**, 012168 (2012).

[9] C. Weber, C. Yee, et al., "Scaling of the transition temperature of hole-doped cuprate superconductors with the charge-transfer energy", Europhysics Letters **100**, 37001 (2012).

[10] D. Kan, A. Yamanaka, et al., "Preparation and optical properties of single-crystalline CaCuO2 thin films with infinite layer structure", Physica C: Superconductivity **412-414**, Proceedings of the 16th International Symposium on Superconductivity (ISS 2003). Advances in Superconductivity XVI. Part I, 298–302 (2004).

[11] M. Yoshida, S. Tajima, et al., "Two-magnon and two-phonon excitations in some parent insulating compounds of the high-$T_c$ cuprates", Phys. Rev. B **46**, 6505–6510 (1992).

[12] R. Neudert, S.-L. Drechsler, et al., "Four-band extended hubbard hamiltonian for the one-dimensional cuprate $Sr_2CuO_3$: distribution of oxygen holes and its relation to strong intersite coulomb interaction", Phys. Rev. B **62**, 10752–10765 (2000).

[13] N. Barišić and D. K. Sunko, "High-$T_c$ Cuprates: a Story of Two Electronic Subsystems", Journal of Superconductivity and Novel Magnetism, 1–19 (2022).

[14] E. Pavarini, I. Dasgupta, et al., "Band-structure trend in hole-doped cuprates and correlation with $T_{c\max}$", Phys. Rev. Lett. **87**, 047003 (2001).

[15] A. S. Botana and M. R. Norman, "Similarities and differences between $LaNiO_2$ and $CaCuO_2$ and implications for superconductivity", Phys. Rev. X **10**, 011024 (2020).

[16] J. H. Jefferson, H. Eskes, and L. F. Feiner, "Derivation of a single-band model for $CuO_2$ planes by a cell-perturbation method", Phys. Rev. B **45**, 7959–7972 (1992).

[17] L. Martinelli, D. Betto, et al., "Fractional spin excitations in the infinite-layer cuprate $CaCuO_2$", Phys. Rev. X **12**, 021041 (2022).

[18] K. Wohlfeld, M. Daghofer, et al., "Intrinsic coupling of orbital excitations to spin fluctuations in mott insulators", Phys. Rev. Lett. **107**, 147201 (2011).

[19] J. Schlappa, K. Wohlfeld, et al., "Spin–orbital separation in the quasi-one-dimensional Mott insulator $Sr_2CuO_3$", Nature **485**, 82–85 (2012).

[20] T. K. Kim, H. Rosner, et al., "Unusual electronic structure of the pseudoladder compound $CaCu_2O_3$", Phys. Rev. B **67**, 024516 (2003).

[21] H. Watanabe, T. Shirakawa, et al., "Unified description of cuprate superconductors using a four-band $d$–$p$ model", Phys. Rev. Res. **3**, 033157 (2021).

[22] K. Bieniasz, S. Johnston, and M. Berciu, "Theory of dispersive optical phonons in resonant inelastic x-ray scattering experiments", Phys. Rev. B **105**, L180302 (2022).

[23] L. N. Bulaevskii, E. L. Nagaev, and D. I. Khomskii, "A New Type of Auto-localized State of a Conduction Electron in an Antiferromagnetic Semiconductor", JETP **27**, 836 (1968).

[24] G. Martínez and P. Horsch, "Spin polarons in the $t$-$J$ model", Phys. Rev. B **44**, 317–331 (1991).

[25] T. P. Devereaux, A. M. Shvaika, et al., "Directly characterizing the relative strength and momentum dependence of electron-phonon coupling using resonant inelastic x-ray scattering", Phys. Rev. X **6**, 041019 (2016).

[26] A. Bohrdt, E. Demler, et al., "Parton theory of angle-resolved photoemission spectroscopy spectra in antiferromagnetic mott insulators", Phys. Rev. B **102**, 035139 (2020).

[27] W. S. Lee, S. Johnston, et al., "Role of lattice coupling in establishing electronic and magnetic properties in quasi-one-dimensional cuprates", Phys. Rev. Lett. **110**, 265502 (2013).

[28] A. Marciniak, S. Marcantoni, et al., "Vibrational coherent control of localized d–d electronic excitation", Nature Physics **17**, 368–373 (2021).



\end{document}